\documentclass[english]{article}%
\usepackage{amssymb}
\usepackage{fontenc}
\usepackage[latin1]{inputenc}
\usepackage{amsmath}
\usepackage{fontenc}
\usepackage{babel}
\usepackage{amsfonts}
\usepackage{graphicx}%
\providecommand{\U}[1]{\protect\rule{.1in}{.1in}}
 
\begin{document}

\title{{\Large Coherent and semiclassical states in magnetic field in the presence of
the Aharonov-Bohm solenoid}}
\author{V. G. Bagrov\thanks{Department of Physics, Tomsk State University, 634050,
Tomsk, Russia. Tomsk Institute of High Current Electronics, SB RAS, 634034
Tomsk, Russia; e-mail: bagrov@phys.tsu.ru} , S. P. Gavrilov\thanks{Institute
of Physics, University of São Paulo, Brazil; On leave from Department of
General and Experimental Physics, Herzen State Pedagogical University of
Russia, Moyka emb. 48, 191186 St. Petersburg, Russia; e-mail:
gavrilovsergeyp@yahoo.com}, D. M. Gitman\thanks{Institute of Physics,
University of São Paulo, CP 66318, CEP 05315-970 São Paulo, SP, Brazil;
e-mail: gitman@dfn.if.usp.br}, D. P. Meira Filho\thanks{Institute of Physics,
University of São Paulo, Brazil; e-mail: dmeira@dfn.if.usp.br}}
\maketitle

\begin{abstract}
A new approach to constructing coherent states (CS) and semiclassical states
(SS) in magnetic-solenoid field is proposed. The main idea is based on the
fact that the AB solenoid breaks the translational symmetry in the $xy$-plane,
this has a topological effect such that there appear two types of trajectories
which embrace and do not embrace the solenoid. Due to this fact, one has to
construct two different kinds of CS/SS, which correspond to such trajectories
in the semiclassical limit. Following this idea, we construct CS in two steps,
first the instantaneous CS (ICS) and the time dependent CS/SS as an evolution
of the ICS. The construction is realized for nonrelativistic and relativistic
spinning particles both in ($2+1)$- and ($3+1)$- dimensions and gives a
non-trivial example of SS/CS for systems with a nonquadratic Hamiltonian. It
is stressed that CS depending on their parameters (quantum numbers) describe
both pure quantum and semiclassical states. An analysis is represented that
classifies parameters of the CS in such respect. Such a classification is used
for the semiclassical decompositions of various physical quantities.


\end{abstract}

\section{Introduction\label{S.1}}

Quantum interaction of charged particles with the field of an infinitely long
and infinitesimally thin magnetic solenoid (further Aharonov-Bohm (AB) field)
was studied theoretically and experimentally already for a long time. In spite
of the fact that particle wave functions vanish on the solenoid line, the
particles feel the presence of AB solenoid \cite{AhaBo59}. This phenomenon is
called the AB effect and is interpreted as a possibility for\textbf{\ }locally
trivial vector potentials to gives rise to observable effects in a nontrivial
topology. A number of theoretical works and convinced experiments were done to
clarify AB effect and to prove its existence. By the middle of the 80's the
AB-effect in low energy physics was becoming a good instrument for
investigating new physical phenomena, principally in condensed matter physics,
where the AB ring has been the mainstay of mesoscopic physics research since
its inception, see \cite{OlaPo85} for a general review. It was discovered that
the effect is relevant to a number of physical problems, e.g. to anyons in
high -$T_{c}$ superconductivity \cite{CWWH89}, electronic excitations in
graphene with topological defects \cite{SitV08,JackMPT09}, nanotubes
\cite{nanotube-review2007}, nonrelativistic scattering in Chern-Simons theory
\cite{BFP95}, theory of unparticles \cite{Kob07}, and so on. AB vacuum
polarization and AB radiation are relevant to cosmic string dynamics, see for
example \cite{AMW89,AW89}.

A splitting of Landau levels in a superposition of the AB field and a parallel
uniform magnetic field gives an example of the AB effect for bound states. In
what follows, we call such a superposition the magnetic-solenoid field (MSF).
Solutions of the Schrödinger equation with MSF were first studied in
\cite{Lewis83}. Solutions of relativistic wave equations (Klein-Gordon and
Dirac ones) with MSF were first obtained in \cite{183} and then used in
\cite{192} to study AB effect in cyclotron and synchrotron radiations. On the
basis of these solutions, Green functions and the problem of the
self-adjointness of Dirac Hamiltonians with MSF was studied
\cite{FP01,205,212,ExnSV02,208}. A complete spectral analysis for all the
self-adjoint nonrelativistic and relativistic Hamiltonians with MSF was
performed in \cite{GitTySV09}. Recently, the interest to MSF (and related
multivortex examples) has been renewed in connection with planar physics
problems and quantum Hall effect \cite{N00}. It is important to stress that in
contrast to the pure AB field case, where particles interact with the solenoid
for a finite short time, moving in MSF the particles interact with solenoid
permanently. This opens up more possibilities to study such an interaction and
correspond a number of real physical situations. For example, recent
fabrication of a graphene allows one an experimental observation and
application of effects with relativistic spinning particles under usual
laboratory conditions \cite{CNetetalRevMP2009}.

In some cases, it is enough and, moreover, more adequate (and convenient) to
use a semiclassical description of a physical system. Semiclassical states
(SS) of the system provide such a description. Usually, such states are
identified with different kinds of the so-called coherent states (CS).
However, such a formal identification is known for systems with quadratic
Hamiltonians, in other case both SS and CS construction and their
identification is problematic. In addition to well-known applications of SS/CS
in quantum theory \cite{CSQT72}, there appear recently new important
applications to quantum computations, see, e.g. \cite{quantcomput}.
Constructing SS/CS for particles in the AB field and in MSF is a nontrivial
problem (which was an open problem until present), in particular, due to the
nonquadratic structure of particle Hamiltonians with such fields. Besides of
numerous possible practical applications constructing such states gives an
important example of SS/CS for nonquadratic Hamiltonians and may, for example,
answer an important theoretical question: to what extend the AB effect is of a
pure quantum nature. In a sense constructing SS/CS is a complimentary task to
the path integral construction, which also is an open problem in the case of
the particle in the MSF. One can suppose that SS/CS in MSF are in a sense
analogous to ones in pure magnetic field. In the latter case SS are identified
with CS that are well-known, see e.g. \cite{MalMa68,BagGi90}. In such CS, mean
values of particle coordinates move along classical trajectories. The latter
trajectories are circles whose radii and center position (quantum numbers)
label these quantum CS. Constructing CS for particles in MSF, we will try to
maintain basic properties of already known CS for quadratic systems. In
particular, such CS have to minimize uncertainty relations for some physical
quantities (e.g. coordinates and momenta) at a fixed time instant and means of
particle coordinates, calculated with respect to time-dependent CS, have to
move along the corresponding classical trajectories. In addition, CS have to
be labeled by quantum numbers that have a direct classical analog, let say by
phase-space coordinates. It is also desirable for time-dependent CS to
maintain their form under the time evolution.

It should be mentioned that some attempts to construct SS/CS for particles in
the MSF are presented in the works \cite{CS-old}. However, the states
constructed there do not obey the principle requirement for SS/CS, the
corresponding means do not move along classical trajectories. In our recent
article \cite{265}, we succeeded to construct a principally different kind of
CS for nonrelativistic spinless particles in the MSF. In the present work we
extend this construction to the case of relativistic spinless and
nonrelativistic and relativistic spinning particles both in $3+1$- and $2+1$-
dimensions (dim.). In addition, developing a semiclassical approximation
techniques, we constructed SS in MSF on the base of the CS. The progress is
related to a nontrivial observation that in the problem under consideration
there are two kind of SS/CS those which correspond classical trajectories
which embrace the solenoid and those which do not. It should be stressed that
the identification of SS and CS in MSF depends essentially on quantum numbers
that label these states (on types and positions of the corresponding classical
trajectories). Particles in constructed SS/CS move along classical
trajectories, the states maintain their form under the time evolution, and
form a complete set of functions, which can be useful in semiclassical
calculations. In the absence of the AB field these states are reduced to the
well known in the case of a uniform magnetic field Malkin-Man'ko CS
\cite{MalMa68}. The constructed states give a non-trivial example of SS/CS for
systems with a nonquadratic Hamiltonian. In addition, they allow one to treat
the AB effect on the classical language, revealing an influence of AB field on
parameters of classical trajectories in magnetic field. It should be noted
that quantum motion of spin $1/2$ Dirac fermions is qualitative different in
$3+1$- and in $2+1$- dim. Since Dirac fermions in $2+1$- dim. (in particular
massless ones) describe single-electron dynamics in graphene, we have devoted
a part of our study to their SS and CS in MSF.

The article is organized as follows. In sec. \ref{S2}, we start our
consideration with classical description of particle motion in the MSF. We
recall that MSF is a collinear superposition of a constant uniform magnetic
field of strength $B$ and the AB field (field of an infinitely long and
infinitesimally thin solenoid with a finite constant internal magnetic flux
$\Phi$). Setting the $z$ axis along the AB solenoid, the MSF strength takes
the form $\mathbf{B}=\left(  0,0,B_{z}\right)  ,$%
\begin{equation}
B_{z}=B+\Phi\delta\left(  x\right)  \delta\left(  y\right)  =B+\frac{\Phi}{\pi
r}\delta\left(  r\right)  ,\,B=\mathrm{const},\,\Phi=\mathrm{const}..
\label{a1}%
\end{equation}
We use the following electromagnetic potentials\footnote{We accept the
following notations for four- and three-vectors: $a=\left(  a^{\mu}%
,\mu=0,i\right)  =\left(  a^{0},\mathbf{a}\right)  $,$\ $
\par
$\mathbf{a}=\left(  a^{i},i=1,2,3\right)  \mathbf{=}\left(  a^{1}=a_{x}%
,a^{2}=a_{y},a^{3}=a_{z}\right)  $,$\ a_{i}=-a^{i}$, in particular, for the
space-time coordinates: $x^{\mu}=(x^{0}=ct,\ x^{1}=x,\ x^{2}=y,\ x^{3}=z)$, as
well as cylindrical coordinates $r,\varphi$, in the $xy$ plane, such
that$\ x=r\cos\varphi$, $y=r\sin\varphi,$ and $r^{2}=x^{2}+y^{2}$. Besides,
$dx=dx^{0}d\mathbf{x}\,$,$\;d\mathbf{x}\,=dx^{1}dx^{2}dx^{3}\,$, \ and
Minkowski tensor $\eta_{\mu\nu}=\mathrm{diag}\left(  1,-1,-1,-1\right)  $.}
$A^{\mu}$, assigned to MSF (\ref{a1}): $A^{0}=A^{3}=0,$ and%
\begin{equation}
\ A^{1}=-y\left(  \frac{\Phi}{2\pi r^{2}}+\frac{B}{2}\right)  ,\ A^{2}%
=x\left(  \frac{\Phi}{2\pi r^{2}}+\frac{B}{2}\right)  . \label{a2}%
\end{equation}
In Sec. \ref{S3}, we briefly outline relativistic quantum mechanics of
spinning particles in MSF, introducing important for our purposes physical
quantities.\emph{ }Here we use so-called natural self-adjoint extensions of
the corresponding Hamiltonians, which correspond to zero-radius limit of the
regularized case of a finite-radius solenoid.{\LARGE \ }Explicit forms of
relativistic and nonrelativistic quantum stationary states of spinning
particles in MSF are placed in the Appendix A. In Sec. \ref{S4}, subsec.
\ref{SS4.2}, we build instantaneous CS for nonrelativistic and relativistic
spinning particles both in ($2+1)$- and ($3+1)$-dim., using some universal
constructions. In subsec. \ref{SS4.4}, we study the semiclassical
approximation. On the base of the CS, we construct SS, developing a techniques
of semiclassical calculations for means of various physical quantities. In
Sec. \ref{S5}, we construct time-dependent CS for different kinds of
particles, find and analyze trajectories of means. Some details of these
calculations are placed in the Appendix B. We summarize and discuss the
obtained results in Sec. \ref{S6}.

\section{Classical motion in MSF\label{S2}}

As was mentioned in the Introduction, our intension is to construct CS in the
MSF. Basic expected properties of the CS are described in terms of the
classical motion in the MSF. That is why, we start our exposition with this
section, where we present a brief description of classical motion of a charge
$q=\pm e\ $with a mass $M$ in the MSF. Trajectories $x^{\mu}\left(  s\right)
$ are parametrized by the Minkowski interval $s$ and obey the Lorentz
equations:%
\begin{equation}
Mc^{2}{\ddot{x}}^{\nu}=qF^{\nu\mu}\dot{x}_{\mu},\ \label{b3}%
\end{equation}
where ${\dot{x}}^{\nu}=dx^{\mu}/ds$, $F_{\mu\nu}=\partial_{\mu}A_{\nu
}-\partial_{\nu}A_{\mu}$. As it follows from (\ref{a1}), the only nonzero
components of $F^{\nu\mu}$ are $F^{21}=-F^{12}=B$. For trajectories that do
not intersect the axis $z$, we obtain from (\ref{b3}):%
\begin{equation}
P_{0}=\mathrm{const},\ P_{3}=\mathrm{const},\ {\dot{P}}_{1}=\epsilon\varkappa
P_{2},\ {\dot{P}}_{2}=-\epsilon\varkappa P_{1},\ P_{1}^{2}+P_{2}%
^{2}=\mathbf{P}_{\bot}^{2}=\mathrm{const}, \label{b4a}%
\end{equation}
where $P^{\nu}=Mc\dot{x}^{\nu}=p^{\nu}-\frac{q}{c}A^{\nu},$ is the kinetic
momentum, and $p^{\,\nu}$ is the generalized particle momentum, $P^{\nu}%
P_{\nu}=(Mc)^{2}$,$\ \varkappa=|qB|/Mc^{2},\ \epsilon=\mathrm{sign}\left(
qB\right)  $. Thus, the total particle energy $\mathcal{E}=cP^{0}$ is also an
integral of motion. One can see that the general solution of (\ref{b4a}) reads%
\begin{align}
&  ct=\frac{p_{0}}{Mc}s,\ z=-\frac{p_{3}}{Mc}s+z_{0}=-\frac{cp_{3}}{p_{0}%
}t+z_{0},\ x=x_{0}+R\cos\psi,\nonumber\\
&  y=y_{0}-\epsilon R\sin\psi;\ \ \psi=\text{$\varkappa$}s+\psi_{0}=\omega
t+\psi_{0},\,\,\omega=\frac{|qB|}{p_{0}},\ \label{b7}%
\end{align}
where $x_{0},\,y_{0},\,z_{0},\,p_{\,0},\,p_{\,3},\,R,$ and $\,\psi_{0}$ are
integration constants. It follows from (\ref{b7}) that%
\begin{align}
&  \ (x-x_{0})^{2}+(y-y_{0})^{2}=R^{2},\ x_{0}=R_{c}\cos\alpha,\ \ y_{0}%
=R_{c}\sin\alpha\ ,\nonumber\\
&  r^{2}=x^{2}+y^{2}=R^{2}+R_{c}^{2}+2RR_{c}\cos(\psi+\epsilon\alpha
),\ R_{c}=\sqrt{x_{0}^{2}+y_{0}^{2}}\ . \label{b8a}%
\end{align}
Projections of particle trajectories on $xy$-plane are circles of the radii
$R$ with central points ($x_{0},\,y_{0})$ placed on the distance $R_{c}$ from
the origin. Particle images on the $xy$-plane are rotating with the
synchrotron frequency $\omega$. For an observer which is placed near the
solenoid with $z>0,$ the rotation of the particle with $\epsilon=1$ is
clockwise, and for the particle with $\epsilon=-1$ is anticlockwise.\emph{\ }%
Thus, equations of motion for the charge $-q$ can be obtained from equations
of motion (\ref{b7}) with the charge $q$ by the substitution $p_{0}$ by
$-p_{0}$. Along the axis $z,$ the particle has a constant velocity
$dz/dt=-cp_{3}/p_{0}$. We denote by $r_{max}=R+R_{c}$ the maximal possible
moving off and by $r_{min}=|R-R_{c}|$ the minimal possible moving off of the
particle from the $z$-axis.

We note that equations of motion for a nonrelativistic particle ($\mathbf{P}%
^{2}\ll\left(  Mc\right)  ^{2})$ in the MSF follow from (\ref{b7}) setting
$p_{0}=Mc$. Then $ct=s$ and $\omega=\omega_{\mathrm{NR}}=|qB|/Mc$, where
$\omega_{\mathrm{NR}}$ is the cyclotron frequency.

The square of the particle rotation energy is $E_{\bot}^{2}=c^{2}%
\mathbf{P}_{\bot}^{2}$ and determines the radius $R$\ as follows%
\begin{equation}
R^{2}=E_{\bot}^{2}\left(  qB\right)  ^{-2}. \label{17a}%
\end{equation}
Using (\ref{b8a}), one can calculate angular momentum projection $L_{z}$,
\begin{equation}
L_{z}=yp_{1}-xp_{2}=\frac{\epsilon Mc\varkappa}{2}(R_{c}^{2}-R^{2}%
)+\frac{q\Phi}{2\pi c}, \label{b13}%
\end{equation}
which is an (dependent) integral of motion.

\begin{figure}[ptb]%
\centering
\includegraphics[
height=2.6394in,
width=2.7665in
]%
{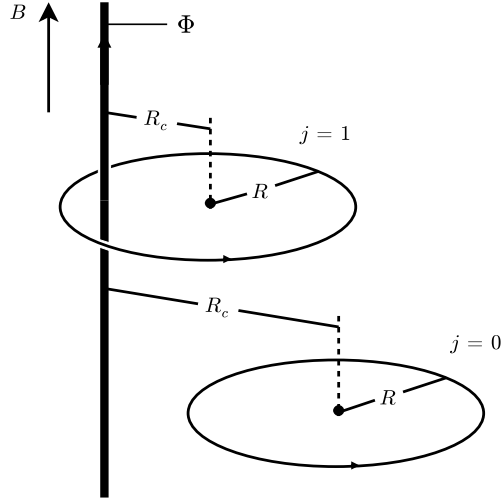}%
\caption{Two types of trajectories in the MSF}%
\label{twotypes}%
\end{figure}

The presence of AB solenoid (the magnetic flux $\Phi$) breaks the
translational symmetry in the $xy$-plane. In classical theory, this fact has
only a topological effect; there appear two types of trajectories, we label
them by an index $j=0,1$ such that $j=1$ corresponds to $(R^{2}-R_{c}^{2})>0$
(embraces the solenoid), and $j=0$ corresponds to $(R^{2}-R_{c}^{2})<0$ (does
not embrace the solenoid), see Fig. \ref{twotypes}.%

Already in classical theory, it is convenient, to introduce dimensionless
complex quantities $a_{1}$ and $a_{2}$ (containing the constant $\hbar$) as
follows:
\begin{align}
a_{1}  &  =\frac{iP_{1}-\epsilon P_{2}}{\sqrt{2\hbar Mc\varkappa}}%
=-\sqrt{\gamma/2}Re^{-i\psi},\ \ \gamma=\frac{|qB|}{c\hbar\,}\ ,\nonumber\\
a_{2}  &  =\frac{Mc\varkappa(x-i\epsilon y)-iP_{1}-\epsilon P_{2}}%
{\sqrt{2\hbar Mc\varkappa}}=\sqrt{\gamma/2}R_{c}e^{-i\epsilon\alpha}.
\label{b14}%
\end{align}
They define physical quantities $R,R_{c},x,y,\mathbf{P}_{\bot}^{2}$ and
$L_{z}$ as follows:%
\begin{align}
&  R^{2}=\frac{2}{\gamma}a_{1}^{\ast}a_{1},\;R_{c}^{2}=2\gamma^{-1}a_{2}%
^{\ast}a_{2},\ (x-i\epsilon y)=\sqrt{2\gamma^{-1}}\left(  a_{2}-a_{1}^{\ast
}\right)  ,\label{20}\\
&  \mathbf{P}_{\bot}^{2}=2\gamma\hbar^{2}a_{1}^{\ast}a_{1},\ L_{z}%
=\epsilon\hbar\left(  a_{2}^{\ast}a_{2}-a_{1}^{\ast}a_{1}\right)  +\frac
{q\Phi}{2\pi c}\ . \label{21}%
\end{align}
One can see that $a_{1}{\exp(i\omega t)}$ and $a_{2}$ are complex (dependent)
integrals of motion.

Another important dimensionless integral of motion $\lambda$ (in classical
theory $\lambda>0$) reads:
\begin{equation}
\lambda=\frac{p_{0}+p_{3}}{Mc}\ .\, \label{b18}%
\end{equation}
Thus, we can chose the set $x_{0},\,y_{0},\,z_{0},\,\lambda,\,R,$
and$\,\psi_{0}$ as six independent integrals of motion.

Often, it is convenient to use the light-cone variables $x_{\pm}$ ,
\begin{equation}
x_{-}=ct-z,\ \ x_{+}=ct+z\Longleftrightarrow ct=\frac{x_{+}+x_{-}}%
{2},\ \ z=\frac{x_{+}-x_{-}}{2}\,. \label{b17}%
\end{equation}
In terms of such variables, the general solution (\ref{b7}) takes the form:%
\begin{align}
&  ct=\frac{1+(\varkappa R)^{2}+\lambda^{2}}{2{\lambda}^{2}}x_{-}%
,\ z=\frac{1+(\varkappa R)^{2}-\lambda^{2}}{2{\lambda}^{2}}x_{-}%
+z_{0},\nonumber\\
&  x=x_{0}+R\cos\psi,\ y=y_{0}-\epsilon R\sin\psi;\ \psi=\widetilde{\omega
}x_{-}+\psi_{0},\,\,\widetilde{\omega}=\varkappa/\lambda,\ s=\lambda^{-1}%
x_{-}\ , \label{b20}%
\end{align}
where $x_{-}$ plays the role of the time.

\section{Quantum mechanics with MSF\label{S3}}

In quantum theory, it is convenient to represent the magnetic flux $\Phi$ of
the AB solenoid via the Dirac's fundamental magnetic flux $\Phi_{0}=2\pi
c\hbar/e$ as follows:%
\begin{equation}
\left(  \Phi/\Phi_{0}\right)  \mathrm{sign}B=l_{0}+\mu\Longrightarrow
l_{0}=\left[  \left(  \Phi/\Phi_{0}\right)  \mathrm{sign}B\right]
\in\mathbb{Z},\ 0\leq\mu=\left(  \Phi/\Phi_{0}\right)  \mathrm{sign}B-l_{0}<1,
\label{c.2}%
\end{equation}
where $l_{0}$ is an integer and the quantity $\mu$ is called the mantissa of
the magnetic flux. In fact, $\mu$ determines all the quantum effects in the AB
and MSF, see e.g. \cite{192}\textrm{. }We note that the definition (\ref{c.2})
differs from the one $\tilde{\mu}=\left(  \Phi/\Phi_{0}\right)  -\left[
\left(  \Phi/\Phi_{0}\right)  \right]  $ for the mantissa of $\Phi$, which was
used in some earlier works, and which does not contain the factor
$\mathrm{sign}B.$ The quantities $\mu$ and $\tilde{\mu}$ are related as
follows: $\mu=\tilde{\mu}$,$\ B>0$;$\ \mu=1-\tilde{\mu}$,$\ B<0$. It turns out
that the definition (\ref{c.2}) is very convenient and allows one to write
universal expressions for any mutual orientations of the uniform magnetic
field and the AB flux.

The quantum behavior of spinning (spin $1/2$) relativistic particles in the
MSF is described by Dirac wave functions $\Psi$ that obeys the Dirac equation
with the electromagnetic potentials (\ref{a2}),
\begin{equation}
i\hbar\partial_{t}\Psi=\hat{H}\Psi,\;\hat{H}=c\gamma^{0}\left(
\mbox{\boldmath$\gamma$\unboldmath}\mathbf{\hat{P}}+Mc\right)  \,,
\label{abe1}%
\end{equation}
where $\gamma^{\nu}=\left(  \gamma^{0}%
,\mbox{\boldmath$\gamma$\unboldmath}\right)  $%
,$\;\mbox{\boldmath$\gamma$\unboldmath}=\left(  \gamma^{k}\right)  $ are
$\gamma$-matrices; $\hat{P}^{k}=\hat{p}^{k}-\frac{q}{c}A^{k}$, $\hat{p}%
^{k}=-i\hbar\partial_{k}$. Below, we consider the Dirac equation in $\left(
2+1\right)  $-dim., where $k=1,2$, and in $\left(  3+1\right)  $-dim., where
$k=1,2,3$.

It is natural that the solutions of the Dirac equation in $\left(  2+1\right)
$-dim. and in $\left(  3+1\right)  $-dim. have\emph{ }much in common.\emph{
}Nevertheless, the algebra of the Dirac $\gamma$-matrices for these cases is
different as well as the spin description, all this implies, e.g., the well
known fact that quantum mechanics of spinning particle (both nonrelativistic
and relativistic) in the presence of an uniform magnetic field is essentially
different in $(2+1)$-dim. and in $(3+1)$-dim.\emph{ }That is why, we consider
the problem under consideration both in $(2+1)$-dim. and $(3+1)$-dim.
separately (the former case cannot be extracted from the latter one in a
trivial manner).

It is also known that AB effect in condensed matter physics, in particular,
planar physics is important in the nonrelativistic case, $\mathcal{E}%
_{\perp\left(  \sigma\right)  }^{2}\ll Mc^{2}$. Therefore, we pay a special
attention to such a limit.\emph{ }In addition, the massless (which is, in a
sense, equivalent to the ultrarelativistic case) Dirac equation in
$(2+1)$-dim. describes under some conditions the graphene physics. In such a
case, the Fermi velocity $v_{F}\approx c/300$ plays the role of the effective
velocity of light and has to substitute $c$ in all the corresponding
expressions. In this connection, we consider the ultrarelativistic limit in detail.

In $(2+1)$-dim., a Dirac wave function $\Psi$ is a spinor dependent on
$x^{0},x^{1},$ and $x^{2},$ and there are two nonequivalent representations
for $\gamma$-matrices:
\[
\gamma^{0}=\sigma^{3},\;\gamma^{1}=i\sigma^{2},\;\gamma^{2}=-i\sigma^{1}%
\zeta,\;\;\zeta=\pm1\,,
\]
where $\mbox{\boldmath$\sigma$\unboldmath}=\left(  \sigma^{i}\right)  $ are
Pauli matrices. Choosing the \textquotedblright
polarizations\textquotedblright\ $\zeta=+1$, we describe \textquotedblright
spin up\textquotedblright\ particles, and choosing $\zeta=-1$, we describe
\textquotedblright spin down\textquotedblright\ particles. In $(2+1)$-dim.
these are different particles. There exist \textquotedblright spin
up\textquotedblright\ and \textquotedblright spin down\textquotedblright%
\ antiparticles. In contrast to $(3+1)$-dim. case, particles and antiparticles
in $(2+1)$-dim. have only one spin polarization state.

Stationary states of the Dirac equation with MSF in $(2+1)$-dim. have the
form:%
\begin{equation}
\Psi=\exp\left[  -\frac{i}{\hbar}(cp_{0}t)\right]  \psi_{p_{0}}^{(\zeta
)}\left(  x_{\perp}\right)  \,,\;\zeta=\pm1,\;x_{\perp}=\left(  0,x^{1}%
,x^{2}\right)  \,, \label{abe6}%
\end{equation}
where spinors $\psi_{p_{0}}^{(\zeta)}\left(  x_{\perp}\right)  $ are subjected
to the equations:%
\begin{align}
&  \left(  \mbox{\boldmath$\sigma$\unboldmath}\mathbf{\hat{P}}_{\perp
}+Mc\sigma^{3}\right)  \psi_{p_{0}}^{(1)}(x_{\perp})=p_{0}\psi_{p_{0}}%
^{(1)}(x_{\perp}),\;\mathbf{\hat{P}}_{\perp}=\left(  \hat{P}_{1},\hat{P}%
_{2}\right)  \,;\label{abe7a}\\
&  \left(  \sigma^{1}\mbox{\boldmath$\sigma$\unboldmath}\mathbf{\hat{P}%
}_{\perp}\sigma^{1}+Mc\sigma^{3}\right)  \psi_{p_{0}}^{(-1)}(x_{\perp}%
)=p_{0}\psi_{p_{0}}^{(-1)}(x_{\perp})\,. \label{abe7b}%
\end{align}
We note that $cp_{0}=\mathcal{E}>0$ for particles, and $cp_{0}=-\mathcal{E}<0$
for antiparticle states.

One can see that
\begin{equation}
\psi_{p_{0}}^{(-1)}(x_{\perp})=\sigma^{2}\psi_{-p_{0}}^{(1)}(x_{\perp})\,.
\label{abe8}%
\end{equation}
That is the reason why we are going to consider only the case $\zeta=1$ in
what follows. In such a case, a self-adjoint Hamiltonian $\hat{H}^{\vartheta}$
has the form
\begin{equation}
\hat{H}^{\vartheta}=c\left(  \mbox{\boldmath$\sigma$\unboldmath}\mathbf{\hat
{P}}_{\perp}+Mc\sigma^{3}\right)  . \label{abe8a}%
\end{equation}
Its domain $D_{H}^{\vartheta}$ depends essentially on the sign $\vartheta
=\mathrm{sign}\Phi=\pm1$ of the magnetic flux, that is why the Hamiltonian has
a label $\vartheta.$

In $(3+1)$-dim., a Dirac wave function $\Psi$ is a bispinor dependent on
$x^{0},$ $x^{1},$ $x^{2}$, and $x^{3}$. Then unlike $(2+1)$-dim. we are not
restricted in the choice of evolution parameter of time $x^{0}$, but we can
also use the light-cone variable $x_{-}$. There is also an opportunity to
build differently spinors adapting them to the nonrelativistic or
ultrarelativistic limit. All this is described in detail in the Appendix A.

It should be reminded that all self-adjoint extensions of $2+1$ -\ and $3+1$ -
Dirac Hamiltonians in MSF were constructed in \cite{205,212,GitTySV09}, see,
also \cite{FP01}. The domains of $3+1$ - Dirac Hamiltonian in MSF are trivial
extensions of the corresponding domains mentioned in ($2+1$) case, that is why
we retain for them the same notation $D_{H}^{\vartheta}$ and we use for
self-adjoint $3+1$ - Dirac Hamiltonian the same notation $\hat{H}^{\vartheta}%
$. Of course, in this case $\hat{H}^{\vartheta}=c\gamma^{0}\left(
\sum_{k=1,2,3}\gamma^{k}\hat{P}^{k}+Mc\right)  $. In addition, considering a
regularized case of a finite-radius solenoid, it was demonstrated that
zero-radius limit yields two (depending on $\vartheta$) of self-adjoint
extensions, with domains $D_{H}^{\vartheta}$. In contrast to the spinless
case\footnote{The domain of a spinless particle Hamiltonian in MSF involves
only regular radial functtions as $r\rightarrow0$, see \cite{212}. Here we use
the terms \textquotedblright regular\textquotedblright\ and \textquotedblright
irregular\textquotedblright\ as $r\rightarrow0$ in the following sense. We
call a function to be regular if it behaves as $r^{c}$ as $r\rightarrow0$ with
$c\geq0$, and irregular if $c<0$.} both domains $D_{H}^{\vartheta}$ involve
irregular but still square-integrable radial functions that do not vanish as
$r\rightarrow0$. In fact, this means that any wave function is completely
determined by its values for $r>0$. Its value in the point $r=0$ can be set
arbitrary. In the Appendix A, we represent solutions of eqs. (\ref{abe7a}) and
eqs. (\ref{abe1}) in $\left(  3+1\right)  $-dim. for both values of
$\vartheta$ in the domains $D_{H}^{\vartheta}$.

Some important remarks should be made.

1. In case of spinning particles, some results essentially depend both on the
mantissa of the magnetic flux $\mu$ and on the direction of the flux
$\vartheta$. The latter dependence appears due to the spin presence and is
specific only for states with irregular radial functions. In such states there
is a superstrong contact interaction between the magnetic moment of the
particle and the solenoid flux. Namely this interaction dependent on
$\vartheta$ can be repulsive or attractive. Clear that irregular radial
functions appear in the attractive case.

2. In the spinless case (and even in spinning case) in states with regular
radial functions, there is a certain translation invariance with respect to a
change of the integer number $l_{0}$ (see (\ref{c.2})) by an arbitrary integer
$k.$ Such an invariance means that physics depends only on the mantissa of the
magnetic flux $\mu$. In spinning case, and for $\mu\neq0,$\ this invariance
turns out to be partially broken in states with irregular radial functions
(however, a translation invariance with respect to the change of $l_{0}$ by
integers $k,$ obeying the condition \textrm{sign}$l_{0}=$\textrm{sign}$\left(
l_{0}+k\right)  $, still holds).

3. In states with irregular radial functions, we cannot say that particles do
not penetrate the AB solenoid. However, even in this case, a locally trivial
vector potential gives rise to observable effects in a nontrivial topology and
this is a manifestation of the AB effect for such states.

In $(3+1)$-dim., the operators $\,\hat{P}_{3}$, $z$-component of the total
angular momentum operator%
\begin{equation}
\hat{J}_{z}=\hat{L}_{z}+\Sigma_{z}/2,\; \label{k13b}%
\end{equation}
and $({\mbox{\boldmath$\alpha$}}_{\perp}\mathbf{\hat{P}}_{\perp})^{2}$, where
$\hat{L}_{z}=x\hat{p}_{y}-y\hat{p}_{x}=-i\hbar\partial_{\varphi}%
$,$\ \ {\mbox{\boldmath$\alpha$}}_{\perp}=(\alpha^{1},\,\alpha^{2}%
,\,0)$,$\ \mathbf{\hat{P}}_{\perp}=(\hat{P}^{1},\,\hat{P}^{2},\,0)$, are
self-adjoint on the domain $D_{H}^{\vartheta}$ and mutually commuting
integrals of motion (all these operators commute with the Hamiltonian $\hat
{H}^{\vartheta}$) \cite{205,212}. In $(2+1)$-dim., the total angular momentum
operator $\hat{J}=-i\hbar\partial_{\varphi}+\hbar\sigma^{3}/2$, which is a
dimensional reduction of the operator $\hat{J}_{z}$ in ($3+1)$-dim., and
$\left(  \mbox{\boldmath$\sigma$\unboldmath}\mathbf{\hat{P}}_{\perp}\right)
^{2}$ are self-adjoint on $D_{H}^{\vartheta}$ and mutually commuting integrals
of motion \cite{205,212}. One can say that $c^{2}\left(
\mbox{\boldmath$\sigma$\unboldmath}\mathbf{\hat{P}}_{\perp}\right)  ^{2}$ and
$\hat{J}$ in $(2+1)$-dim. and $c^{2}({\mbox{\boldmath$\alpha$}}_{\perp
}\mathbf{\hat{P}}_{\perp})^{2}$ and $\hat{J}_{z}$ in ($3+1)$-dim. are
integrals of motion which play the role of a square of the transverse kinetic
energy ($E_{\bot}^{2}$ in (\ref{17a})) and $z$-component of angular momentum
($L_{z}$ in (\ref{b13})) in the case of a spinning particle, respectively.
Then it is useful to define self-adjoint operators $\hat{R}^{2}$ and$\;\hat
{R}_{c}^{2}$ by analogy with corresponding classical relations (\ref{17a}) and
(\ref{b13}) as follows:%
\begin{align}
\hat{R}^{2}  &  =c^{2}\left(  \mbox{\boldmath$\sigma$\unboldmath}\mathbf{\hat
{P}}_{\perp}\right)  ^{2}\left(  qB\right)  ^{-2},\ \hat{R}_{c}^{2}-\hat
{R}^{2}=-2c\left[  \left(  l_{0}+\mu\right)  \hbar-\epsilon\hat{J}\right]
\left\vert qB\right\vert ^{-1}\;\mathrm{in\;}2+1\;\mathrm{\dim.},\nonumber\\
\hat{R}^{2}  &  =c^{2}({\mbox{\boldmath$\alpha$}}_{\perp}\mathbf{\hat{P}%
}_{\perp})^{2}\left(  qB\right)  ^{-2},\ \hat{R}_{c}^{2}-\hat{R}%
^{2}=-2c\left[  \left(  l_{0}+\mu\right)  \hbar-\epsilon\hat{J}_{z}\right]
\left\vert qB\right\vert ^{-1}\;\mathrm{in\;}3+1\;\mathrm{\dim.} \label{spinR}%
\end{align}

One can find two types ($j=0,1$) of solutions of the Dirac equation which are
common eigenvectors of operators $\hat{R}^{2}$ and $\hat{J}$ \ in $(2+1)$-dim.
and operators $\hat{R}^{2}$ and $\hat{J}_{z}$ in $(3+1)$-dim., see
(\ref{abe10}), (\ref{k15}), and (\ref{k21}) in the Appendix A, respectively.
Such solutions have two quantum numbers $n_{1}$ and $n_{2}$\emph{ }in common
then we may be using to them the general notation $\Psi_{n_{1},\,n_{2}%
}^{(j\,)}$. Note that eigenvalues of the operators $\hat{J}$ \ and $\hat
{J}_{z}$ are the same, $J_{z}=J=\epsilon\hbar(l_{0}-l+1/2)$, where $l$ is an
integer. Then, using an appropriate inner product on $xy$-plane, see
(\ref{Dinn}) in $(2+1)$-dim. and (\ref{ipxy}) in ($3+1)$-dim., we obtain the
mean of the operator $\hat{R}^{2}-\hat{R}_{c}^{2},$%
\begin{align}
&  \left(  \Psi_{n_{1},\,n_{2}}^{(j\,)},\,\left(  \hat{R}^{2}-\hat{R}_{c}%
^{2}\right)  \Psi_{n_{1},\,n_{2}}^{(j\,)}\right)  \left(  \Psi_{n_{1},\,n_{2}%
}^{(j\,)},\,\Psi_{n_{1},\,n_{2}}^{(j\,)}\right)  ^{-1}=\frac{2}{\gamma}\left(
l+\mu\right)  ,\nonumber\\
&  \left(  \Psi,\,\Psi\right)  =\left(  \Psi,\,\Psi\right)  _{D}%
\;\mathrm{in\;}2+1\;\mathrm{\dim.},\;\;\left(  \Psi,\,\Psi\right)  =\left(
\Psi,\,\Psi\right)  _{D}^{\perp}\;\mathrm{in\;}3+1\;\mathrm{\dim.} \label{j}%
\end{align}
In the semiclassical limit the sign of the mean allows one to interpret the
corresponding states as particle trajectories that embrace and do not embrace
the solenoid. Namely, an orbit embraces the solenoid for $l\geq0$ (type $j=1$)
and do not for $l\leqslant-1$ (type $j=0$). This classification corresponds to
the classical one introduced in the previous section, see eq. (\ref{b13}) and
Fig. \ref{twotypes}. Trajectories with $l=0$,$-1$ are situated most close to
the solenoid.

For ${\mu}=0$ there is no any impact of AB solenoid on the energy spectrum and
the energy spectrum is given by the Landau formula. For $\mu\neq0$, energies
of states with $j=1$ differ from the Landau levels, whereas energies with
$j=0$ coincide with the Landau levels, see (\ref{ab14}) in the Appendix A.

\section{Instantaneous CS on $xy-$plane\label{S4}}

\subsection{Quantum states\label{SS4.2}}

In spite of the differences between stationary states of spinning particle in
($2+1)$-dim. and in ($3+1)$-dim., see (\ref{ab11}), (\ref{k18}), and
(\ref{k20}), in both dimensions one can build CS on $xy$-plane in a similar
manner, using some universal constructions. Let us introduce operators
$\hat{a}_{1}$, $\hat{a}_{2}$, and $\hat{a}_{1}^{\dagger},\hat{a}_{2}^{\dagger
}$ that correspond to classical quantities $a_{1}$, $a_{2}$, and $a_{1}^{\ast
}$, $a_{2}^{\ast}$ \textrm{,}%
\begin{align}
\hat{a}_{1}  &  =\frac{i\hat{P}_{1}-\epsilon\hat{P}_{2}}{\sqrt{2\hbar
\,Mc\,\varkappa}},\ \ \hat{a}_{2}=\frac{Mc\,\varkappa\,(x-i\epsilon
y)-i\hat{P}_{1}-\epsilon\hat{P}_{2}}{\sqrt{2\hbar\,Mc\,\varkappa}%
}\ ;\nonumber\\
\hat{a}_{1}^{\dagger}  &  =-\frac{i\hat{P}_{1}+\epsilon\hat{P}_{2}}%
{\sqrt{2\hbar\,Mc\,\varkappa}},\ \ \hat{a}_{2}^{\dagger}=\frac{Mc\,\varkappa
\,(x+i\epsilon y)+i\hat{P}_{1}-\epsilon\hat{P}_{2}}{\sqrt{2\hbar
\,Mc\,\varkappa}}\ . \label{c1a}%
\end{align}
It should be noted that operators $\hat{P}_{1}$ and $\hat{P}_{2}$ are
symmetric but not self-adjoint on the domain $D_{H}^{\vartheta}$. That is why,
one cannot consider $\hat{a}_{1}^{\dagger}$ and $\hat{a}_{2}^{\dagger}$ as
adjoint to $\hat{a}_{1}$ and $\hat{a}_{2}$ respectively. Nevertheless, the
operators (\ref{c1a}) play an important role in the further constructions.

Using properties of Laguerre functions, one can find the action of these
operators on the functions (\ref{ab12}):
\begin{align}
\hat{a}_{1}\Phi_{n_{1},\,n_{2},\sigma}^{(j)}(\varphi,\,\rho)  &  =\sqrt{n_{1}%
}\,\Phi_{n_{1}-1,\,n_{2},\sigma}^{(j)}(\varphi,\,\rho)\,,\ \ \hat{a}%
_{1}^{\dagger}\Phi_{n_{1},\,n_{2},\sigma}^{(j)}(\varphi,\,\rho)=\sqrt{n_{1}%
+1}\,\Phi_{n_{1}+1,\,n_{2},\sigma}^{(j)}(\varphi,\,\rho)\,,\nonumber\\
\hat{a}_{2}\Phi_{n_{1},\,n_{2},\sigma}^{(j)}(\varphi,\,\rho)  &  =\sqrt{n_{2}%
}\,\Phi_{n_{1},\,n_{2}-1,\sigma}^{(j)}(\varphi,\,\rho)\,,\ \ \hat{a}%
_{2}^{\dagger}\Phi_{n_{1},\,n_{2},\sigma}^{(j)}(\varphi,\,\rho)=\sqrt{n_{2}%
+1}\,\Phi_{n_{1},\,n_{2}+1,\sigma}^{(j)}(\varphi,\,\rho)\,, \label{s0}%
\end{align}
where possible values of $n_{1}$ and$\ n_{2}$ depend on $m,l,\sigma$, and $j$
according to (\ref{ab12}) and the functions $\Phi_{n_{1}+s_{1},\,n_{2}%
+s_{2},\sigma}^{(j)}$ are defined in (\ref{dopf}).

Formal commutators between the operators $\hat{a}_{1}^{\dagger},\hat{a}_{1},$
and$\,\,\hat{a}_{2}^{\dagger},\hat{a}_{2}$ have the form:%
\[
\left[  \hat{a}_{1},\hat{a}_{1}^{\dagger}\right]  =1+f,\ \ \left[  \hat{a}%
_{2},\hat{a}_{2}^{\dagger}\right]  =1-f,\ \ \left[  \hat{a}_{1},\hat{a}%
_{2}\right]  =f,\ \ \left[  \hat{a}_{1},\hat{a}_{2}^{\dagger}\right]  =0,
\]
with a singular function $f=\Phi\left(  \pi Br\right)  ^{-1}\delta
(r)=2(l_{0}+\mu)\delta(\rho)$. All the solutions of the Dirac equation in
$\left(  2+1\right)  $- and $\left(  3+1\right)  $-dim. introduced in the
Appendix A are expressed via the functions (\ref{ab12}) and describe states of
spinning particles out of the solenoid, $r>0$. In such a case, the function
$f$ gives zero contributions and can be neglected. In turn, this means that on
the domains $D_{H}^{\vartheta}$ operators $\hat{a}_{1}^{\dagger},\hat{a}%
_{2}^{\dagger},$ and $\hat{a}_{1},\hat{a}_{2}$ behave as creation and
annihilation operators. Then quantities $x-i\epsilon y$ and $\hat{L}_{z}$ can
be expressed in terms of the operators $\hat{a}_{1}^{\dagger}$, $\hat{a}_{1}$
and$\,\,\hat{a}_{2}^{\dagger}$, $\hat{a}_{2}$ as follows:%
\begin{equation}
\ (x-i\epsilon y)=\sqrt{2\gamma^{-1}}\left(  \hat{a}_{2}-\hat{a}_{1}^{\dagger
}\right)  ,\ -\frac{\epsilon}{\hbar}\hat{L}_{z}+l_{0}+\mu=\left(  \hat{N}%
_{1}-\hat{N}_{2}\right)  ,\;\hat{N}_{s}=\hat{a}_{s}^{\dagger}\hat{a}%
_{s},\;s=1,2, \label{RRO}%
\end{equation}
which is the same in $2+1$ and $3+1$ dim. The operator $c^{2}\left(
\mbox{\boldmath$\sigma$\unboldmath}\mathbf{\hat{P}}_{\perp}\right)  ^{2}$ in
$2+1$ dim. case and $c^{2}({\mbox{\boldmath$\alpha$}}_{\perp}\mathbf{\hat{P}%
}_{\perp})^{2}$ in $3+1$ dim. case have the following representations:%
\begin{align}
c^{2}\left(  \mbox{\boldmath$\sigma$\unboldmath}\mathbf{\hat{P}}_{\perp
}\right)  ^{2}  &  =2\hbar c\left\vert qB\right\vert \left[  \hat{N}%
_{1}+\left(  1-\sigma^{3}\epsilon\right)  /2\right]  ,\;\mathrm{in\;}%
2+1\;\mathrm{\dim.}\nonumber\\
c^{2}({\mbox{\boldmath$\alpha$}}_{\perp}\mathbf{\hat{P}}_{\perp})^{2}  &
=2\hbar c\left\vert qB\right\vert \left[  \hat{N}_{1}+\left(  1-\Sigma
_{z}\epsilon\right)  /2\right]  ,\;\mathrm{in\;}3+1\;\mathrm{\dim.}
\label{s0.1}%
\end{align}
where relations (\ref{s0}) and expression (\ref{ab14}) are used.{\LARGE \ }%
Then the operators $\hat{R}^{2}$ and $\hat{R}_{c}^{2}$ have the following form
(see the definitions (\ref{spinR})):
\begin{align}
\hat{R}^{2}  &  =\gamma^{-1}\left(  2\hat{N}_{1}+1-\sigma^{3}\epsilon\right)
\;\mathrm{in\;}2+1\;\mathrm{\dim.},\;\hat{R}^{2}=\gamma^{-1}\left(  2\hat
{N}_{1}+1-\Sigma_{z}\epsilon\right)  \;\mathrm{in\;}3+1\;\mathrm{\dim
.},\nonumber\\
\hat{R}_{c}^{2}  &  =\gamma^{-1}\left(  2\hat{N}_{2}+1\right)  \;\mathrm{in\;}%
2+1\;\mathrm{and\;in\ }3+1\;\mathrm{\dim.} \label{s0.2}%
\end{align}

As was already mentioned in the Introduction, our aim is to construct CS. One
can formulate a definition of CS for systems with quadratic Hamiltonians, see
\cite{CSQT72}. Unfortunately, no general definition of CS for arbitrary
quantum system exists. In our case, with a nonquadratic Hamiltonian, defining
CS, we would like to maintain basic properties of already known CS for
quadratic systems. First of all, these states have to minimize uncertainty
relations for some physical quantities (e.g. coordinates and momenta) at any
fixed time instant. Second, means of these quantities, calculated with respect
to time-dependent CS, have to move along classical trajectories. It is also
desirable for time-dependent CS to maintain their form under the time
evolution. Here, it is supposed that time-dependent CS are solutions of the
corresponding wave equation, Dirac or Pauli (and Klein-Gordon or Schrödinger
equation in the case of spinless particle). Thus, the problem of constructing
the CS states is in main reduced to a suitable choice of the form of the CS at
a fixed time instant. We call such CS instantaneous CS (ICS) in what follows.
In the case of quadratic systems, e.g. a non-relativistic particle in the
magnetic field, ICS are eigenvectors of the corresponding annihilation
operators, let say the operators $\hat{a}_{1},\hat{a}_{2}$ from (\ref{c1a})
without the magnetic flux. In the case of MSF where particle Hamiltonians are
nonquadratic, these operators are not exactly annihilation operators for both
types of the functions $\Phi_{n_{1},\,n_{2},\sigma}^{(j)}$. Nevertheless, as
is demonstrated below, one can construct some kind of ICS that have the above
described properties. These states are very close to eigenvectors of the
introduced operators $\hat{a}_{1},\hat{a}_{2}$ from (\ref{c1a}) . At the same
time, these states maintain their form under the time evolution.

The presence of the AB flux breaks the translational symmetry in the
$xy$-plane. That is why in the problem under consideration, there appear two
types of CS, those which correspond classical trajectories which embrace the
solenoid and those which do not. Taking into account classification of quantum
states according to types $j=1$ and $j=0$, which depends on the sign of the
mean value (\ref{j}), we see that each of these CS must be constructed using
stationary states of the same type.

It is convenient to pass from the functions $\Phi_{n_{1},\,n_{2},\sigma}%
^{(j)}$ (\ref{ab12}) to new functions $\mathbf{\Phi}_{z_{1},\,z_{2},\sigma
}^{(j)}$ as follows:
\begin{equation}
\mathbf{\Phi}_{z_{1},\,z_{2},\sigma}^{(j)}(\varphi,\,\rho)=\sum_{\tilde{l}%
}\mathbf{\Phi}_{z_{1},\,z_{2},\sigma}^{(j)\tilde{l}}(\varphi,\,\rho
),\;\;\mathbf{\Phi}_{z_{1},\,z_{2},\sigma}^{(j)\tilde{l}}(\varphi,\,\rho
)=\sum_{m}\frac{z_{1}^{n_{1}}z_{2}^{n_{2}}\Phi_{n_{1},\,n_{2},\sigma}%
^{(j)}(\varphi,\,\rho)}{\sqrt{\Gamma(1+n_{1})\,\Gamma(1+n_{2})}}. \label{s1}%
\end{equation}
Here $z_{1}$ and $z_{2}$ are complex parameters, possible values of $n_{1}$
and$\ \ n_{2}$ depend on $m,\tilde{l},\sigma,$ and $j$ according to eq.
(\ref{ab12}), and we set $\mathcal{N}=1$. The functions $\mathbf{\Phi}%
_{z_{1},\,z_{2},\sigma}^{(j)\tilde{l}}$ can be expressed via special functions
$Y_{\alpha}$,%
\begin{align}
\mathbf{\Phi}_{z_{1},\,z_{2},\sigma}^{(0)\tilde{l}}(\varphi,\,\rho)  &
=\exp\left\{  i\epsilon\left[  l_{0}-l+\left(  1-\epsilon\sigma\right)
/2\right]  \varphi\right\}  Y_{-\alpha_{\tilde{l}}}\left(  z_{1},z_{2}%
,\rho\right)  ,\ \alpha_{\tilde{l}}=\tilde{l}-\left(  1-\epsilon\sigma\right)
/2+\mu,\nonumber\\
\;\mathbf{\Phi}_{z_{1},\,z_{2},\sigma}^{(1)\tilde{l}}(\varphi,\,\rho)  &
=\exp\left\{  i\epsilon\left[  l_{0}-l+\left(  1-\epsilon\sigma\right)
/2\right]  \varphi+\pi\left[  l-\left(  1-\epsilon\right)  \left(
1+\sigma\right)  /4\right]  \right\}  Y_{\alpha_{\tilde{l}}}\left(
z_{2},z_{1},\rho\right)  ,\nonumber\\
Y_{\alpha}(z_{1},z_{2};\rho)  &  =\sum_{m=0}^{\infty}\frac{z_{1}^{m}%
\,z_{2}^{m+\alpha}I_{m+\alpha,\,m}(\rho)}{\sqrt{\Gamma(1+m)\Gamma(1+m+\alpha
)}}. \label{s1.1a}%
\end{align}
By the help of the well-known sum,
\[
\sum_{m=0}^{\infty}\frac{z^{m}\,I_{\alpha+m,\,m}(x)}{\sqrt{\Gamma
(1+m)\Gamma(1+\alpha+m)}}=z^{-\frac{\alpha}{2}}\exp\left(  z-x/2\right)
\,J_{\alpha}(2\sqrt{xz}\,),
\]
where $J_{\alpha}$ are the Bessel functions of the first kind, one can obtain
the following representation for $Y_{\alpha}$:
\begin{equation}
Y_{\alpha}(z_{1},z_{2};\rho)=\exp\left(  z_{1}z_{2}-\frac{\rho}{2}\right)
\left(  \sqrt{z_{2}/z_{1}}\right)  ^{\alpha}J_{\alpha}(2\sqrt{z_{1}z_{2}\rho
}). \label{g3}%
\end{equation}

Using the functions $\Phi_{z_{1},\,z_{2},\sigma}^{(j)}$, one can construct ICS
on $xy-$plane. In ($2+1)$-dim. and in ($3+1)$-dim., ICS are constructed by the
help of spinors\ described in the Appendix A by substituting the functions
$\Phi_{n_{1},\,n_{2},\sigma}^{(j)}$ for the function $\Phi_{z_{1}%
,\,z_{2},\sigma}^{(j)}$.

Thus, using eqs. (\ref{ab11}), (\ref{ab11a}), and (\ref{s0.1}),\ we obtain ICS
for massive, $\zeta=+1,$ spinning particles on $xy$-plane and in ($2+1$)-dim.:%
\begin{align}
&  \mathbf{\psi}_{\pm,z_{1},\,z_{2}}^{(j)}(\varphi,\,\rho)=\left\{  \sigma
^{3}\left[  \pm\Pi_{0}\left(  \hat{N}_{1}\right)
-\mbox{\boldmath$\sigma$\unboldmath}\mathbf{\hat{P}}_{\perp}\right]
+Mc\right\}  \mathbf{u}_{z_{1},\,z_{2},\pm1}^{(j)}(\varphi,\rho),\nonumber\\
&  \mathbf{u}_{z_{1},\,z_{2},\sigma}^{(j)}(\varphi,\rho)=\mathbf{\Phi}%
_{z_{1},\,z_{2},\sigma}^{(j)}(\varphi,\,\rho)\upsilon_{\sigma}\,,\nonumber\\
&  \Pi_{0}\left(  \hat{N}_{1}\right)  =\hat{\Pi}_{0}^{2}\frac{2}{\sqrt{\pi}%
}\int_{0}^{\infty}e^{-\hat{\Pi}_{0}^{2}\tau^{2}}d\tau,\;\hat{\Pi}_{0}%
^{2}=M^{2}c^{2}+2\hbar\left\vert qB\right\vert /c\left[  \hat{N}_{1}+\left(
1-\sigma\epsilon\right)  /2\right]  . \label{s5}%
\end{align}
For such states, we have%
\begin{equation}
\left(  \mathbf{\psi}_{\pm,z_{1},\,z_{2}}^{(j)},\,\mathbf{\psi}_{\pm
,z_{1}^{\prime},\,z_{2}^{\prime}}^{(j^{\prime})}\right)  _{D}=2Mc\left(
\mathbf{\Phi}_{z_{1},z_{2},,\pm1}^{(j\,)},\left[  \Pi_{0}\left(  \hat{N}%
_{1}\right)  +Mc\right]  \mathbf{\Phi}_{z_{1}^{\prime},z_{2}^{\prime},\pm
1}^{(j\,^{\prime})}\right)  _{\bot}\ , \label{s6}%
\end{equation}
where the inner product $\left(  ,\right)  _{\bot}$ is defined by (\ref{inpr}).

According to (\ref{ab17}), (\ref{ab18}), and (\ref{ab19}), nonrelativistic ICS
for $2+1$ spin up\ particles have the form:
\begin{align}
&  \mathbf{\Psi}_{\pm,\,z_{1},z_{2}}^{(j)\mathrm{up}}(\varphi,\,\rho
)=\mathbf{u}_{z_{1},\,z_{2},\pm1}^{(j)}(\varphi,\rho),\nonumber\\
&  \left(  \mathbf{\Psi}_{\pm,\,z_{1},z_{2}}^{(j)\mathrm{up}},\,\mathbf{\Psi
}_{\pm,\,z_{1}^{\prime},z_{2}^{\prime}}^{(j^{\prime})\mathrm{up}}\right)
_{D}=\left(  \mathbf{\Phi}_{z_{1},z_{2},,\pm1}^{(j\,)},\mathbf{\Phi}%
_{z_{1}^{\prime},z_{2}^{\prime},\pm1}^{(j^{\prime})}\right)  _{\bot},
\label{s7a}%
\end{align}
whereas for the spin down particles they read:%
\begin{align}
&  \mathbf{\Psi}_{\pm,\,z_{1},z_{2}}^{(j)\mathrm{down}}(\varphi,\,\rho
)=\mathbf{u}_{z_{1},\,z_{2},\mp1}^{(j)}(\varphi,\rho),\nonumber\\
&  \left(  \mathbf{\Psi}_{\pm,\,z_{1},z_{2}}^{(j)\mathrm{down}},\,\mathbf{\Psi
}_{\pm,\,z_{1}^{\prime},z_{2}^{\prime}}^{(j^{\prime})\mathrm{down}}\right)
_{D}=\left(  \mathbf{\Phi}_{z_{1},z_{2},,\mp1}^{(j\,)},\mathbf{\Phi}%
_{z_{1}^{\prime},z_{2}^{\prime},\mp1}^{(j^{\prime})}\right)  _{\bot},
\label{s7b}%
\end{align}

According to (\ref{k23}), ICS on $xy-$plane for $3+1$ nonrelativistic
particles with a given spin polarization $s=\pm1$, have the form:
\begin{align}
&  \mathbf{\Psi}_{\pm,\,z_{1},z_{2,}+1}^{(j)\mathrm{NR}}(\varphi
,\,\rho)=\binom{\mathbf{\Psi}_{\pm,\,z_{1},z_{2}}^{(j)\mathrm{up}}%
(\varphi,\,\rho)}{0},\;\;\mathbf{\Psi}_{\pm,\,z_{1},z_{2,}-1}^{(j)\mathrm{NR}%
}(\varphi,\,\rho)=\binom{0}{\mathbf{\Psi}_{\pm,\,z_{1},z_{2}}%
^{(j)\mathrm{down}}(\varphi,\,\rho)},\nonumber\\
&  \left(  \mathbf{\Psi}_{\pm,\,z_{1},z_{2,}s}^{(j)\mathrm{NR}},\,\mathbf{\Psi
}_{\pm,\,z_{1}^{\prime},z_{2,}^{\prime}s^{\prime}}^{(j^{\prime})\mathrm{NR}%
}\right)  _{D}^{\perp}=\delta_{s,s^{\prime}}\left(  \mathbf{\Phi}_{z_{1}%
,z_{2},,\pm s}^{(j\,)},\mathbf{\Phi}_{z_{1}^{\prime},z_{2}^{\prime},\pm
s}^{(j^{\prime})}\right)  _{\bot}, \label{s7}%
\end{align}
where the inner product of four-component spinors $\Psi$ and$\,\Psi^{\prime}$
on $xy$-plane is defined in (\ref{ipxy}).

According to (\ref{ab21}) - (\ref{ab22}), ICS for $2+1$ massless $\zeta=+1$
fermions are
\begin{align}
&  \mathbf{\Psi}_{\pm,\,z_{1},z_{2}}^{(j,+1)}(\varphi,\,\rho)=\mathbf{u}%
_{\pm,z_{1},\,z_{2}}^{(j,+1)}(\varphi,\rho),\;\mathbf{u}_{\pm,z_{1},\,z_{2}%
}^{(1,+1)}(\varphi,\rho)=\binom{\mathbf{\Phi}_{z_{1},z_{2},+1}^{(1)}%
(\varphi,\rho)}{\mp i\epsilon\mathbf{\Phi}_{z_{1},z_{2},-1}^{(1)}(\varphi
,\rho)},\nonumber\\
&  \mathbf{u}_{\pm,z_{1},\,z_{2}}^{(0,+1)}(\varphi,\rho)=\binom{\mathbf{\tilde
{\Phi}}_{z_{1},z_{2},+1}^{(0)}(\varphi,\rho)}{\pm i\epsilon\mathbf{\tilde
{\Phi}}_{z_{1},z_{2},-1}^{(0)}(\varphi,\rho)}+\mathbf{u}_{0,z_{1},\,z_{2}%
}^{(0,+1)}(\varphi,\rho),\nonumber\\
&  \mathbf{u}_{0,z_{1},\,z_{2}}^{(0,+1)}(\varphi,\rho)=\sum_{\tilde{l}%
}c_{\epsilon}\left.  \frac{z_{2}^{n_{2}}\Phi_{0,\,n_{2},\epsilon}%
^{(0)}(\varphi,\,\rho)}{\sqrt{\,\Gamma(1+n_{2})}}\right\vert _{m=0}%
\upsilon_{\epsilon}\,,\;\mathbf{\tilde{\Phi}}_{z_{1},z_{2},,\sigma}%
^{(0)}(\varphi,\rho)=\sum_{\tilde{l}}\mathbf{\tilde{\Phi}}_{z_{1}%
,\,z_{2},\sigma}^{(0)\tilde{l}}(\varphi,\,\rho),\nonumber\\
&  \mathbf{\tilde{\Phi}}_{z_{1},\,z_{2},\sigma}^{(0)\tilde{l}}(\varphi
,\,\rho)=\mathbf{\Phi}_{z_{1},\,z_{2},\sigma}^{(0)\tilde{l}}(\varphi
,\,\rho)-\frac{1}{2}\left(  1+\sigma\epsilon\right)  \left.  \frac
{z_{2}^{n_{2}}\Phi_{0,\,n_{2},\sigma}^{(0)}(\varphi,\,\rho)}{\sqrt
{\,\Gamma(1+n_{2})}}\right\vert _{m=0}. \label{s8}%
\end{align}
The inner products of such states have the form%
\begin{equation}
\left(  \mathbf{\Psi}_{\pm,\,z_{1},z_{2}}^{(j,+1)},\,\mathbf{\Psi}%
_{\pm,\,z_{1}^{\prime},z_{2}^{\prime}}^{(j^{\prime},+1)}\right)  _{D}%
=\sum_{\sigma=\pm1}\left(  \mathbf{\Phi}_{z_{1},z_{2},,\sigma}^{(j\,)}%
,\mathbf{\Phi}_{z_{1}^{\prime},z_{2}^{\prime},\sigma}^{(j\,^{\prime})}\right)
_{\bot}. \label{s9}%
\end{equation}
In the same manner, by the help of (\ref{ab24}), one can obtain ICS for $2+1$
massless $\zeta=-1$ fermions.

According to eq. (\ref{k18}), for $3+1$ relativistic spinning particles, ICS
on $xy$-plane have the form%
\begin{equation}
\mathbf{U}_{z_{1},z_{2},\sigma}^{(j)}(\varphi,\,\rho)=\mathbf{\Phi}%
_{z_{1},z_{2},\sigma}^{(j\,)}(\varphi,\,\rho)\left(
\begin{array}
[c]{c}%
\upsilon_{\sigma}\\
-\sigma\upsilon_{\sigma}%
\end{array}
\right)  . \label{s10}%
\end{equation}
The inner product of such states on $xy$-plane reads:%
\begin{equation}
\left(  \mathbf{U}_{z_{1},z_{2},\sigma}^{(j)},\,\mathbf{U}_{z_{1}^{\prime
},z_{2}^{\prime},\sigma}^{(j^{\prime})}\right)  _{D}^{\perp}=2\left(
\mathbf{\Phi}_{z_{1},z_{2},,\sigma}^{(j\,)},\mathbf{\Phi}_{z_{1}^{\prime
},z_{2}^{\prime},\sigma}^{(j\,^{\prime})}\right)  _{\bot}. \label{s11}%
\end{equation}

One can see that in all the cases, the inner product of ICS on $xy$-plane is
expressed via the matrix elements:
\begin{align}
&  \left(  \mathbf{\Phi}_{z_{1},z_{2},,\sigma}^{(j\,)},\mathbf{\Phi}%
_{z_{1}^{\prime},z_{2}^{\prime},,\sigma}^{(j^{\prime})}\right)  _{\bot}%
=\delta_{jj^{\prime}}\mathcal{R}_{\sigma}^{(j)};\ \ \mathcal{R}_{\sigma}%
^{(0)}=Q_{1-\mu_{\sigma}}\left(  \sqrt{z_{1}^{\ast}z_{1}^{\prime}},\sqrt
{z_{2}^{\ast}z_{2}^{\prime}}\right)  ,\ \ \nonumber\\
&  \mathcal{R}_{\sigma}^{(1)}=Q_{\mu_{\sigma}}\left(  \sqrt{z_{2}^{\ast}%
z_{2}^{\prime}},\sqrt{z_{1}^{\ast}z_{1}^{\prime}}\right)  ,\;\mu_{\sigma}%
=\mu-\frac{1}{2}\vartheta\epsilon\left(  1-\vartheta\sigma\right)
;\nonumber\\
&  Q_{\alpha}(u,\,v)=Q_{\alpha}^{-}(u,\,v)+\left(  v/u\right)  ^{\alpha
}I_{\alpha}(2uv),\;Q_{\alpha}^{-}(u,\,v)=\sum_{l=1}^{\infty}\left(
v/u\right)  ^{\alpha+l}I_{\alpha+l}(2uv), \label{s12}%
\end{align}
where $I_{\alpha}$ are the modified Bessel functions of the first kind. We
note that in contrast to the spinless case \cite{265}, subindex $\alpha$ in
the functions $Q_{\alpha}(u,\,v)$ can take also negative values $-1<\alpha<0$.

It should be noted the importance of the obtained result. It turns out that
all the means and matrix elements with respect to the ICS are expressed only
via two functions\emph{ }$Q_{\alpha}(u,\,v)$ and $Q_{\alpha}^{-}%
(u,\,v)$.\emph{ }That is why the further study of such physical quantities is
reduced to the analysis of these functions.

It follows from (\ref{s0}) that:%
\begin{equation}
\hat{N}_{k}\mathbf{\Phi}_{z_{1},\,z_{2},\sigma}^{(j)}(\varphi,\,\rho
)=z_{k}\partial_{z_{k}}\mathbf{\Phi}_{z_{1},\,z_{2},\sigma}^{(j)}%
(\varphi,\,\rho),\;k=1,2, \label{s13}%
\end{equation}
and
\begin{align}
&  a_{1}\mathbf{\Phi}_{z_{1},\,z_{2},\sigma}^{(j)}(\varphi,\,\rho
)=z_{1}\left[  \mathbf{\Phi}_{z_{1},\,z_{2},\sigma}^{(j)}(\varphi
,\,\rho)-\left(  -1\right)  ^{j}\left.  \mathbf{\Phi}_{z_{1},\,z_{2},\sigma
}^{(j)\tilde{l}}(\varphi,\,\rho)\right\vert _{\tilde{l}=-\left(
1+\vartheta\epsilon\right)  /2}\right]  ,\nonumber\\
&  a_{2}\mathbf{\Phi}_{z_{1},\,z_{2},\sigma}^{(j)}(\varphi,\,\rho
)=z_{2}\left[  \mathbf{\Phi}_{z_{1},\,z_{2},\sigma}^{(j)}(\varphi
,\,\rho)+\left(  -1\right)  ^{j}\left.  \mathbf{\Phi}_{z_{1},\,z_{2},\sigma
}^{(j)\tilde{l}}(\varphi,\,\rho)\right\vert _{\tilde{l}=\left(  1-\vartheta
\epsilon\right)  /2}\right]  . \label{s14}%
\end{align}

Eqs. (\ref{s14}) allow one to calculate the matrix elements%
\begin{equation}
\left(  \mathbf{\Phi}_{z_{1},z_{2},\sigma}^{(j)},a_{k}\mathbf{\Phi}%
_{z_{1}^{\prime},\,z_{2}^{\prime},\sigma}^{(j^{\prime})}\right)  _{\bot
}=\left(  a_{k}\right)  _{z_{1},\,z_{2};\,z_{1}^{\prime},\,z_{2}^{\prime
},\sigma}^{(j,\,j^{\prime})}\,,\ \ k=1,\,2. \label{s15}%
\end{equation}
Results of such calculations are, for example:
\begin{align}
(a_{1})_{z_{1},\,z_{2};\,z_{1}^{\prime},\,z_{2}^{\prime},\sigma}^{(0,\,0)}  &
=z_{1}^{\prime}Q_{1-\mu_{\sigma}}^{-}(\sqrt{z_{1}^{\ast}z_{1}^{\prime}%
},\,\sqrt{z_{2}^{\ast}z_{2}^{\prime}}\,),\;(a_{2})_{z_{1},\,z_{2}%
;\,z_{1}^{\prime},\,z_{2}^{\prime},\sigma}^{(0,\,0)}=z_{2}^{\prime}%
Q_{1-\mu_{\sigma}}(\sqrt{z_{1}^{\ast}z_{1}^{\prime}},\,\sqrt{z_{2}^{\ast}%
z_{2}^{\prime}}\,),\nonumber\\
(a_{1})_{z_{1},\,z_{2};\,z_{1}^{\prime},\,z_{2}^{\prime},\sigma}^{(1,\,1)}  &
=z_{1}^{\prime}Q_{\mu_{\sigma}}(\sqrt{z_{2}^{\ast}z_{2}^{\prime}}%
,\,\sqrt{z_{1}^{\ast}z_{1}^{\prime}}\,),\;(a_{2})_{z_{1},\,z_{2}%
;\,z_{1}^{\prime},\,z_{2}^{\prime},\sigma}^{(1,\,1)}=z_{2}^{\prime}%
Q_{\mu_{\sigma}}^{-}(\sqrt{z_{2}^{\ast}z_{2}^{\prime}},\,\sqrt{z_{1}^{\ast
}z_{1}^{\prime}}\,). \label{s16}%
\end{align}

Using an appropriate inner product, see above, we define means of an operator
$\hat{F}$\ with respect of the ICS on $xy$-plane, $\overline{(F)}_{(j)}%
$\emph{.} Then, we consider important cases when a matrix operator $\hat{F}$
is either the identity matrix $I$ multiplied by a differential operator
$\hat{f}$, $\hat{F}=\hat{f}I$, or $\hat{F}=c^{2}\left(
\mbox{\boldmath$\sigma$\unboldmath}\mathbf{\hat{P}}_{\perp}\right)  ^{2}$ in
$2+1$ dim., and $\hat{F}=c^{2}({\mbox{\boldmath$\alpha$}}_{\perp}%
\mathbf{\hat{P}}_{\perp})^{2}$ in $3+1$ dim. Here, we can express
${\overline{(F)}}_{(j)}$ via the means $\left(  \mathbf{\Phi}_{z_{1}%
,z_{2},\sigma}^{(j)},\hat{F}_{\sigma}\mathbf{\Phi}_{z_{1},\,z_{2},\sigma
}^{(j)}\right)  _{\bot}$, where either $\hat{F}_{\sigma}=\hat{f}$, or $\hat
{F}_{\sigma}=c^{2}\mathbf{\hat{P}}_{\perp}^{2}-\epsilon\hbar c\left\vert
qB\right\vert \sigma$. Thus, we obtain, for example, for$\;3+1$ relativistic
and $2+1$ nonrelativistic spin up particles the following expression (with the
corresponding interpretations of the number $\sigma$)%
\begin{equation}
{\overline{(F)}}_{(j)}=\frac{\left(  \mathbf{\Phi}_{z_{1},z_{2},\sigma}%
^{(j)},\hat{F}_{\sigma}\mathbf{\Phi}_{z_{1},\,z_{2},\sigma}^{(j)}\right)
_{\bot}}{\left(  \mathbf{\Phi}_{z_{1},z_{2},\sigma}^{(j)},\mathbf{\Phi}%
_{z_{1},\,z_{2},\sigma}^{(j)}\right)  _{\bot}}. \label{s18}%
\end{equation}
For $2+1$ massive $\zeta=+1$ relativistic particles we obtain%
\begin{equation}
{\overline{(F)}}_{(j)}=\frac{\left.  \left(  \mathbf{\Phi}_{z_{1},z_{2}%
,\sigma}^{(j)},\hat{F}_{\sigma}\left[  \Pi_{0}\left(  z_{1}^{\prime}%
\partial_{z_{1}^{\prime}}\right)  +Mc\right]  \mathbf{\Phi}_{z_{1}^{\prime
},\,z_{2},\sigma}^{(j)}\right)  _{\bot}\right\vert _{z_{1}=z_{1}^{\prime}}%
}{\left.  \left(  \mathbf{\Phi}_{z_{1},z_{2},\sigma}^{(j)},\left[  \Pi
_{0}\left(  z_{1}^{\prime}\partial_{z_{1}^{\prime}}\right)  +Mc\right]
\mathbf{\Phi}_{z_{1}^{\prime},\,z_{2},\sigma}^{(j)}\right)  _{\bot}\right\vert
_{z_{1}=z_{1}^{\prime}}}, \label{s19}%
\end{equation}
whereas for $2+1$ massless $\zeta=+1$ fermions, we have:
\begin{equation}
{\overline{(F)}}_{(j)}=\frac{\sum_{\sigma=\pm1}\left(  \mathbf{\Phi}%
_{z_{1},z_{2},\sigma}^{(j)},\hat{F}_{-\vartheta}\mathbf{\Phi}_{z_{1}%
,\,z_{2},\sigma}^{(j)}\right)  _{\bot}}{\sum_{\sigma=\pm1}\left(
\mathbf{\Phi}_{z_{1},z_{2},\sigma}^{(j)},\mathbf{\Phi}_{z_{1},\,z_{2},\sigma
}^{(j)}\right)  _{\bot}}. \label{s20}%
\end{equation}
Note that means ${\overline{(F)}}_{(j)}$ for $\;3+1$ nonrelativistic spinning
particles and antiparticles at given $s$ are expressed via means (\ref{s18})
for $2+1$ nonrelativistic case according to (\ref{s7}).

Then, using (\ref{s13}) and notation (\ref{s18}), we obtain the means of
operators $\hat{N}_{k}$, $\;k=1,2,$ for example,%
\begin{equation}
{\overline{(N_{k})}}_{(j)}=z_{k}\left.  \partial_{z_{k}^{\prime}}%
\ln\mathcal{R}_{\sigma}^{(j)}\right\vert _{z_{k}^{\prime}=z_{k}}\; \label{s21}%
\end{equation}
for$\;3+1$ particle and for$\;$nonrelativistic $2+1$ spin up\ particles;
\begin{equation}
{\overline{(N_{k})}}_{(j)}=z_{k}\left.  \partial_{z_{k}^{\prime}}\ln
\sum_{\sigma=\pm1}\mathcal{R}_{\sigma}^{(j)}\right\vert _{z_{k}^{\prime}%
=z_{k}}\; \label{s22}%
\end{equation}
for $2+1$ massless $\zeta=+1$ fermions;%
\begin{equation}
{\overline{(N_{k})}}_{(j)}=z_{k}\left.  \partial_{z_{k}^{\prime}}\ln\left\{
\left[  \Pi_{0}\left(  z_{1}^{\prime}\partial_{z_{1}^{\prime}}\right)
+Mc\right]  \mathcal{R}_{\sigma}^{(j)}\right\}  \right\vert _{z_{k}^{\prime
}=z_{k}}\; \label{s23}%
\end{equation}
for $2+1$ relativistic massive $\zeta=+1$ particles.

Using (\ref{s16}), we find that%
\begin{equation}
{\overline{(x-i\epsilon y)}}_{(j)}=\sqrt{2\gamma^{-1}}\left[  {\overline
{(a_{2})}}_{(j)}-{\overline{(a_{1})}}_{(j)}^{\ast}\right]  , \label{s24}%
\end{equation}
where, for example,
\begin{align}
&  {\overline{(a_{1})}}_{(0)}=z_{1}\Delta_{1-\mu_{\sigma}}(|z_{1}%
|,|z_{2}|),\ \ \;{\overline{(a_{2})}}_{(0)}=z_{2},\nonumber\\
&  {\overline{(a_{1})}}_{(1)}=z_{1},\ \ {\overline{(a_{2})}}_{(1)}=z_{2}%
\Delta_{\mu_{\sigma}}(|z_{2}|,|z_{1}|),\ \ \Delta_{\alpha}(u,v)=\frac
{Q_{\alpha}^{-}(u,\,v)}{Q_{\alpha}(u,\,v)} \label{s25}%
\end{align}
for$\;3+1$ particle and for$\;$nonrelativistic $2+1$ spin up\ particle
($\sigma=+1$) and antiparticle ($\sigma=-1$), and
\begin{align}
&  {\overline{(a_{1})}}_{(0)}=z_{1}\frac{\sum_{\sigma=\pm1}Q_{1-\mu_{\sigma}%
}^{-}(|z_{1}|,|z_{2}|)}{\sum_{\sigma=\pm1}Q_{1-\mu_{\sigma}}(|z_{1}|,|z_{2}%
|)},\ \ {\overline{(a_{2})}}_{(0)}=z_{2},\nonumber\\
&  {\overline{(a_{1})}}_{(1)}=z_{1},\ \ {\overline{(a_{2})}}_{(1)}=z_{2}%
\frac{\sum_{\sigma=\pm1}Q_{\mu_{\sigma}}^{-}(|z_{2}|,|z_{1}|)}{\sum
_{\sigma=\pm1}Q_{\mu_{\sigma}}(|z_{2}|,|z_{1}|)}, \label{s26}%
\end{align}
for $2+1$ massless $\zeta=+1$ fermions.

Note that one can get the means ${\overline{(N_{k})}}_{(j)}$ and
${\overline{(a_{k})}}_{(j)}$ for the case of spinless particle from expression
(\ref{s21}), (\ref{s25}), and (\ref{s12}) at $\mu_{\sigma}=\mu$, see
\cite{265}.

\subsection{Semiclassical approximation\label{SS4.4}}

Representations (\ref{s21}) - (\ref{s23}) allow us to connect means of
$\hat{R}^{2}$ and $\hat{R}_{c}^{2}$ with the parameters $z_{1}$ and $z_{2}$.
It follows from (\ref{s0.2}) that%
\begin{equation}
{\overline{(R^{2})}}_{(j)}=\gamma^{-1}\left[  2{\overline{(N_{1})}}%
_{(j)}+1-\sigma\epsilon\right]  ,\;{\overline{(R_{c}^{2})}}_{(j)}=\gamma
^{-1}\left[  2{\overline{(N_{2})}}_{(j)}+1\right]  . \label{RRmean}%
\end{equation}
Note that these relations are valid in the case of spinless particle at
$\sigma=0$. We expect that in the semiclassical limit ${\overline{(N_{k})}%
}_{(j)}\approx\left\vert z_{k}\right\vert ^{2}$. At the same time length
scales defined by means of ${\overline{(R^{2})}}_{(j)}$ and ${\overline
{(R_{c}^{2})}}_{(j)}$ have to be sufficiently large, which implies $\left\vert
z_{k}\right\vert ^{2}\gg1$ in the semiclassical limit.

We note that in the pure quantum case, as a characteristic quantum scale of
the rotational motion we can take the quantity%
\[
\mathcal{E}_{\mathrm{quant}}^{2}=2\left\vert qB\right\vert \hbar c=2M^{2}%
c^{4}\left\vert B\right\vert /B_{0},\;B_{0}=M^{2}c^{3}/\left\vert q\right\vert
\hbar,
\]
where $B_{0}=m_{e}^{2}c^{3}/e\hbar\simeq4,4\cdot10^{13}G$ is the critical
magnetic field above which the nonlinearity of QED becomes actual. The
corresponding length scale is
\begin{equation}
R_{\mathrm{quant}}=\sqrt{2\gamma^{-1}}=\sqrt{2B_{0}/\left\vert B\right\vert
}\lambda_{\mathrm{C}},\ \lambda_{\mathrm{C}}=\hbar/Mc. \label{Rquant}%
\end{equation}
For the angular momentum projection $J_{z}$\emph{\ }a characteristic quantum
scale\emph{\ }is obviously\emph{\ }$\hbar$\emph{. }For a given energy, i.e.,
for a given ${\overline{(R^{2})}}_{(j)}$\emph{, }the quantity ${\overline
{(J_{z})}}_{(j)}$\emph{\ }is proportional to ${\overline{(R_{c}^{2})}}_{(j)}$
due to (\ref{spinR}),\emph{\ }and, therefore,\emph{\ }can be characterized by
the corresponding length scale $R_{\mathrm{quant}}$. Note that the
$R_{\mathrm{quant}}$ \ is much larger than the Compton length $\lambda
_{\mathrm{C}}$ if the magnetic field $B$ is weak, $B_{0}/\left\vert
B\right\vert \gg1$. Thus, the conditions $\left\vert z_{k}\right\vert ^{2}%
\gg1$, correspond to ones%
\[
{\overline{(R^{2})}}_{(j)},{\overline{(R_{c}^{2})}}_{(j)}\gg R_{\mathrm{quant}%
}^{2}\ .
\]
At the same time, in the quantum case, the dimensionless quantities
$\left\vert z_{k}\right\vert ^{2}$ are of the order $1$. We see that the
semiclassical decompositions are adequate namely in case of strong enough
magnetic fields (e.g., pulsar magnetic fields $B$ for which $B_{0}/\left\vert
B\right\vert \sim10^{-2}$ and $R_{\mathrm{quant}}\ll\lambda_{\mathrm{C}}$).

We expect that the sign of the difference%
\begin{equation}
d_{\left(  j\right)  }=\sqrt{{\overline{(R^{2})}}_{(j)}}-\sqrt{{\overline
{(R_{c}^{2})}}_{(j)}} \label{distance}%
\end{equation}
is related to the trajectory type in the classical limit. One can see that
such a limit implies the following conditions:%
\[
\left\vert d_{\left(  j\right)  }\right\vert \gg R_{\mathrm{quant}}%
\sim\left\vert \left\vert z_{1}\right\vert -\left\vert z_{2}\right\vert
\right\vert \gg1.
\]
In particular, for states with $j=0,$ we have $\left\vert z_{1}\right\vert
\ll\left\vert z_{2}\right\vert $, and for states with $j=1$, we have
$\left\vert z_{1}\right\vert \gg\left\vert z_{2}\right\vert $. We note that in
both cases the corresponding functions $Q_{\alpha}(u,v)$ are calculated at
$\left\vert v\right\vert >\left\vert u\right\vert \gg1$.

There exist all the derivatives $\partial_{v}\left[  \left(  v/u\right)
^{\alpha+l}I_{\alpha+l}(2uv)\right]  $, the series $Q_{\alpha}^{-}(u,\,v)$
(\ref{s12}) converges and the series $\sum_{l=1}^{\infty}\partial_{v}\left[
\left(  v/u\right)  ^{\alpha+l}I_{\alpha+l}(2uv)\right]  $ converges uniformly
on the half-line $0<\operatorname{Re}v<\infty$. Thus, one can write a
differential equation for $Q_{\alpha}^{-}(u,\,v)$,
\[
\partial_{v}Q_{\alpha}^{-}(u,\,v)=2v\left[  \left(  v/u\right)  ^{\alpha
}I_{\alpha}(2uv)+Q_{\alpha}^{-}(u,\,v)\right]  .
\]
A solution of this equation, which corresponds to (\ref{s12}), reads%
\[
Q_{\alpha}^{-}(u,\,v)=2e^{v^{2}}\int_{0}^{v}e^{-\tilde{v}^{2}}\left(
\tilde{v}/u\right)  ^{\alpha}I_{\alpha}(2u\tilde{v})\tilde{v}d\tilde{v}\ .
\]
Using eq. (6.631.4) \cite{GR94}, we represent this solution as follows:
\begin{align}
&  Q_{\alpha}^{-}(u,\,v)=e^{u^{2}+v^{2}}\tilde{Q}_{\alpha}^{-}(u,\,v),\;\tilde
{Q}_{\alpha}^{-}(u,\,v)=\left[  1-T(u,\,v)\right]  ,\nonumber\\
&  T(u,\,v)=2e^{-u^{2}}\int_{v}^{\infty}e^{-\tilde{v}^{2}}\left(  \tilde
{v}/u\right)  ^{\alpha}I_{\alpha}(2u\tilde{v})\tilde{v}d\tilde{v}. \label{z4}%
\end{align}
Then
\begin{equation}
Q_{\alpha}(u,\,v)=e^{u^{2}+v^{2}}\tilde{Q}_{\alpha}(u,\,v),\;\;\tilde
{Q}_{\alpha}(u,\,v)=\left[  1-T(u,\,v)+e^{-u^{2}-v^{2}}\left(  v/u\right)
^{\alpha}I_{\alpha}(2uv)\right]  , \label{z5}%
\end{equation}
and we can calculate, for example, the means (\ref{s21}):%
\begin{align}
&  {\overline{(N_{k})}}_{(j)}=\left\vert z_{k}\right\vert ^{2}+z_{k}\left.
\partial_{z_{k}^{\prime}}\ln\mathcal{\tilde{R}}_{\sigma}^{(j)}\right\vert
_{z_{k}^{\prime}=z_{k}},\nonumber\\
&  \mathcal{\tilde{R}}_{\sigma}^{(0)}=\tilde{Q}_{1-\mu_{\sigma}}\left(
\sqrt{z_{1}^{\ast}z_{1}^{\prime}},\sqrt{z_{2}^{\ast}z_{2}^{\prime}}\right)
,\;\mathcal{\tilde{R}}_{\sigma}^{(1)}=\tilde{Q}_{\mu_{\sigma}}\left(
\sqrt{z_{2}^{\ast}z_{2}^{\prime}},\sqrt{z_{1}^{\ast}z_{1}^{\prime}}\right)  .
\label{z6}%
\end{align}
These means can be represented explicitly in an real form, taking into account
that%
\begin{align*}
z_{k}\left.  \partial_{z_{k}^{\prime}}\ln\mathcal{\tilde{R}}_{\sigma}%
^{(0)}\right\vert _{z_{k}^{\prime}=z_{k}}  &  =\left.  \frac{\delta
_{k,1}\left\vert z_{1}\right\vert \partial_{u}\tilde{Q}_{1-\mu_{\sigma}%
}(u,\,v)+\delta_{k,2}\left\vert z_{2}\right\vert \partial_{v}\tilde{Q}%
_{1-\mu_{\sigma}}(u,\,v)}{2\tilde{Q}_{1-\mu_{\sigma}}(u,\,v)}\right\vert
_{u=\left\vert z_{1}\right\vert ,\ v=\left\vert z_{2}\right\vert }\ ,\\
z_{k}\left.  \partial_{z_{k}^{\prime}}\ln\mathcal{\tilde{R}}_{\sigma}%
^{(1)}\right\vert _{z_{k}^{\prime}=z_{k}}  &  =\left.  \frac{\delta
_{k,1}\left\vert z_{1}\right\vert \partial_{v}\tilde{Q}_{\mu_{\sigma}%
}(u,\,v)+\delta_{k,2}\left\vert z_{2}\right\vert \partial_{u}\tilde{Q}%
_{\mu_{\sigma}}(u,\,v)}{2\tilde{Q}_{\mu_{\sigma}}(u,\,v)}\right\vert
_{u=\left\vert z_{2}\right\vert ,\ v=\left\vert z_{1}\right\vert }\ .
\end{align*}

We stress that means (\ref{z6}) allow the limit $\mu_{\sigma}\rightarrow0$.
Thus, the contribution due to the AB field can be easily isolated.

Using power decomposition of the function $\tilde{v}^{\alpha+1}I_{\alpha
}(2u\tilde{v})e^{-2u\tilde{v}}$ near the point $\tilde{v}=v$ for an estimation
of the integral $T(u,\,v)$ in (\ref{z4})\ and asymptotics of the function
$I_{\alpha}(2uv)$, one can see that $\left\vert z_{k}\right\vert ^{2}\gg
z_{k}\left.  \partial_{z_{k}^{\prime}}\ln\mathcal{\tilde{R}}_{\sigma}%
^{(j)}\right\vert _{z_{k}^{\prime}=z_{k}}$ for $\left\vert v\right\vert
\gtrsim\left\vert u\right\vert \gg1$. Thus, we obtain the semiclassical
expansions:%
\begin{equation}
\left\vert z_{1}\right\vert ^{2}=\frac{\gamma}{2}{\overline{(R^{2})}}%
_{(j)}+\ldots,\;\;\left\vert z_{2}\right\vert ^{2}=\frac{\gamma}{2}%
{\overline{(R_{c}^{2})}}_{(j)}+\ldots,\ \ \left\vert z_{k}\right\vert ^{2}%
\gg1, \label{z7}%
\end{equation}
which connect means of $\hat{R}^{2}$ and $\hat{R}_{c}^{2}$ with the parameters
$z_{1}$ and $z_{2}$. Thus if $\left\vert \left\vert z_{1}\right\vert
-\left\vert z_{2}\right\vert \right\vert \gg1,$ eqs. (\ref{z7})\ imply the
following relations%
\begin{equation}
\left\vert z_{1}\right\vert \ll\left\vert z_{2}\right\vert ,\ j=0;\ \left\vert
z_{1}\right\vert \gg\left\vert z_{2}\right\vert ,\ j=1. \label{z7a}%
\end{equation}

It should be noted that relations (\ref{z7a}) have nothing to do with
conditions of the applicability of the semiclassical expansions (\ref{z7}).
Obtaining the latter expansions we have supposed that $\left\vert
z_{1}\right\vert \lesssim\left\vert z_{2}\right\vert $ for states with $j=0,$
and $\left\vert z_{1}\right\vert \gtrsim\left\vert z_{2}\right\vert $ for
states with $j=1$. Therefore, relations (\ref{z7})\emph{\ }between means of
$\hat{R}^{2}$ and $\hat{R}_{c}^{2}$ and parameters $z_{1}$ and $z_{2}$ take
place even if a definite relation between \textrm{sign}$\left(  \left\vert
z_{1}\right\vert -\left\vert z_{2}\right\vert \right)  $ and $j$ is absent.

Retaining only leading terms in decompositions (\ref{z7}), we reproduce the
corresponding classical relations (\ref{20}) with $\left\vert z_{1}\right\vert
=\left\vert a_{1}\right\vert $\ and $\left\vert z_{2}\right\vert =\left\vert
a_{2}\right\vert $. In other words, one can say that the classical relations
(\ref{20}) correspond to the leading approximation for sufficiently large
radii. Thus, the leading approximation in the semiclassical expansions
corresponds to the classical limit. Next-to-leading terms define physical
quantities in the semiclassical approximation. These terms depend on the space
dimension and particle spin.

Let us consider semiclassical approximation retaining next-to-leading and
next-next-to-leading terms. If $\left\vert v\right\vert \gg\left\vert
u\right\vert \gg1$, one can approximate the integral $T(u,\,v)$ in
(\ref{z4})\ by a power series in $u/v$ as follows:%
\[
T(u,\,v)=\left(  v/u\right)  ^{\alpha}I_{\alpha}(2uv)e^{-u^{2}-v^{2}}\left(
1+u/v+\ldots\right)  .
\]
Then, using asymptotic of $I_{\alpha}(2uv)$, we obtain from (\ref{z5}):%
\begin{equation}
\tilde{Q}_{\alpha}(u,\,v)=1-\frac{\left(  u/v\right)  ^{1-\alpha}}{2\sqrt{\pi
uv}}e^{-\left(  v-u\right)  ^{2}}, \label{z6.1}%
\end{equation}
which implies%
\begin{equation}
\frac{\partial_{v}\tilde{Q}_{\alpha}(u,\,v)}{\tilde{Q}_{\alpha}(u,\,v)}%
\approx-\frac{\partial_{u}\tilde{Q}_{\alpha}(u,\,v)}{\tilde{Q}_{\alpha
}(u,\,v)}\approx\frac{\left(  u/v\right)  ^{1/2-\alpha}}{\sqrt{\pi}%
}e^{-\left(  v-u\right)  ^{2}},\;\left\vert v\right\vert \gg\left\vert
u\right\vert \gg1. \label{z6.2}%
\end{equation}
Thus, for semiclassical states corresponding to orbits situated far enough
from the solenoid, i.e., for $\left\vert \left\vert z_{1}\right\vert
-\left\vert z_{2}\right\vert \right\vert \gg1$, the terms $z_{k}\left.
\partial_{z_{k}^{\prime}}\ln\mathcal{\tilde{R}}_{\sigma}^{(j)}\right\vert
_{z_{k}^{\prime}=z_{k}}$ are small as $\exp\left(  -\left\vert \left\vert
z_{1}\right\vert ^{2}-\left\vert z_{2}\right\vert ^{2}\right\vert \right)  $.
Then the semiclassical expansions (\ref{z7}) in the next-to-leading
approximation reads:%
\[
\left\vert z_{1}\right\vert ^{2}\approx\frac{\gamma}{2}{\overline{(R^{2})}%
}_{(j)}-\left(  1-\sigma\epsilon\right)  /2,\;\;\left\vert z_{2}\right\vert
^{2}=\frac{\gamma}{2}{\overline{(R_{c}^{2})}}_{(j)}-1/2,\ \ \left\vert
\left\vert z_{1}\right\vert -\left\vert z_{2}\right\vert \right\vert \gg1.
\]

In the most interesting case when a semiclassical orbit is situated near the
solenoid, such that the condition $\left\vert \left\vert v\right\vert
-\left\vert u\right\vert \right\vert \ll1$ holds, the influence of AB solenoid
(due to $\mu\neq0$) on the orbits are not small. In such a case%
\begin{align}
&  T(u,\,v)=\frac{1}{2}-\frac{v-u}{\sqrt{\pi}}+\frac{\alpha+1/2}{2\sqrt{\pi}%
u}+O\left(  \left\vert v-u\right\vert ^{3}\right)  +O\left(  \left\vert
u\right\vert ^{-2}\right)  ,\nonumber\\
&  \tilde{Q}_{\alpha}(u,\,v)\approx\frac{1}{2}+\frac{v-u}{\sqrt{\pi}}%
-\frac{\alpha-1/2}{2\sqrt{\pi}u},\nonumber\\
&  \ \frac{\partial_{v}\tilde{Q}_{\alpha}(u,\,v)}{\tilde{Q}_{\alpha}%
(u,\,v)}\approx-\frac{\partial_{u}\tilde{Q}_{\alpha}(u,\,v)}{\tilde{Q}%
_{\alpha}(u,\,v)}\approx\frac{2}{\sqrt{\pi}}\left(  1-2\frac{v-u}{\sqrt{\pi}%
}+\frac{\alpha-1/2}{\sqrt{\pi}u}\right)  , \label{z6.4}%
\end{align}
such that, for example, means (\ref{z6}) are
\begin{equation}
{\overline{(N_{k})}}_{(j)}\approx\left\vert z_{k}\right\vert ^{2}+\left(
-1\right)  ^{k}\left\{  \frac{\left\vert z_{k}\right\vert }{\sqrt{\pi}}\left[
\left(  -1\right)  ^{j}+2\frac{\left\vert z_{1}\right\vert -\left\vert
z_{2}\right\vert }{\sqrt{\pi}}\right]  +\frac{1-2\mu_{\sigma}}{2\pi}\right\}
,\;\left\vert \left\vert z_{1}\right\vert -\left\vert z_{2}\right\vert
\right\vert \ll1. \label{Nj}%
\end{equation}
Thus, for $\left\vert \left\vert z_{1}\right\vert -\left\vert z_{2}\right\vert
\right\vert \ll1$ all $\mu$-dependent contributions to the means
${\overline{(N_{k})}}_{(j)}$ are of the order one, which is natural for a pure
quantum case. Next-to-leading contributions to the means (of order $\left\vert
z_{k}\right\vert $) that does not depend on $\mu$ are much bigger. These
semiclassical contributions appear since each of $j$-type ICS includes only a
half of eigenfunctions of the operator $\hat{J}_{z}$. It follows from
(\ref{Nj}) that in the leading approximation%
\begin{equation}
{\overline{(N_{1})}}_{(j)}-{\overline{(N_{2})}}_{(j)}\approx\left(  -1\right)
^{j+1}\frac{\left\vert z_{1}\right\vert +\left\vert z_{2}\right\vert }%
{\sqrt{\pi}},\;\left\vert \left\vert z_{1}\right\vert -\left\vert
z_{2}\right\vert \right\vert \ll1. \label{Ndif}%
\end{equation}

At the same time, relations (\ref{z7}) yield in the semiclassical
approximation:
\begin{align*}
\left\vert z_{1}\right\vert ^{2}  &  \approx\frac{\gamma}{2}{\overline
{(R^{2})}}_{(j)}+\left(  -1\right)  ^{j}\sqrt{\frac{\gamma}{2\pi}%
{\overline{(R^{2})}}_{(j)}},\\
\left\vert z_{2}\right\vert ^{2}  &  \approx\frac{\gamma}{2}{\overline
{(R_{c}^{2})}}_{(j)}-\left(  -1\right)  ^{j}\sqrt{\frac{\gamma}{2\pi
}{\overline{(R_{c}^{2})}}_{(j)}},\ \ \left\vert \left\vert z_{1}\right\vert
-\left\vert z_{2}\right\vert \right\vert \ll1.
\end{align*}
Then, using (\ref{Ndif}), we obtain
\[
{\overline{(R^{2})}}_{(j)}-{\overline{(R_{c}^{2})}}_{(j)}\approx\left(
-1\right)  ^{j+1}\sqrt{\frac{2}{\pi\gamma}}\left(  \sqrt{{\overline{(R^{2})}%
}_{(j)}}+\sqrt{{\overline{(R_{c}^{2})}}_{(j)}}\right)  ,\ \ \left\vert
\left\vert z_{1}\right\vert -\left\vert z_{2}\right\vert \right\vert \ll1.
\]
which implies%
\begin{equation}
{\overline{(J_{z})}}_{(j)}-\frac{q\Phi}{2\pi c}\approx\epsilon\left(
-1\right)  ^{j}\frac{\hbar}{\sqrt{\pi}}\left(  \sqrt{{\overline{(R^{2})}%
}_{(j)}}+\sqrt{{\overline{(R_{c}^{2})}}_{(j)}}\right)  R_{\mathrm{quant}}%
^{-1}\ . \label{minL}%
\end{equation}
Then the quantity $d_{\left(  j\right)  }$ (\ref{distance}) is:%
\begin{equation}
d_{\left(  j\right)  }\approx\left(  -1\right)  ^{j+1}\sqrt{\frac{2}{\pi
\gamma}},\ \ \left\vert \left\vert z_{1}\right\vert -\left\vert z_{2}%
\right\vert \right\vert \ll1. \label{Rdif}%
\end{equation}

Thus, for $\left\vert \left\vert z_{1}\right\vert -\left\vert z_{2}\right\vert
\right\vert \ll1,$ and in the semiclassical approximation, the mean minimal
possible moving off $\left\vert d_{\left(  j\right)  }\right\vert $ of the
particle from solenoid line is of order $R_{\mathrm{quant}}$, in particular,
$d_{\left(  j\right)  }<0$ for states with $j=0,$ and $d_{\left(  j\right)
}>0$ for states with $j=1$, independently on the sign of the difference
$\left\vert z_{1}\right\vert -\left\vert z_{2}\right\vert $.

Eqs. (\ref{s13}) and (\ref{z5}) allow us to calculate variances of the
operators $\hat{N}_{k}$,%
\[
\mathrm{Var}_{j}\left(  N_{k}\right)  ={\overline{(N_{k}^{2})}}_{(j)}-\left(
{\overline{(N_{k})}}_{(j)}\right)  ^{2}.
\]
In the semiclassical approximation, we have%
\begin{equation}
\mathrm{Var}_{j}\left(  N_{k}\right)  \approx\left\vert z_{k}\right\vert
^{2},\;\left\vert \left\vert z_{1}\right\vert -\left\vert z_{2}\right\vert
\right\vert \gg1;\;\mathrm{Var}_{j}\left(  N_{k}\right)  \approx\left(
1-1/\pi\right)  \left\vert z_{k}\right\vert ^{2},\;\left\vert \left\vert
z_{1}\right\vert -\left\vert z_{2}\right\vert \right\vert \ll1. \label{VarN}%
\end{equation}

Thus, standard deviations of $\hat{R}^{2}$ and $\hat{R}_{c}^{2}$ in the
semiclassical ICS are of the same order for any value $\left\vert \left\vert
z_{1}\right\vert -\left\vert z_{2}\right\vert \right\vert $, namely:%
\[
\delta_{j}\left(  R^{2}\right)  =\sqrt{\mathrm{Var}_{j}\left(  R^{2}\right)
}\sim R_{\mathrm{quant}}\sqrt{{\overline{(R^{2})}}_{(j)}},\;\delta_{j}\left(
R_{c}^{2}\right)  =\sqrt{\mathrm{Var}_{j}\left(  R_{c}^{2}\right)  }\sim
R_{\mathrm{quant}}\sqrt{{\overline{(R_{c}^{2})}}_{(j)}}.
\]
In this case, the typical spread of the radii $R$ and $R_{c}$ are given by the
standard deviations
\begin{equation}
\delta_{j}\left(  R\right)  =\delta_{j}\left(  R^{2}\right)  \left[
{\overline{(R^{2})}}_{(j)}\right]  ^{-1/2}\sim R_{\mathrm{quant}},\;\delta
_{j}\left(  R_{c}\right)  =\delta_{j}\left(  R_{c}^{2}\right)  \left[
{\overline{(R_{c}^{2})}}_{(j)}\right]  ^{-1/2}\sim R_{\mathrm{quant}}.
\label{sdR}%
\end{equation}

For $\left\vert \left\vert z_{1}\right\vert -\left\vert z_{2}\right\vert
\right\vert \ll1,$ the difference ${\overline{(R^{2})}}_{(j)}-{\overline
{(R_{c}^{2})}}_{(j)}$ is of the order of the standard deviation of $\hat
{R}^{2}-\hat{R}_{c}^{2}$, which is $\delta_{j}\left(  R^{2}\right)
+\delta_{j}\left(  R_{c}^{2}\right)  $. Therefore, the mean angular momentum
(\ref{minL}) is of the order of $\hat{J}_{z}$ standard deviation, and
for\emph{\ }$\left\vert \left\vert z_{1}\right\vert -\left\vert z_{2}%
\right\vert \right\vert \ll1,$\ the quantum scale of angular momentum is much
greater than\emph{\ }$\hbar$\emph{.} In this case $\left\vert d_{\left(
j\right)  }\right\vert $ is of the order $\delta_{j}\left(  R\right)
+\delta_{j}\left(  R_{c}\right)  \sim R_{\mathrm{quant}}$.

We note that ICS \textit{à la} Malkin-Man'ko \cite{MalMa68} (the case $\Phi
=0$) can be associated with the superposition of the $j=1$ and $j=0$ states
for $\mu=0$, which includes already all the eigenfunctions of $\hat{J}_{z}$.
The inner product on the $xy$-plane between such states is the sum
$\mathcal{R}_{\sigma}^{(0)}+\mathcal{R}_{\sigma}^{(1)}$ at $\mu_{\sigma}=0$.
Using eqs. (8.511.1) from \cite{GR94}, we find%
\[
\left.  \left(  \mathcal{R}_{\sigma}^{(0)}+\mathcal{R}_{\sigma}^{(1)}\right)
\right\vert _{\mu_{\sigma}=0}=\exp\left(  z_{1}^{\ast}z_{1}^{\prime}%
+z_{2}^{\ast}z_{2}^{\prime}\right)  \Longrightarrow\left.  \left(
\mathcal{\tilde{R}}_{\sigma}^{(0)}+\mathcal{\tilde{R}}_{\sigma}^{(1)}\right)
\right\vert _{\mu_{\sigma}=0}=1.
\]
Therefore, in such ICS ${\overline{(N_{k})}}=\left\vert z_{k}\right\vert ^{2}%
$. Similar mutual compensations take place in expressions for $\mathcal{\tilde
{R}}_{\sigma}^{(0)}$and $\mathcal{\tilde{R}}_{\sigma}^{(1)}$ for $\mu\neq0$ in
the semiclassical limit, $\left\vert z_{k}\right\vert ^{2}\gg1$. E.g., for
$\left\vert \left\vert z_{1}\right\vert -\left\vert z_{2}\right\vert
\right\vert \ll1,$ we obtain:%
\[
\mathcal{R}_{\sigma}^{(0)}+\mathcal{R}_{\sigma}^{(1)}=1+O\left(  \left\vert
z_{k}\right\vert ^{-2}\right)  .
\]
One can see that in states that are superpositions between different $j$,
leading corrections to means ${\overline{(N_{k})}}=\left\vert z_{k}\right\vert
^{2}$ (\ref{Nj}) disappear in the semiclassical approximation.

If $\left\vert z_{1}\right\vert $ and $\left\vert z_{2}\right\vert $ differ
essentially, i.e., $\left\vert \left\vert z_{1}\right\vert -\left\vert
z_{2}\right\vert \right\vert \gg1$, one may believe that next-to-leading terms
in $\mathcal{R}_{\sigma}^{(j)}$, given by (\ref{z6.1}), remain uncompensated.
That is not true. To see this, one has to take into account that
next-to-leading terms in $\mathcal{R}_{\sigma}^{(0)}+\mathcal{R}_{\sigma
}^{(1)}$ are due to contributions from (\ref{z6.1}) and from leading terms in
$\tilde{Q}_{\alpha}(u,\,v)$ for $\left\vert u\right\vert \gg\left\vert
v\right\vert $,
\begin{equation}
\tilde{Q}_{\alpha}(u,\,v)=\frac{1}{2\sqrt{\pi uv}}\left(  v/u\right)
^{\alpha}e^{-\left(  v-u\right)  ^{2}},\;\left\vert u\right\vert \gg\left\vert
v\right\vert \gg1. \label{z6.5}%
\end{equation}
We recall that for ICS, the domain $\left\vert u\right\vert >\left\vert
v\right\vert $ is not classical one even if $\left\vert z_{k}\right\vert
^{2}\gg1$.

Using (\ref{s24}), (\ref{s25}), and the representation
\[
\Delta_{\alpha}(u,v)=1-\varpi_{\alpha}(u,v),\ \ \varpi_{\alpha}(u,v)=\left.
\left(  v/u\right)  ^{\alpha}I_{\alpha}(2uv)e^{-u^{2}-v^{2}}\right/  \tilde
{Q}_{\alpha}(u,\,v),
\]
we find:%
\begin{align}
&  {\overline{(x-i\epsilon y)}}_{(j)}=\sqrt{2\gamma^{-1}}\left[
{\overline{(a_{2})}}_{(j)}-{\overline{(a_{1})}}_{(j)}^{\ast}\right]
,\nonumber\\
&  {\overline{(a_{2})}}_{(0)}-{\overline{(a_{1})}}_{(0)}^{\ast}=z_{2}%
-z_{1}^{\ast}\Delta_{1-\mu}(|z_{1}|,|z_{2}|)=z_{2}-z_{1}^{\ast}+z_{1}^{\ast
}\varpi_{1-\mu}(|z_{1}|,|z_{2}|),\nonumber\\
&  {\overline{(a_{2})}}_{(1)}-{\overline{(a_{1})}}_{(1)}^{\ast}=z_{2}%
\Delta_{\mu}(|z_{2}|,|z_{1}|)-z_{1}^{\ast}=z_{2}-z_{1}^{\ast}-z_{2}\varpi
_{\mu}(|z_{2}|,|z_{1}|).\label{z9}%
\end{align}
By the help of eqs. \ (\ref{s25}) and (\ref{Nj}) we calculate the variance of
$\left(  x+y\right)  $ in $j$-type states:
\begin{align}
&  \mathrm{Var}_{j}\left(  x+y\right)  ={\overline{\left(  \left\vert
x-i\epsilon y\right\vert ^{2}\right)  }}_{(j)}-\left\vert {\overline
{(x-i\epsilon y)}}_{(j)}\right\vert ^{2}\nonumber\\
&  \ =2\gamma^{-1}\left[  {\overline{(N_{1})}}_{(j)}+{\overline{(N_{2})}%
}_{(j)}+1-\left\vert {\overline{(a_{1})}}_{(j)}\right\vert ^{2}-\left\vert
{\overline{(a_{2})}}_{(j)}\right\vert ^{2}\right]  .\label{z9b}%
\end{align}

Let us consider the semiclassical limit $\left\vert z_{k}\right\vert ^{2}\gg1$
for ICS with $j=0$ and $\left\vert z_{1}\right\vert \lesssim\left\vert
z_{2}\right\vert $ and for ICS with $j=1$ and $\left\vert z_{1}\right\vert
\gtrsim\left\vert z_{2}\right\vert ,$ in both such cases $\left\vert
v\right\vert \gtrsim\left\vert u\right\vert \gg1$. In this case, using the
above results, one can verify that corrections to the classical expression
$\left[  {\overline{(a_{2})}}_{(j)}-{\overline{(a_{1})}}_{(j)}^{\ast}\right]
=z_{2}-z_{1}^{\ast}$ are small. In particular, using eqs. (\ref{z6.1}) and
(\ref{z6.4}), and asymptotics of $I_{\alpha}(2uv)$, in the next-to-leading
approximation we obtain the following result%
\begin{align}
d_{\alpha}(u,v) &  =\frac{1}{2\sqrt{\pi uv}}\left(  v/u\right)  ^{\alpha
}e^{-\left(  v-u\right)  ^{2}},\;\left\vert v\right\vert \gg\left\vert
u\right\vert ;\nonumber\\
d_{\alpha}(u,v) &  =\frac{1}{\sqrt{\pi u}}\left(  1-2\frac{v-u}{\sqrt{\pi}%
}+\frac{\alpha-1/2}{\sqrt{\pi}u}\right)  ,\;\left\vert v-u\right\vert
\ll1.\label{z10}%
\end{align}
Thus, eqs. (\ref{z9}) match with once (\ref{z6}) in the classical limit, and
due to (\ref{z6}), (\ref{z6.2}), and (\ref{z6.4}), in the semiclassical
approximation, the variances\emph{\ }(\ref{z9b}) are relatively small,%
\begin{align}
\mathrm{Var}_{j}\left(  x+y\right)   &  \approx2\gamma^{-1},\;\left\vert
z_{2}\right\vert \gg\left\vert z_{1}\right\vert \;\mathrm{for\;}%
j=0,\;\left\vert z_{1}\right\vert \gg\left\vert z_{2}\right\vert
\;\mathrm{for\;}j=1;\nonumber\\
\mathrm{Var}_{0}\left(  x+y\right)   &  \approx\frac{4|z_{1}|}{\sqrt{\pi
}\gamma},\;\mathrm{Var}_{1}\left(  x+y\right)  \approx\frac{4|z_{2}|}%
{\sqrt{\pi}\gamma},\;\left\vert \left\vert z_{1}\right\vert -\left\vert
z_{2}\right\vert \right\vert \ll1.\label{z10b}%
\end{align}
However, near the AB solenoid, where $\left\vert \left\vert z_{1}\right\vert
-\left\vert z_{2}\right\vert \right\vert \ll1,$\ the variances increase
significantly. Thus, it turns out that quantum length scale for $\sqrt
{{\overline{(R^{2})}}_{(j)}}$\ and $\sqrt{{\overline{(R_{c}^{2})}}_{(j)}}$\ is
essentially different from quantum length scale for $\overline{(x)}_{(j)}%
$\ and\ $\overline{(y)}_{(j)}$.

Note that in this case, the principal part of next-to-leading contributions to
$\overline{(x)}_{(j)}$ and \ $\overline{(y)}_{(j)}$, given by (\ref{z10}), do
not depend on $\mu$, which is quite similar to the behavior of ${\overline
{(N_{k})}}_{(j)}$ given by (\ref{Nj}). However, here a continuous limit to the
case $\mu=0$ does not exist. This can be checked considering means
$\overline{(x)}$ and $\overline{(y)}$ in the superposition of the $j=1$ and
$j=0$ states. For example, the mean $\overline{(x)}$ in the latter
superposition includes both means $\overline{(x)}_{(j)}$ and some interference
terms given by (\ref{s15}). The latter terms are absent only for $\mu=0$. That
is why the means $\overline{(x)}$ and \ $\overline{(y)}$ in the Malkin-Man'ko
CS cannot be obtained in the limit $\mu\rightarrow0$. Thus, namely means
$\overline{(x)}_{(j)}$,\ $\overline{(y)}_{(j)}$, and $\overline{(x)}%
$,$\ \overline{(y)}$ are especially sensitive to the topological effect of
breaking of the translational symmetry in the $xy$-plane due to the presence
of AB solenoid.

In the pure quantum case, the mean values depend significantly on the particle
spin and on the mantissa $\mu$ and are quite different from the corresponding
classical values. E.g., for small $\left\vert u\right\vert $ and $\left\vert
v\right\vert $ the functions $Q_{\alpha}(u,\,v)$ with positive and negative
$\alpha$ behave essentially different. Using the representation (8.445)
\cite{GR94} for the function $I_{\alpha}(2uv)$, we obtain:%
\begin{align*}
&  Q_{\alpha}(u,\,v)\approx v^{2\alpha}/\Gamma\left(  1+\alpha\right)
,\ \left\vert uv\right\vert \ll1\\
&  ~\Longrightarrow\left\{
\begin{array}
[c]{c}%
Q_{\alpha}(u,\,v)\overset{\left\vert uv\right\vert \rightarrow
0}{\longrightarrow}0,\ \ \alpha>0,\\
Q_{\alpha}(u,\,v)\overset{\left\vert uv\right\vert \rightarrow
0}{\longrightarrow}\infty,\ \ \alpha<0.
\end{array}
\right.
\end{align*}
Note that in the case of spinless particle $\alpha>0$ (see \cite{265}), while
$\alpha$ can take also negative values $-1<\alpha<0$ in the case of spinning
particle. Thus, e.g., for $|z_{1}z_{2}|\ll1$, ${\overline{(a_{2})}}%
_{(j)}-{\overline{(a_{1})}}_{(j)}^{\ast}$ differs essentially from
$z_{2}-z_{1}^{\ast}$. We have $\Delta_{\alpha}(u,v)=v^{2}\left(
\alpha+1\right)  ^{-1}$. Then,%
\begin{equation}
{\overline{(a_{2})}}_{(0)}-{\overline{(a_{1})}}_{(0)}^{\ast}\approx
z_{2},\;{\overline{(a_{2})}}_{(1)}-{\overline{(a_{1})}}_{(1)}^{\ast}%
\approx-z_{1}^{\ast}. \label{z11}%
\end{equation}
In this case, using (\ref{s13}) and (\ref{s12}), we obtain $\mathrm{Var}%
_{j}\left(  x+y\right)  \approx2\gamma^{-1}$. However, it is big in comparison
with small $\left\vert z_{k}\right\vert ^{2}$.

If $\left\vert u\right\vert \gg\left\vert v\right\vert $, we deal with the
quantum case even for big $\left\vert z_{k}\right\vert ^{2}$. Here
$\Delta_{\alpha}(u,v)=v/u\rightarrow0$ which gives a justification for
relations (\ref{z11}). In addition, in the quantum case, we have
${\overline{(N_{k})}}_{(j)}\sim1,$ and, at the same time, contributions to
${\overline{(N_{k})}}_{(j)}$ that depend on $z_{k}$ are much smaller than $1$.
That is why means ${\overline{(R^{2})}}_{(j)}$ and ${\overline{(R_{c}^{2})}%
}_{(j)}$, which are expressed via ${\overline{(N_{k})}}_{(j)}$ by eq.
(\ref{RRmean}), depend only slightly on $z_{k}$. In this case the variances%
\[
\mathrm{Var}_{0}\left(  x+y\right)  \approx2\gamma^{-1}\left\vert
z_{1}\right\vert ^{2},\ \mathrm{Var}_{1}\left(  x+y\right)  \approx
2\gamma^{-1}\left\vert z_{2}\right\vert ^{2}%
\]
are much bigger than squares of the corresponding means $\left\vert
{\overline{(x-i\epsilon y)}}_{(j)}\right\vert ^{2}$.

Let us consider uncertainty relations in the semiclassical ICS. Let $\hat
{F}_{1}$ and $\hat{F}_{2}$ be two self-adjoint operators satisfying the
commutation relation $\left[  \hat{F}_{1},\hat{F}_{2}\right]  =i\hat{F}_{3}$,
where $\hat{F}_{3}$ is a symmetric operator with a real mean $\overline
{\left(  F_{3}\right)  }$. In this case the uncertainty relation
\[
\mathrm{Var}\left(  F_{1}\right)  \mathrm{Var}\left(  F_{2}\right)  \geq
\frac{1}{4}\overline{\left(  F_{3}\right)  }%
\]
holds, see e.g. \cite{Dav76}. Adopting this general relation to our particular
cases, we obtain:
\begin{equation}
\mathrm{Var}_{j}\left(  \mathbf{P}_{\bot}^{2}\right)  \mathrm{Var}_{j}\left(
x+y\right)  \geq\hbar^{2}\left\vert {\overline{(P_{1}+i\epsilon P_{2})}}%
_{(j)}\right\vert ^{2},\;\mathrm{Var}_{j}\left(  L_{z}\right)  \mathrm{Var}%
_{j}\left(  x+y\right)  \geq\frac{\hbar^{2}}{4}\left\vert {\overline
{(x-i\epsilon y)}}_{(j)}\right\vert ^{2}. \label{NR}%
\end{equation}
Here ${\overline{(x-i\epsilon y)}}_{(j)}$ is given by (\ref{z9})
and$\left\vert {\overline{(P_{1}+i\epsilon P_{2})}}_{(j)}\right\vert ^{2}$ can
be represented by the help of (\ref{c1a}) as
\[
{\overline{(P_{1}+i\epsilon P_{2})}}_{(j)}=2\gamma\hbar^{2}\left\vert
{\overline{(a_{1})}}_{(j)}\right\vert ^{2}.
\]
The variances $\mathrm{Var}_{j}\left(  \mathbf{P}_{\bot}^{2}\right)  $ and
$\mathrm{Var}_{j}\left(  L_{z}\right)  $ can be expressed via the
$\mathrm{Var}_{j}\left(  N_{k}\right)  $ (\ref{VarN}),
\[
\mathrm{Var}_{j}\left(  \mathbf{P}_{\bot}^{2}\right)  =\left(  2\gamma
\hbar^{2}\right)  ^{2}\mathrm{Var}_{j}\left(  N_{1}\right)  ,\;\mathrm{Var}%
_{j}\left(  L_{z}\right)  =\hbar^{2}\mathrm{Var}_{j}\left(  N_{1}%
-N_{2}\right)  ,
\]
and $\mathrm{Var}_{j}\left(  x+y\right)  $ are given by (\ref{z10b}).

We note that $\left\vert {\overline{(a_{1})}}_{(j)}\right\vert ^{2}%
\approx\left\vert z_{1}\right\vert ^{2}$ for any $\left\vert \left\vert
z_{1}\right\vert -\left\vert z_{2}\right\vert \right\vert ,$ and for
definiteness sake, we suppose that $\left\vert z_{1}^{\ast}-z_{2}\right\vert
=\left\vert \left\vert z_{1}\right\vert -\left\vert z_{2}\right\vert
\right\vert $ for $\left\vert \left\vert z_{1}\right\vert -\left\vert
z_{2}\right\vert \right\vert \ll1$. Then, using (\ref{VarN}) and (\ref{z10b}),
we see that for $\left\vert \left\vert z_{1}\right\vert -\left\vert
z_{2}\right\vert \right\vert \gg1$\ the products of the variances from
(\ref{NR}) are close to their possible minimal values,
\[
\mathrm{Var}_{j}\left(  \mathbf{P}_{\bot}^{2}\right)  \mathrm{Var}_{j}\left(
x+y\right)  \approx4\hbar^{2}\left\vert {\overline{(P_{1}+i\epsilon P_{2})}%
}_{(j)}\right\vert ^{2},\;\mathrm{Var}_{j}\left(  L_{z}\right)  \mathrm{Var}%
_{j}\left(  x+y\right)  \approx\hbar^{2}\left\vert {\overline{(x-i\epsilon
y)}}_{(j)}\right\vert ^{2},
\]
and for $\left\vert \left\vert z_{1}\right\vert -\left\vert z_{2}\right\vert
\right\vert \ll1$ these variances are much bigger than the means $\hbar
^{2}\left\vert {\overline{(P_{1}+i\epsilon P_{2})}}_{(j)}\right\vert ^{2}$ and
$\left(  \hbar^{2}/4\right)  \left\vert {\overline{(x-i\epsilon y)}}%
_{(j)}\right\vert ^{2}$, respectively.

Of course, the AB effect is global. However, there exists a difference how
this effect manifests itself in the pure AB field and in the MSF. In the
latter case, there exists a possibility to characterize especially constructed
quantum states with respect to their "closeness" to the AB solenoid. Namely
the CS have, in a sense, such characteristics. The more close are such states
to the solenoid, the more they are affected by it.

\section{Time dependent CS\label{S5}}

On the base of ICS discussed above, one can construct already time-dependent
CS (we call them simply CS in what follows) as solutions of the corresponding
nonstationary wave equations. One ought to mention that CS for nonrelativistic
spinless particle in the MSF were constructed in our recent work \cite{265}.
Below, we are going to construct CS for nonrelativistic and relativistic
spinning particles in $2+1$ and $3+1$ dim.

\subsection{Nonrelativistic particles\label{SS5.1}}

In $2+1$ dim. the quantum behavior of nonrelativistic spin up\ particle
(antiparticle) is governed by the Pauli equation (\ref{Pauli}), where the
Hamiltonian can be represented as follows%
\[
\hat{H}_{\pm}^{\mathrm{NR}}=\hbar\omega_{\mathrm{NR}}\left.  \left[  \hat
{N}_{1}+\left(  1-\sigma\epsilon\right)  /2\right]  \right\vert _{\sigma=\pm
1}.
\]
Solutions $\Psi_{\pm1}^{\mathrm{up}}(t,\,\mbox{\boldmath$r$})$ of such an
equation are%

\[
\Psi_{\sigma}^{\mathrm{up}}(t,\,\mbox{\boldmath$r$})=\mathcal{N}\exp\left\{
-i\left[  \omega_{\mathrm{NR}}\left(  \sigma-\epsilon\right)  /2\right]
t\right\}  \Phi_{\sigma}(t,\,\varphi,\,\rho)\upsilon_{\sigma},
\]
respectively for $\pm$ cases, where $\mathcal{N}$ is normalization constant,
$\upsilon_{\sigma}$ is given by (\ref{t-s}), and functions $\Phi_{\sigma}$ are
solutions of the following equation:
\begin{equation}
i\partial_{t}\Phi_{\sigma}(t,\,\varphi,\,\rho)=\sigma\omega_{\mathrm{NR}}%
\hat{N}_{1}\Phi_{\sigma}(t,\,\varphi,\,\rho)\ . \label{NR1}%
\end{equation}
One can obey (\ref{NR1}) setting $\Phi_{\sigma}(t,\,\varphi,\,\rho)=\left.
\mathbf{\Phi}_{z_{1},\,z_{2},\sigma}^{(j)}(\varphi,\,\rho)\right\vert
_{z_{1}=z_{1}\left(  t\right)  },$ where $z_{1}\left(  t\right)  $ is a
complex function of time $t$. Then
\begin{equation}
i\partial_{t}\mathbf{\Phi}_{z_{1},\,z_{2},\sigma}^{(j)}(\varphi,\,\rho
)=i{\dot{z}}_{1}\partial_{z_{1}}\mathbf{\Phi}_{z_{1},\,z_{2},\sigma}%
^{(j)}(\varphi,\,\rho),\ \ \dot{z}_{1}=dz_{1}/dt\ . \label{NR2}%
\end{equation}
Substituting (\ref{NR2}) into (\ref{NR1}), we find $i{\dot{z}}_{1}%
=\sigma\omega_{\mathrm{NR}}z_{1}$, where (\ref{s13}) is used. It is convenient
to write a solution for $z_{1}(t)$ as follows:
\begin{equation}
z_{1}(t)=-|z_{1}|\exp(-i\sigma\psi),\ \ \psi=\omega_{\mathrm{NR}}t+\psi_{0},
\label{NR3}%
\end{equation}
where $|z_{1}|$ is\ a given constant. Thus the functions
\begin{equation}
\Psi_{\mathrm{CS},\sigma}^{(j)\mathrm{up}}%
(t,\,\mbox{\boldmath$r$})=\mathcal{N}\exp\left\{  -i\left[  \omega
_{\mathrm{NR}}\left(  \sigma-\epsilon\right)  /2\right]  t\right\}
\mathbf{\Phi}_{z_{1}\left(  t\right)  ,\,z_{2},\sigma}^{(j)}(\varphi
,\rho)\upsilon_{\sigma} \label{NR4}%
\end{equation}
are solutions of the $2+1$ Pauli equation for spin up\ particle. At the same
time they have special properties that allow us to treat them as CS and even
SS under certain conditions.

Consider the $2+1$ Pauli equation (\ref{Pauli}) for spin down\ particles. The
corresponding Hamiltonian reads:%
\[
\hat{H}_{\pm}^{\mathrm{NR}}=\hbar\omega_{\mathrm{NR}}\left.  \left[  \hat
{N}_{1}+\left(  1-\sigma\epsilon\right)  /2\right]  \right\vert _{\sigma=\mp
1}.
\]
Solutions $\Psi_{\sigma}^{\mathrm{down}}(t,\,\mbox{\boldmath$r$})$ of such an
equation have the form:%

\[
\Psi_{\sigma}^{\mathrm{down}}(t,\,\mbox{\boldmath$r$})=\mathcal{N}\exp\left\{
-i\left[  \omega_{\mathrm{NR}}\left(  \sigma+\epsilon\right)  /2\right]
t\right\}  \Phi_{-\sigma}(t,\,\varphi,\,\rho)\upsilon_{-\sigma}\ ,
\]
where $\sigma=+1$ for particle and $\sigma=-1$ for antiparticle. Similar to
spin up case, one can construct CS as follows%
\begin{equation}
\Psi_{\mathrm{CS},\sigma}^{(j)\mathrm{down}}%
(t,\,\mbox{\boldmath$r$})=\mathcal{N}\exp\left\{  -i\left[  \omega
_{\mathrm{NR}}\left(  \sigma+\epsilon\right)  /2\right]  t\right\}
\mathbf{\Phi}_{z_{1}\left(  t\right)  ,\,z_{2},-\sigma}^{(j)}(\varphi
,\,\rho)\upsilon_{-\sigma}. \label{NR5}%
\end{equation}

We note that means ${\overline{(F)}}_{(j)}$ in such CS are reduced to
${\overline{(F)}}_{(j)}$ given by (\ref{s18}).

In ($3+1)$-$\dim.,$ one can find CS for nonrelativistic spinning particles,
with a given spin polarization $s$. Such CS obey the nonrelativistic Dirac
equation with the Hamiltonian, $\hat{H}_{\pm}^{\mathrm{NR}}=\left[
({\mbox{\boldmath$\alpha$}}_{\perp}\mathbf{\hat{P}}_{\perp})^{2}+\hat{P}%
_{3}^{2}\right]  /2M$, see the Appendix A, and have the form:%
\begin{align}
&  \Psi_{\mathrm{CS},\sigma,s}^{(j)\mathrm{NR}}(x)=\exp\left\{  -\frac
{i}{\hbar}\left[  \frac{\left(  p_{3}\right)  ^{2}\sigma t}{2M}+p_{3}z\right]
\right\}  \Psi_{\mathrm{CS},\sigma,s}^{(j)\mathrm{NR}}%
(t,\,\mbox{\boldmath$r$}),\nonumber\\
&  \Psi_{\mathrm{CS},\sigma,+1}^{(j)\mathrm{NR}}%
(t,\,\mbox{\boldmath$r$})=\binom{\Psi_{\mathrm{CS},\sigma}^{(j)\mathrm{up}%
}(t,\,\mbox{\boldmath$r$})}{0},\;\Psi_{\mathrm{CS},\sigma,-1}^{(j)\mathrm{NR}%
}(t,\,\mbox{\boldmath$r$})=\binom{0}{\Psi_{\mathrm{CS},\sigma}%
^{(j)\mathrm{down}}(t,\,\mbox{\boldmath$r$})}, \label{NR6}%
\end{align}
where representation (\ref{s7}) is used. If to consider only physical
observables $\hat{F}$ that do not depend on $z$, then means ${\overline{(F)}%
}_{(j)}$ in CS (\ref{NR6}) are expressed via the corresponding means on the
$xy$-plane, i.e., via the corresponding means (\ref{s18}) for ($2+1)$-dim. particles.

\subsection{Relativistic particles in $\left(  3+1\right)  $%
-dimensions\label{SS5.2}}

In $3+1$ dim., we consider the Dirac equation in the light-cone variables,
such that $x_{-}$ plays the role of time. Solutions $\Psi$ of such an equation
have the form:%
\begin{align*}
&  \Psi_{\lambda,\sigma}(x)=\mathcal{N}\exp\left\{  -\frac{i}{2\hbar}\left[
\lambda Mcx_{+}+\left(  \frac{Mc}{\lambda}+\hbar\widetilde{\omega}\left(
1-\sigma\epsilon\right)  \right)  x_{-}\right]  \right\} \\
&  \times\Phi_{\lambda,\sigma}(x_{-,}\varphi,\,\rho)\left(
\begin{array}
[c]{c}%
\upsilon_{\sigma}\\
-\sigma\upsilon_{\sigma}%
\end{array}
\right)  ,
\end{align*}
where%
\begin{equation}
i\frac{\partial\Phi_{\lambda,\sigma}(x_{-},\varphi,\,\rho)}{\partial x_{-}%
}=\tilde{\omega}\hat{N}_{1}\Phi_{\lambda,\sigma}(x_{-},\varphi,\,\rho)\ ,
\label{i7}%
\end{equation}
see the Appendix A.

One can obey (\ref{i7}) setting%
\begin{equation}
\Phi_{\lambda,\sigma}(x_{-},\varphi,\,\rho)=\left.  \mathbf{\Phi}%
_{z_{1},\,z_{2},\sigma}^{(j)}(\varphi,\,\rho)\right\vert _{z_{1}=z_{1}\left(
x_{-}\right)  }, \label{118}%
\end{equation}
where $z_{1}\left(  x_{-}\right)  $ is a complex function of the time $x_{-}$.
Substituting (\ref{118}) into (\ref{i7}), taking into account%
\[
i\frac{\partial\mathbf{\Phi}_{z_{1},\,z_{2},\sigma}^{(j)}(\varphi,\,\rho
)}{\partial x_{-}}=i\frac{dz_{1}}{dx_{-}}\partial_{z_{1}}\mathbf{\Phi}%
_{z_{1},\,z_{2},\sigma}^{(j)}(\varphi,\,\rho),
\]
and (\ref{s13}), we find $idz_{1}/dx_{-}=\tilde{\omega}z_{1}$. A solution of
the latter equation has the form:
\begin{equation}
z_{1}\left(  x_{-}\right)  =-|z_{1}|\exp(-i\psi),\ \ \psi=\tilde{\omega}%
x_{-}+\psi_{0}, \label{119}%
\end{equation}
where $|z_{1}|$ and $\psi_{0}$\ are assumed to be some constants. Thus, we
have a set of solutions of the Dirac equation in the following form
\begin{align}
&  \Psi_{\mathrm{CS},\sigma}^{(j)}(x)=\mathcal{N}\exp\left\{  -\frac{i}%
{2\hbar}\left[  \lambda Mcx_{+}+\left(  \frac{Mc}{\lambda}+\hbar
\widetilde{\omega}\left(  1-\sigma\epsilon\right)  \right)  x_{-}\right]
\right\} \nonumber\\
&  \times\mathbf{\Phi}_{z_{1}(x_{-}),\,z_{2},\sigma}^{(j)}(\varphi
,\,\rho)\left(
\begin{array}
[c]{c}%
\upsilon_{\sigma}\\
-\sigma\upsilon_{\sigma}%
\end{array}
\right)  . \label{121}%
\end{align}
We interpret these solutions as CS with light-cone time $x_{-}$ evolution.

Suppose, we deal with physical observables $\hat{F}$ that do not depend of
$x_{+}$, which is natural for the axial symmetry of the problem under
consideration. Matrix elements of such observables in CS (\ref{121}) (we use
the inner product (\ref{ip}) on the hypersurface $x_{-}=\mathrm{const}$) take
the form:%
\[
\left(  \Psi_{\mathrm{CS},\sigma}^{(j)},\hat{F}\Psi_{\mathrm{CS}%
,\sigma^{\prime}}^{\prime(j^{\prime})}\right)  _{x_{-}}=\frac{\left(
4\pi\right)  ^{2}\hbar}{\gamma Mc}\delta_{\sigma^{\prime},\sigma}\delta\left(
\lambda^{\prime}-\lambda\right)  \left(  \mathbf{\Phi}_{z_{1},\,z_{2},\sigma
}^{(j)},\hat{F}_{\sigma}\mathbf{\Phi}_{z_{1}^{\prime},\,z_{2}^{\prime},\sigma
}^{(j^{\prime})}\right)  _{\bot},
\]
where the inner product $\left(  ,\right)  _{\bot}$ is given by eq.
(\ref{inpr}). That is why means ${\overline{(F)}}_{(j)}$ in such CS are
expressed via ${\overline{(F)}}_{(j)}$ given by (\ref{s18}).

Following the same way in the spinless case, one can construct CS that are
solutions of the Klein-Gordon equation.

\subsection{$t$ and $x_{-}$ evolution of mean values\label{SS5.3}}

Let us calculate means ${\overline{(x)}}_{(j)}$ and ${\overline{(y)}}_{(j)}$
in the nonrelativistic CS constructed above. These means are expressed via the
means ${\overline{(x-i\epsilon y)}}_{(j)}$, which have the form (\ref{s24}),
(\ref{s25}). Taking into account eq. (\ref{NR3}), one can see that means
${\overline{(x)}}_{(j)}$ and ${\overline{(y)}}_{(j)}$ are moving along a
circle on the $xy$-plane with the cyclotron frequency $\omega_{\mathrm{NR}}$,
i.e., the trajectory of the means has the classical form. The same equations
allows one to find a mean radius ${\overline{(R)}}_{(j)}$ of such a circle and
the distance ${\overline{(R_{c})}}_{(j)}$ between its center and the origin,%
\begin{align}
{\overline{(R)}}_{(0)}  &  =\sqrt{2\gamma^{-1}}\left\vert z_{1}\right\vert
\Delta_{1-\mu_{\sigma}}(|z_{1}|,|z_{2}|),\ \ {\overline{(R_{c})}}_{(0)}%
=\sqrt{2\gamma^{-1}}\left\vert z_{2}\right\vert ,\nonumber\\
{\overline{(R)}}_{(1)}  &  =\sqrt{2\gamma^{-1}}\left\vert z_{1}\right\vert
,\ \ {\overline{(R_{c})}}_{(1)}=\sqrt{2\gamma^{-1}}\left\vert z_{2}\right\vert
\Delta_{\mu_{\sigma}}(|z_{2}|,|z_{1}|). \label{mv2}%
\end{align}
Note that for the spinless particle $\mu_{\sigma}=\mu$.

In the general case, the quantities ${\overline{(R)}}_{(j)}^{2}$ and
${\overline{(R_{c})}}_{(j)}^{2}$ do not coincide with the corresponding
quantities ${\overline{(R^{2})}}_{(j)}$ and ${\overline{(R_{c}^{2})}}_{(j)}$
given by eq. (\ref{RRmean}). The latter quantities are expressed in terms of
means of square of the transverse kinetic energy and $\hat{J}_{z}$ according
to (\ref{s0.2}) and (\ref{s13}).

It follows from eq. (\ref{s25}) that $\Delta_{1-\mu_{\sigma}}(|z_{1}%
|,|z_{2}|)<1$\ and $\Delta_{\mu_{\sigma}}(|z_{2}|,|z_{1}|)<1$. This allows us
to give the following interpretation for two types of states with $j=0,1$.
States with $j=1$ correspond to orbits that embrace the AB solenoid (which
corresponds to $\left\vert z_{1}\right\vert \gtrsim\left\vert z_{2}\right\vert
$ in the semiclassical limit). For such orbits $\overline{(R_{c})}_{(1)}%
<R_{c},$ where the quantity $R_{c}=\sqrt{2\hbar/M\omega}\left\vert
z_{2}\right\vert $ is interpreted as a distance between AB solenoid and the
orbit center (see the classical limit of eq. (\ref{z7})). At the same time,
the mean radius of the orbits coincides with the classical radius
$R=\sqrt{2\hbar/M\omega}\left\vert z_{1}\right\vert $. The interpretation of
$R$ as the classical radius follows from eq. (\ref{z7}) in the classical
limit. States with $j=0$ correspond to orbits that do not embraces the AB
solenoid\ (which corresponds to $\left\vert z_{1}\right\vert \lesssim
\left\vert z_{2}\right\vert $ in the semiclassical limit). For such orbits
$\overline{(R_{c})}_{(0)}=R_{c}$ and $\overline{(R)}_{(0)}<R$.

\begin{figure}[ptb]%
\centering
\includegraphics[
height=2.7735in,
width=2.8703in
]%
{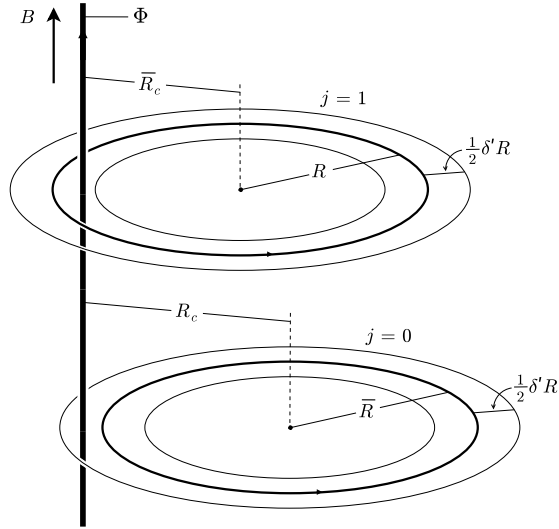}%
\caption{ Spreads of particle positions near classical orbits for $R\approx
R_{c}$.}%
\label{spread}%
\end{figure}

One can see the standard deviations $\delta_{j}\left(  R\right)  $,
$\delta_{j}\left(  R_{c}\right)  $, and $\delta_{j}\left(  x+y\right)
=\sqrt{\mathrm{Var}_{j}\left(  x+y\right)  }$ in CS (\ref{sdR}) and
(\ref{z10b}), are relatively small for the semiclassical orbits situated far
enough from the solenoid, i.e., for $\left\vert \left\vert z_{1}\right\vert
-\left\vert z_{2}\right\vert \right\vert \gg1$. In this case the CS are in
main concentrated near classical orbits. In the most interesting case when a
semiclassical orbit is situated near the solenoid, such that the condition
$\left\vert \left\vert z_{1}\right\vert -\left\vert z_{2}\right\vert
\right\vert \ll1$ holds, the standard deviation $\delta_{j}\left(  x+y\right)
$ increases significantly, $\delta_{j}\left(  x+y\right)  =\delta^{\prime
}\left(  R\right)  \approx2\pi^{-1/4}\gamma^{-1/2}|z_{1}|^{1/2}$, while the
standard deviations $\delta_{j}\left(  R\right)  $ and $\delta_{j}\left(
R_{c}\right)  $ remain relatively small. In this case $R\approx R_{c}$,
however, $\overline{(R_{c})}_{(1)}<R$ and $\overline{(R)}_{(0)}<R_{c}$, as it
has to be for the semiclassical orbits. Thus, the standard deviation
$\delta_{j}\left(  x+y\right)  $ of particle positions near classical orbits
is relatively large at $R\approx R_{c}$, such that $\delta^{\prime}\left(
R\right)  \gg\left\vert R-\overline{(R_{c})}_{(1)}\right\vert ,\left\vert
\overline{(R)}_{(0)}-R_{c}\right\vert $. We show the corresponding spreads on
Fig. \ref{spread} (where $\overline{R}_{c}=\overline{(R_{c})}_{(1)}$ and
$\overline{R}=\overline{(R)}_{(0)}$)%

We stress that for $\mu\neq0,$ relations between CS/SS parameters of particle
trajectory in a constant uniform magnetic field differ from classical ones due
to the presence of the AB solenoid. Above, we have demonstrated this,
considering the radius $R$ (related to the energy of particle rotation) and
the distance $R_{c}$ (related to particle angular momentum). Such relations do
not feel the presence of AB solenoid for $\mu=0$, and, even for $\mu\neq0,$ in
the classical limit.

For relativistic particles in ($3+1)$-dim., we consider means ${\overline
{(x-i\epsilon y)}}_{(j)}$ in CS (\ref{121}) on the hypersurface $x_{-}%
=\mathrm{const}$. Such means are reduced to the means ${\overline{(x-i\epsilon
y)}}_{(j)}$ represented above by expressions (\ref{s24}), (\ref{s25}). The
relations (\ref{mv2}) and (\ref{RRmean}) remain true. Here however, the
evolution is parametrized by the light-cone time $x_{-}$, via the function
$z_{1}(x_{-})$ given by eq. (\ref{119}). One can see that means ${\overline
{(x)}}_{(j)}$ and ${\overline{(y)}}_{(j)}$ are rotating along circles on the
$xy$-plane with the synchrotron frequency $\omega$, i.e., their trajectories
have the classical form (\ref{b20}).

\subsection{Ultrarelativistic particles in $\left(  2+1\right)  $-dimensions
\label{SS5.4}}

In subsec. \ref{SS5.2} we have constructed relativistic CS in 3+1 dim., and in
subsec. \ref{SS5.3}, we have demonstrated that in such CS the means have the
classical form (\ref{b20}). We succeeded to do this using light-cone
parametrization of the evolution via the function $z_{1}(x_{-})$ given by eq.
(\ref{119}). Such a parametrization is possible only in the relativistic case
in $3+1$ dim. Indeed, using (\ref{d11}), we can represent eq. (\ref{k4}) for
eigenfunctions of $\hat{P}_{0}+\hat{P}_{3}$ with the eigenvalues $\lambda$ in
the form of the first order Schrödinger-like equation (\ref{Sch-like}), where
$x_{-}$ plays the role of the time and the operator $\hat{H}_{x_{-}%
}=\mathcal{\hat{Q}}^{2}\left(  2\lambda Mc\right)  ^{-1}$ plays the role of
the Hamiltonian. The Hamiltonian $\hat{H}_{x_{-}}$ is quadratic with respect
to the momentum operators. In the cases in ($2+1)$-dim. considered above, the
light-cone variables $x_{\pm}$ cannot be introduced. Then we have to use the
time $t$ parametrization of the evolution. This is the reason why we cannot
construct CS as exact solutions of the Dirac equation. This is a consequence
of the fact that in the case under consideration, Dirac Hamiltonian is not
quadratic in momenta and the corresponding ICS do not maintain their form in
course of the evolution. Below, we consider an example of such an evolution of
ICS. We take massless $\zeta=+1$ fermions in ($2+1$)-dim. with the Hamiltonian
$\hat{H}^{\vartheta}=c\mbox{\boldmath$\sigma$\unboldmath}\mathbf{\hat{P}%
}_{\perp}$.

One can see (using results of the Appendix A) that ICS (\ref{s8}) obey the
following relation%
\begin{equation}
\hat{H}^{\vartheta}\mathbf{\Psi}_{\pm,\,z_{1},z_{2}}^{(j,+1)}(\varphi
,\,\rho)=\pm c\hat{\Pi}_{0}\mathbf{\Psi}_{\pm,\,z_{1},z_{2}}^{(j,+1)}%
(\varphi,\,\rho), \label{rp1}%
\end{equation}
where $\hat{\Pi}_{0}=\Pi_{0}\left(  \hat{N}_{1}\right)  $ is given by
(\ref{s5}) at $\sigma=-\vartheta$. Then a formal solution of the Dirac
equation, with ICS (\ref{s8}) as an initial condition, reads:%
\begin{equation}
\Psi_{\pm,\,z_{1},z_{2}}^{(j,+1)}(t,\varphi,\,\rho)=\exp\left[  \mp\frac
{ic}{\hbar}\hat{\Pi}_{0}t\right]  \mathbf{\Psi}_{\pm,\,z_{1},z_{2}}%
^{(j,+1)}(\varphi,\,\rho). \label{rp2}%
\end{equation}
We call such solution quasi-CS in what follows. As usual, we define means of
an operator $\hat{F}$ in quasi-CS (\ref{rp2}) by ${\overline{(F)}}_{(j)}$,%
\begin{equation}
{\overline{(F\left(  t\right)  )}}_{(j,\pm)}=\frac{\left(  \Psi_{\pm
,\,z_{1},z_{2}}^{(j,+1)}\left(  t\right)  ,\hat{F}\Psi_{\pm,\,z_{1},z_{2}%
}^{(j,+1)}\left(  t\right)  \right)  _{D}}{\left(  \mathbf{\Psi}_{\pm
,\,z_{1},z_{2}}^{(j,+1)},\mathbf{\Psi}_{\pm,\,z_{1},z_{2}}^{(j,+1)}\right)
_{D}}. \label{rp3}%
\end{equation}

One can see that in the semiclassical limit the means
\begin{equation}
{\overline{(x\left(  t\right)  -i\epsilon y\left(  t\right)  )}}_{(j,\pm
)}=\sqrt{2\gamma^{-1}}\left[  {\overline{(a_{2}\left(  t\right)  )}}_{(j,\pm
)}-{\overline{(a_{1}\left(  t\right)  )}}_{(j,\pm)}^{\ast}\right]  \label{rp4}%
\end{equation}
are moving along the corresponding classical trajectories. Since the operator
$\hat{a}_{2}$ commutes with $\hat{N}_{1}$, the mean ${\overline{(a_{2}\left(
t\right)  )}}_{(j,\pm)}$ does not depend on time and coincides with its
initial value, ${\overline{(a_{2}\left(  t\right)  )}}_{(j,\pm)}%
={\overline{(a_{2})}}_{(j)}$ , the latter is given by eq. (\ref{s26}).
Calculating the mean ${\overline{(a_{1}\left(  t\right)  )}}_{(j,\pm)}^{\ast
},$ we find that
\begin{align}
&  {\overline{(a_{1}\left(  t\right)  )}}_{(0,\pm)}^{\ast}=z_{1}^{\ast}%
\frac{\sum_{\sigma=\pm1}\exp\left[  \pm i\Omega\left(  \frac{\left\vert
z_{1}\right\vert }{2}\frac{d}{d\left\vert z_{1}\right\vert }\right)  t\right]
Q_{1-\mu_{\sigma}}^{-}(|z_{1}|,|z_{2}|)}{\sum_{\sigma=\pm1}Q_{1-\mu_{\sigma}%
}(|z_{1}|,|z_{2}|)},\nonumber\\
&  {\overline{(a_{1}\left(  t\right)  )}}_{(1,\pm)}^{\ast}=z_{1}^{\ast}%
\frac{\sum_{\sigma=\pm1}\exp\left[  \pm i\Omega\left(  \frac{\left\vert
z_{1}\right\vert }{2}\frac{d}{d\left\vert z_{1}\right\vert }\right)  t\right]
Q_{\mu_{\sigma}}(|z_{2}|,|z_{1}|)}{\sum_{\sigma=\pm1}Q_{\mu_{\sigma}}%
(|z_{2}|,|z_{1}|)}\ , \label{rp11}%
\end{align}
see the Appendix B, where the frequency operator $\Omega\left(  \hat{N}%
_{1}\right)  $ is given by eq. (\ref{rp10}).

Let now $\left\vert z_{1}\right\vert ^{2}\gg1,$ which corresponds to the
semiclassical limit. Using relations (\ref{z4}) and (\ref{z5}), we represent
(\ref{rp11}) in the form%
\begin{align*}
&  {\overline{(a_{1}\left(  t\right)  )}}_{(0,\pm)}^{\ast}=z_{1}^{\ast}%
\frac{\sum_{\sigma=\pm1}\exp\left[  \pm i\tilde{\Omega}t\right]  \tilde
{Q}_{1-\mu_{\sigma}}^{-}(|z_{1}|,|z_{2}|)}{\sum_{\sigma=\pm1}\tilde{Q}%
_{1-\mu_{\sigma}}(|z_{1}|,|z_{2}|)},\\
&  {\overline{(a_{1}\left(  t\right)  )}}_{(1,\pm)}^{\ast}=z_{1}^{\ast}%
\frac{\sum_{\sigma=\pm1}\exp\left[  \pm i\tilde{\Omega}t\right]  \tilde
{Q}_{\mu_{\sigma}}(|z_{2}|,|z_{1}|)}{\sum_{\sigma=\pm1}\tilde{Q}_{\mu_{\sigma
}}(|z_{2}|,|z_{1}|)}\ ,\\
&  \tilde{\Omega}=\Omega\left(  \left\vert z_{1}\right\vert ^{2}%
+\frac{\left\vert z_{1}\right\vert }{2}\frac{d}{d\left\vert z_{1}\right\vert
}\right)  .
\end{align*}
The semiclassical expansions of the operator $\tilde{\Omega}$ is:%
\[
\tilde{\Omega}\approx\omega\left(  \left\vert z_{1}\right\vert \right)
\left[  1-\frac{1}{4|z_{1}|}\frac{d}{d|z_{1}|}+O\left(  |z_{1}|^{-2}\right)
\right]  ,\;\omega\left(  \left\vert z_{1}\right\vert \right)  =c\left\vert
qB\right\vert \mathcal{E}^{-1}\left(  \left\vert z_{1}\right\vert \right)
,\ \ \mathcal{E}\left(  \left\vert z_{1}\right\vert \right)  =\sqrt{2\hbar
c\left\vert qB\right\vert }\left\vert z_{1}\right\vert ,
\]
such that the standard deviation of $\tilde{\Omega}$ is of the order
$\omega\left(  \left\vert z_{1}\right\vert \right)  \left\vert z_{1}%
\right\vert ^{-1}$. Here $\mathcal{E}\left(  \left\vert z_{1}\right\vert
\right)  $ is the mean energy in quasi-CS, in the classical limit. Taking into
account next-to-leading corrections to $\exp\left[  \pm i\tilde{\Omega
}t\right]  ,$ and using decompositions (\ref{z6.1}) and (\ref{z6.4}), we
obtain ${\overline{(a_{1}\left(  t\right)  )}}_{(j,\pm)}^{\ast}={\overline
{(a_{1})}}_{(j)}^{\ast}e^{\pm i\bar{\omega}_{j}t}\ $, where%
\begin{align}
&  \bar{\omega}_{j}=\omega\left(  \left\vert z_{1}\right\vert \right)  \left[
1+O\left(  \left\vert z_{1}\right\vert ^{-2}\right)  \right]  ,\ \ \left\vert
\left\vert z_{1}\right\vert -\left\vert z_{2}\right\vert \right\vert
\gg1;\nonumber\\
&  \bar{\omega}_{j}=\omega\left(  \left\vert z_{1}\right\vert \right)  \left[
1+\frac{\left(  -1\right)  ^{j}}{2\sqrt{\pi}|z_{1}|}+O\left(  \left\vert
z_{1}\right\vert ^{-2}\right)  \right]  ,\ \left\vert \left\vert
z_{1}\right\vert -\left\vert z_{2}\right\vert \right\vert \ll1. \label{rp12}%
\end{align}
In such an approximation, $\omega\left(  \left\vert z_{1}\right\vert \right)
$ coincides with the classic synchrotron frequency $\omega$ given by eq.
(\ref{b7}).

Thus, in the classical limit, means ${\overline{(x)}}_{(j,\pm)}$ and
${\overline{(y)}}_{(j,\pm)}$ in the quasi-CS are rotating along circles on the
$xy$-plane with the synchrotron frequency $\omega$. The greater is $\left\vert
z_{1}\right\vert ^{2}$ the smaller is spreading of the mean trajectories. One
can see that additional (due to the next corrections)\ modifications of
quasi-CS are essential only for evolution time that is much greater than the
classical rotation period.

The mean radius of a trajectory is%
\[
\sqrt{{\overline{(R^{2})}}_{(j,\pm)}}=\sqrt{\left[  2{\overline{(N_{1})}%
}_{(j,\pm)}+1+\vartheta\epsilon\right]  \gamma^{-1}}.
\]
The next-to-leading corrections to this relation can be find by the help of
eqs. (\ref{s22}), (\ref{z6.2}), and (\ref{Nj}). We note that in each
approximation the classical relation between the rotation frequency and the
radius holds, such that $\bar{\omega}_{j}\sqrt{{\overline{(R^{2})}}_{(j,\pm)}%
}=c$ (we remind that in the graphene case $c$ means the effective velocity of
light, that is the Fermi velocity $v_{F}$).

\section{ Summary and Discussion\label{S6}}

A new approach to constructing CS/SS in MSF is proposed. The main idea is
based on the fact that the AB solenoid breaks the translational symmetry in
the $xy$-plane, this has a topological effect such that there appear two types
of\emph{\ }circular trajectories which embrace and do not embrace the
solenoid. Due to this fact, one has to construct two different kinds of CS/SS,
which correspond to such trajectories in the semiclassical limit. Following
this idea, we construct CS in two steps, first the instantaneous CS (ICS) and
the time dependent CS/SS as an evolution of the ICS.

The approach is realized for nonrelativistic and relativistic spinning
particles that allows us to build CS both in ($2+1)$- and ($3+1)$-dim., using
some universal constructions, and gives a non-trivial example of SS/CS for
systems with a nonquadratic Hamiltonian.

It is stressed that CS depending on their parameters (quantum numbers)
describe both pure quantum and semiclassical states. An analysis is
represented that classifies parameters of the CS in such respect. Such a
classification is used for the semiclassical decompositions of various
physical quantities.

In the pure quantum case, the mean values depend significantly on the particle
spin and on the mantissa $\mu$ and are quite different from the corresponding
classical values.\emph{ }In the semiclassical approximation, relations between
CS/SS parameters and parameters that characterize classical trajectories are
established. In the general case these relations differ from ones in the pure
magnetic field and such a distinction can be treated as AB effect in the
CS/SS. The classical relations correspond to the leading approximation for
sufficiently large radii. Thus, the leading approximation in the semiclassical
expansions corresponds to the classical limit. Next-to-leading terms define
physical quantities in the semiclassical approximation. These terms depend on
the space dimension and particle spin.

The following properties of the constructed time dependent CS/SS should be stressed:

a) In the nonrelativistic case, both in ($2+1)$- and ($3+1)$-dim., the time
dependent CS in each time instant retain the form of the corresponding ICS.
The mean trajectories in such CS coincide with classical ones, whereas the
particle distributions are concentrated near the classical trajectories in the
semiclassical approximation. In the presence of the AB solenoid, the spread of
particle positions near the classical trajectory depends essentially on the
mutual disposition between the trajectory and the solenoid. Such a spread is
growing for trajectories situated near the AB solenoid. It should be note that
namely due to the bounded character of particle motion in the MSF, particle
positions are essentially sensitive to the topological effect of breaking of
the translational symmetry in the $xy$-plane due to the presence of AB
solenoid. Thus, in spite of the well-known fact that the AB effect is global,
in the MSF quantum states can be classified with respect to their "closeness"
to the AB solenoid.

b) In the relativistic case, in ($3+1)$-dim., CS are constructed in the
light-cone variables, where the evolution is parametrized by the light-cone
time $x_{-}.$ Such time-dependent CS obey all the properties as CS from the
previous item a).

c) In ($2+1)$-dim., we constructed time-dependent SS for massless fermions.
Such a problem can be related to the graphene physics. We call the constructed
SS quasi-CS since they retain ICS form with time evolution in the
next-to-leading semiclassical approximation. In such an approximation, the
classical relation between the rotation frequency and the radius holds, the
rotation frequency coincides with the classic synchrotron frequency in the
leading approximation. We stress a principle difference between ($2+1)$-dim.
and ($3+1)$-dim., in ($2+1)$-dim. Dirac Hamiltonian is not quadratic in the
momenta, whereas in ($3+1)$-dim. it is. Namely this fact is responsible for
the destruction of ICS in course of the evolution.

\subparagraph{{\protect\large Acknowledgement}}

The work of VGB is partially supported by Russian Science and Innovations
Federal Agency under contract No 02.740.11.0238 and Russia President grant
SS-3400.2010.2. The work of SPG is supported by FAPESP/Brasil and the program
Bolsista CAPES/Brasil. SPG thanks University of São Paulo for hospitality. DMG
acknowledges the permanent support of FAPESP and CNPq. DPMF thanks CNPq for a support.

\appendix

\section*{Appendix A. Quantum stationary states}

\subsubsection{$\left(  2+1\right)  $-dimensions}

In $(2+1)$-dim., the total angular momentum operator $\hat{J}=-i\hbar
\partial_{\varphi}+\hbar\sigma^{3}/2$, which is a dimensional reduction of the
operator $\hat{J}_{z}$ in ($3+1)$-dim. ($z$-component of the total angular
momentum operator given by (\ref{k13b})), is self-adjoint on $D_{H}%
^{\vartheta}$ and commutes with $\hat{H}^{\vartheta}$. There exist common
eigenvectors $\psi_{n_{1},\,n_{2}}^{(j)}(\varphi,\,\rho)$ of operators
$\hat{H}^{\vartheta}$ and $\hat{J}$,
\begin{align}
&  \hat{H}^{\vartheta}\psi_{n_{1},\,n_{2}}^{(j)}(\varphi,\,\rho)=cp_{0}%
\psi_{n_{1},\,n_{2}}^{(j)}(\varphi,\,\rho),\;cp_{0}=\pm\mathcal{E}%
,\;\mathcal{E}=\sqrt{\left(  Mc^{2}\right)  ^{2}+\mathcal{E}_{\perp}^{2}%
},\nonumber\\
&  \hat{J}\psi_{n_{1},\,n_{2}}^{(j)}(\varphi,\,\rho)=J\psi_{n_{1},\,n_{2}%
}^{(j)}(\varphi,\,\rho),\;J=\epsilon\hbar\left(  l_{0}-l+1/2\right)  ,\ j=1,2.
\label{abe10}%
\end{align}
It is convenient to use the following representation%
\begin{align}
&  \psi_{n_{1},\,n_{2}}^{(j)}(\varphi,\,\rho)=\left[  \sigma^{3}\left(
p_{0}-\mbox{\boldmath$\sigma$\unboldmath}\mathbf{\hat{P}}_{\perp}\right)
+Mc\right]  u_{n_{1},\,n_{2}}^{(j)}(\varphi,\rho),\nonumber\\
&  u_{n_{1},\,n_{2}}^{(j)}(\varphi,\rho)=\sum_{\sigma=\pm1}c_{\sigma}%
\Phi_{n_{1},\,n_{2},\sigma}^{(j)}(\varphi,\,\rho)\upsilon_{\sigma}\,,
\label{ab11}%
\end{align}
where%
\begin{equation}
\upsilon_{1}=\left(
\begin{array}
[c]{l}%
1\\
0
\end{array}
\right)  ,\;\;\upsilon_{-1}=\left(
\begin{array}
[c]{l}%
0\\
1
\end{array}
\right)  , \label{t-s}%
\end{equation}
and $c_{\sigma}$ are some constants. The columns $u_{n_{1},\,n_{2}}^{(j)}$ are
solutions of the eigenvalue problem%
\begin{equation}
c^{2}\left(  \mbox{\boldmath$\sigma$\unboldmath}\mathbf{\hat{P}}_{\perp
}\right)  ^{2}u_{n_{1},\,n_{2}}^{(j)}(\varphi,\rho)=\mathcal{E}_{\perp}%
^{2}u_{n_{1},\,n_{2}}^{(j)}(\varphi,\rho). \label{ab11a}%
\end{equation}
We note that the relation $\left(
\mbox{\boldmath$\sigma$\unboldmath}\mathbf{\hat{P}}_{\perp}\right)
^{2}=\mathbf{\hat{P}}_{\perp}^{2}-\epsilon\hbar c^{-1}\left\vert qB\right\vert
\sigma^{3}$ holds, which gives a relation to the energy spectrum for the
spinless case.

The functions $\Phi_{n_{1},\,n_{2},\sigma}^{(j)}(\varphi,\,\rho)$ have the
form%
\begin{align}
&  \Phi_{n_{1},\,n_{2},\sigma}^{(0)}(\varphi,\,\rho)=\mathcal{N}\exp\left\{
i\epsilon\left[  l_{0}-l+\left(  1-\epsilon\sigma\right)  /2\right]
\varphi\right\}  I_{n_{2},n_{1}}\left(  \rho\right)  ,\nonumber\\
&  n_{1}=m,\ \ n_{2}=m-\tilde{l}+\left(  1-\epsilon\sigma\right)
/2-\mu,\;\tilde{l}=l-\left(  1+\epsilon\right)  \left(  1+\sigma\right)
/2,\;-\infty<\tilde{l}\leq-\left(  1+\vartheta\epsilon\right)
/2;\ \nonumber\\
&  \Phi_{n_{1},\,n_{2},\sigma}^{(1)}(\varphi,\,\rho)=\mathcal{N}\exp\left\{
i\epsilon\left[  l_{0}-l+\left(  1-\epsilon\sigma\right)  /2\right]
\varphi+i\epsilon\pi\left[  l-\left(  1-\epsilon\right)  \left(
1+\sigma\right)  /4\right]  \right\}  I_{n_{1},n_{2}}\left(  \rho\right)
,\nonumber\\
&  n_{1}=m+\tilde{l}-\left(  1-\epsilon\sigma\right)  /2+\mu,\ \ n_{2}%
=m,\ \left(  1-\vartheta\epsilon\right)  /2\leq\tilde{l}<\infty\ ;\ \rho
=\frac{\gamma}{2}r^{2}. \label{ab12}%
\end{align}
Here $I_{n,m}(\rho)$, $m\geq0$ are Laguerre functions that are related to the
Laguerre polynomials $L_{m}^{\alpha}(\rho)$ (see \cite{GR94}) as follows:%
\begin{equation}
I_{m+\alpha,m}(\rho)=\sqrt{\frac{\Gamma\left(  m+1\right)  }{\Gamma\left(
m+\alpha+1\right)  }}e^{-\rho/2}\rho^{\alpha/2}L_{m}^{\alpha}\left(
\rho\right)  ,\ L_{m}^{\alpha}(\rho)=\frac{1}{m!}e^{\rho}\rho^{-\alpha}%
\frac{d^{m}}{d\rho^{m}}e^{-\rho}\rho^{m+\alpha}\,, \label{2.17a}%
\end{equation}
and $\mathcal{N}$ are normalization constants. For any real $\alpha>-1,$ the
functions $I_{\alpha+m,\,m}(\rho)$ form a complete orthonormal set on the
semiaxis $\rho\geqslant0$,%
\begin{equation}
\int_{0}^{\infty}I_{\alpha+k,\,k}(\rho)I_{\alpha+m,\,m}(\rho)d\rho
=\delta_{k,m}\,,\ \ \sum_{m=0}^{\infty}I_{\alpha+m,\,m}(\rho)I_{\alpha
+m,\,m}(\rho^{\prime})=\delta(\rho-\rho^{\prime})\ . \label{f8}%
\end{equation}
Thus, the domains $D_{H}^{\vartheta}$ are described completely by asymptotic
behavior of the functions from eqs. (\ref{ab12}).

We define the functions $\Phi_{n_{1}+s_{1},\,n_{2}+s_{2},\sigma}^{(j)}$
associated with transformation (\ref{s0}) as follows:%
\begin{align}
&  \Phi_{n_{1}+s_{1},\,n_{2}+s_{2},\sigma}^{(0)}(\varphi,\,\rho)=\mathcal{N}%
\exp\left\{  i\epsilon\left[  l_{0}-l-s_{1}+s_{2}+\left(  1-\epsilon
\sigma\right)  /2\right]  \varphi\right\}  I_{n_{2}+s_{2},n_{1}+s_{1}}\left(
\rho\right)  ,\nonumber\\
&  \Phi_{n_{1}+s_{1},\,n_{2}+s_{2},\sigma}^{(1)}(\varphi,\,\rho)=\mathcal{N}%
\exp\left\{  i\epsilon\left[  l_{0}-l-s_{1}+s_{2}+\left(  1-\epsilon
\sigma\right)  /2\right]  \varphi\right. \nonumber\\
&  \left.  +\pi\left[  l+s_{1}-s_{2}-\left(  1-\epsilon\right)  \left(
1+\sigma\right)  /4\right]  \right\}  I_{n_{1}+s_{1},n_{2}+s_{2}}\left(
\rho\right)  ,\ s_{1}=0,\pm1,\ \ s_{2}=0,\pm1. \label{dopf}%
\end{align}
There appear new functions $\Phi_{n_{1},\,n_{2}-1,\sigma}^{(0)}\,$ with
$n_{2}=m+1+\left(  \vartheta-\sigma\right)  \epsilon/2-\mu$ and $\Phi
_{n_{1}-1,\,n_{2},\sigma}^{(1)}$ with $n_{1}=m+\left(  \sigma-\vartheta
\right)  \epsilon/2+\mu$. Such functions were not defined by eqs.
(\ref{ab12}). In addition, for\emph{\ }$n_{1}=0$\emph{\ }or\emph{\ }$n_{2}=0$,
one has to bear in mind that%
\[
\hat{a}_{1}\left.  \Phi_{n_{1},\,n_{2},\sigma}^{(0)}(\varphi,\,\rho
)\right\vert _{m=0}=0,\;\hat{a}_{2}\left.  \Phi_{n_{1},\,n_{2},\sigma}%
^{(1)}(\varphi,\,\rho)\right\vert _{m=0}=0.
\]
This allows as to interpret $\left.  \Phi_{n_{1},\,n_{2},\sigma}%
^{(0)}\right\vert _{m=0}$ and $\left.  \Phi_{n_{1},\,n_{2},\sigma}%
^{(1)}\right\vert _{m=0}$ as vacuum states.

Let us define an inner product of two functions $f(\varphi,\,\rho)$
and$\,g(\varphi,\,\rho)$ as
\begin{equation}
(f,\,g)_{\bot}=\frac{1}{2\pi}\int_{0}^{\infty}d\rho\int_{0}^{2\pi}%
d\varphi\,f^{\ast}(\varphi,\rho)g(\varphi,\rho).\ \label{inpr}%
\end{equation}
With respect to such an inner product, the functions (\ref{ab12}) form an
orthogonal set,
\begin{equation}
\left(  \Phi_{n_{1}^{\prime},\,n_{2}^{\prime},\sigma}^{(j\,^{\prime})}%
,\Phi_{n_{1},\,n_{2},\sigma}^{(j)}\right)  _{\bot}=\left\vert \mathcal{N}%
\right\vert ^{2}\,\delta_{n_{1}^{\prime},\,n_{1}}\,\delta_{n_{2}^{\prime
},\,n_{2}}\,\delta_{j\,^{\prime},\,j}. \label{f23}%
\end{equation}
Moreover, these functions form a complete orthogonal set in $L^{2}\left(
\mathbb{R}^{2}\right)  $.

The energy spectrum of the self adjoint Hamiltonians (\ref{abe8a}) can be
represented with the help of the eigenvalues $\mathcal{E}_{\perp}^{2}$ as%
\begin{equation}
\mathcal{E}_{\perp\left(  \sigma\right)  }^{2}=2\hbar c\left\vert
qB\right\vert \left[  n_{1}+\left(  1-\sigma\epsilon\right)  /2\right]  ,
\label{ab14}%
\end{equation}
where, depending of the $\vartheta$ and $\epsilon$, the quantum number $n_{1}$
takes its possible values according to\ (\ref{ab12}). In the general case,
eigenvalues\emph{\ }$\mathcal{E}$ of the Hamiltonian $\hat{H}^{\vartheta}$ are
expressed via $\mathcal{E}_{\perp}^{2}$ for $\sigma=+1$ or $\sigma=-1$,
according to eq. (\ref{abe10}),\emph{\ }that is why $\mathcal{E}_{\perp\left(
\sigma\right)  }^{2}$ are labeled by the subscript $\sigma.$\emph{\ }Irregular
at the origin radial functions appear in the domain $D_{H}^{+}$ for $\tilde
{l}=0$ and $\sigma=-1,$ and in the domain $D_{H}^{-}$ for $\tilde{l}=0$ and
$\sigma=+1$.

We note that depending on $\vartheta\epsilon$, energy levels of states having
irregular radial functions coincide or differ from the Landau levels. In any
case the difference always depends on $\mu$ only. Energies of states with
$j=1$ differ from the Landau levels, whereas energies with $j=0$ coincide with
the Landau levels. If $\mathcal{E}_{\perp\left(  \sigma\right)  }^{2}\neq0$,
the complete set of eigenvectors $\psi_{n_{1},\,n_{2}}^{(j)}$ is given by eq.
(\ref{ab11}), where constants $c_{\sigma}$ are arbitrary, e.g., either
$c_{+1}\neq0$ or $c_{-1}\neq0$. However, if $\mathcal{E}_{\perp\left(
\sigma\right)  }^{2}=0$, a completeness of the eigenvectors implies a special
choice of $c_{\sigma}$, namely: $c_{-1}\neq0$ if $\epsilon=-1$ and $c_{+1}%
\neq0$ if $\epsilon=+1$. In this case the only negative energy solutions
(antiparticles with $cp_{0}=-M$) are possible if $\epsilon=-1$ and only
positive energy solutions (particles with $cp_{0}=+M$) are possible if
$\epsilon=+1$. They coincide with the corresponding spinors $u_{n_{1},\,n_{2}%
}^{(j)}(\varphi,\rho)$ up to a normalization constant. This is a manifestation
of the well-known asymmetry of the energy spectrum of $2+1$ Dirac particles in
the uniform magnetic field. We see that the same asymmetry takes place in the
presence of the AB-field. For particles, we select $c_{+1}\neq0$, $c_{-1}=0$,
then their energy spectrum is $\mathcal{E}=\sqrt{\left(  Mc^{2}\right)
^{2}+\mathcal{E}_{\perp\left(  +1\right)  }^{2}}$ and for antiparticles
$c_{+1}=0$, $c_{-1}\neq0$, then their energy spectrum is $\mathcal{E}%
=\sqrt{\left(  Mc^{2}\right)  ^{2}+\mathcal{E}_{\perp\left(  -1\right)  }^{2}%
}$.

If we define the inner product of spinors $\psi(\varphi,\,\rho)$
and$\,\psi^{\prime}(\varphi,\,\rho)$ as follows:
\begin{equation}
\left(  \psi,\,\psi^{\prime}\right)  _{D}=\frac{1}{2\pi}\int_{0}^{\infty}%
d\rho\int_{0}^{2\pi}d\varphi\,\psi^{\dag}(\varphi,\rho)\,\psi^{\prime}%
(\varphi,\rho), \label{Dinn}%
\end{equation}
then the inner product of eigenvectors $\psi_{n_{1},\,n_{2}}^{(j)}$ has the
form%
\begin{equation}
\left(  \psi_{n_{1}^{\prime},\,n_{2}^{\prime}}^{(j^{\prime})},\,\psi
_{n_{1},\,n_{2}}^{(j)}\right)  _{D}=2Mc\left\vert c_{\sigma}\right\vert
^{2}\left(  \sigma p_{0}+Mc\right)  \left(  \Phi_{n_{1}^{\prime}%
,\,n_{2}^{\prime},\sigma}^{(j\,^{\prime})},\Phi_{n_{1},\,n_{2},\sigma}%
^{(j)}\right)  _{\bot}\ , \label{ab15}%
\end{equation}
where the inner product $\left(  ,\right)  _{\bot}$ is given by eq.
(\ref{f23}). With respect to the introduced inner product, eigenvectors
(\ref{ab11}) form an orthogonal set for any $\vartheta$.

By the help of eigenvectors (\ref{ab11}), we obtain the following solutions of
the Dirac equation with a given energy $cp_{0}=\pm\mathcal{E}$, in $(2+1)
$-dim.:%
\begin{equation}
\Psi_{p_{0},\,m,\,\tilde{l}}^{(j)}(t,\mathbf{r})=\exp\left[  -\frac{i}{\hbar
}(cp_{0}t)\right]  \psi_{n_{1},\,n_{2}}^{(j)}(\varphi,\,\rho). \label{ab16}%
\end{equation}

We believe that nonrelativistic motion is described by the corresponding Pauli
equation in $(2+1)$-dim.,%
\begin{equation}
i\hbar\partial_{t}\Psi_{\pm}=\pm\hat{H}_{\pm}^{\mathrm{NR}}\Psi_{\pm}%
,\ \ \hat{H}_{\pm}^{\mathrm{NR}}=\left(
\mbox{\boldmath$\sigma$\unboldmath}\mathbf{\hat{P}}_{\perp}\right)  ^{2}/2M.
\label{Pauli}%
\end{equation}
Solutions of such an equation can be obtained from (\ref{ab16}) in the
nonrelativistic limit. They have only one component, taking into account that
in $(2+1)$-dim. there is only one spin polarization. Let us consider, for
example, spin up particles ($\zeta=+1$). Then we obtain from (\ref{ab11}):%
\begin{equation}
\Psi_{\pm,\,m,\,\tilde{l}}^{(j)\mathrm{up}}(t,\mathbf{r})=\exp\left[  \mp
\frac{i}{\hbar}\mathcal{E}_{\left(  \pm\right)  }^{\mathrm{NR}}t\right]
\Phi_{n_{1},\,n_{2},\pm1}^{(j)}(\varphi,\,\rho)\upsilon_{\pm1}, \label{ab17}%
\end{equation}
where $\mathcal{E}_{\left(  \pm\right)  }^{\mathrm{NR}}\mathcal{=E}%
_{\perp\left(  \pm1\right)  }^{2}/2Mc^{2}$. The corresponding inner product of
these solutions reads%
\begin{equation}
\left(  \Psi_{\pm,\,m^{\prime},\,\tilde{l}^{\prime}}^{(j^{\prime})\mathrm{up}%
},\,\Psi_{\pm,\,m,\,\tilde{l}}^{(j)\mathrm{up}}\right)  _{D}=\left(
\Phi_{n_{1}^{\prime},\,n_{2}^{\prime},\pm1}^{(j\,^{\prime})},\Phi
_{n_{1},\,n_{2},\pm1}^{(j)}\right)  _{\bot}. \label{ab18}%
\end{equation}
By the help of relation (\ref{abe8}), we obtain solutions that describe spin
down\ particles:%
\begin{align}
&  \Psi_{\pm,\,m,\,\tilde{l}}^{(j)\mathrm{down}}(t,\mathbf{r})=\exp\left[
\mp\frac{i}{\hbar}\mathcal{E}_{\left(  \mp\right)  }^{\mathrm{NR}}t\right]
\Phi_{n_{1},\,n_{2},\mp1}^{(j)}(\varphi,\,\rho)\upsilon_{\mp1}\ ,\nonumber\\
&  \ \left(  \Psi_{\pm,\,m^{\prime},\,\tilde{l}^{\prime}}^{(j^{\prime
})\mathrm{down}},\,\Psi_{\pm,\,m,\,\tilde{l}}^{(j)\mathrm{down}}\right)
_{D}=\left(  \Phi_{n_{1}^{\prime},\,n_{2}^{\prime},\mp1}^{(j\,^{\prime})}%
,\Phi_{n_{1},\,n_{2},\mp1}^{(j)}\right)  _{\bot}. \label{ab19}%
\end{align}

It is worthwhile to make the following remark: Usually, in the nonrelativistic
limit, negative energy solutions (those which after the charge conjugation
operation represent wave functions of particles with opposite charge $-q,$ and
which are, in fact, antiparticles) of the Dirac equation are not considered.
It is supposed that all the information about the quantum motion of the
antiparticles can be extracted from particle motion. The latter is not true in
the case under consideration (for Dirac eq. with MSF in ($2+1$)-dim.). Here
energy spectra for particles and antiparticles are quite different. This
asymmetry was the reason for us to represent explicitly, even in the
nonrelativistic limit, the negative energy solutions $\Psi_{-,\,m,\,\tilde{l}%
}^{(j)\mathrm{up}}(t,\mathbf{r})$ and $\Psi_{-,\,m,\,\tilde{l}}%
^{(j)\mathrm{down}}(t,\mathbf{r})$, which correspond to spin up and down antiparticles.

For massless $\zeta=+1$ particles and antiparticles (in what follows we call
all such particles simply fermions) in $(2+1)$-dim., the self-adjoint Dirac
Hamiltonian is $\hat{H}^{\vartheta}%
=c\mbox{\boldmath$\sigma$\unboldmath}\mathbf{\hat{P}}_{\perp}$. Its
eigenvalues are $cp_{0}=\pm\mathcal{E}$ , where $\mathcal{E}=\mathcal{E}%
_{\perp\left(  -\vartheta\right)  }$ (the eigenvalues $\mathcal{E}%
_{\perp\left(  \sigma\right)  }$ are given by eq. (\ref{ab14})). The
corresponding eigenvectors of $\hat{H}^{\vartheta}$ have the form:%
\begin{align}
&  \Psi_{\pm,\,m,\,\tilde{l}}^{(j,+1)}(t,\mathbf{r})=\exp\left[  \mp\frac
{i}{\hbar}(\mathcal{E}t)\right]  u_{\pm,n_{1},\,n_{2}}^{(j,+1)}(\varphi
,\rho),\nonumber\\
&  u_{\pm,n_{1},\,n_{2}}^{(0,+1)}(\varphi,\rho)=\binom{\Phi_{n_{1}+\left(
1+\epsilon\right)  /2,\,n_{2},+1}^{(0)}(\varphi,\,\rho)}{\pm i\epsilon
\Phi_{n_{1}+\left(  1-\epsilon\right)  /2,\,n_{2},-1}^{(0)}(\varphi,\,\rho
)},\nonumber\\
&  u_{\pm,n_{1},\,n_{2}}^{(1,+1)}(\varphi,\rho)=\binom{\Phi_{n_{1},\,n_{2}%
,+1}^{(1)}(\varphi,\,\rho)}{\mp i\epsilon\Phi_{n_{1},\,n_{2},-1}^{(1)}%
(\varphi,\,\rho)},\mathrm{\;}\mathcal{E}\neq0, \label{ab21}%
\end{align}
where $\Phi_{n_{1},\,n_{2},\sigma}^{(j)}$ are given by eqs. (\ref{ab12}). The
inner product of such solutions reads%
\begin{equation}
\left(  \Psi_{\pm,\,m^{\prime},\,\tilde{l}^{\prime}}^{(j^{\prime},+1)}%
,\Psi_{\pm,\,m,\,\tilde{l}}^{(j,+1)}\right)  _{D}=\sum_{\sigma=\pm1}\left(
\Phi_{n_{1}^{\prime}+\left(  1-j^{\prime}\right)  \left(  1+\sigma
\epsilon\right)  /2,n_{2}^{\prime},\sigma}^{(j^{\prime})},\Phi_{n_{1}+\left(
1-j\right)  \left(  1+\sigma\epsilon\right)  /2,n_{2},\sigma}^{(j)}\right)
_{\bot}. \label{ab23}%
\end{equation}

In addition, there are nontrivial zero-mode ($\mathcal{E}=0$) solutions:
\begin{align}
&  u_{0,n_{1},\,n_{2}}^{(0,+1)}(\varphi,\rho)=c_{\epsilon}\left.  \Phi
_{n_{1},\,n_{2},\epsilon}^{(0)}(\varphi,\,\rho)\right\vert _{m=0}%
\upsilon_{\epsilon}\,,\nonumber\\
&  \left(  u_{0,n_{1}^{\prime},\,n_{2}^{\prime}}^{(0,+1)},u_{0,n_{1},\,n_{2}%
}^{(0,+1)}\right)  _{D}=\left.  \left(  \Phi_{n_{1}^{\prime},\,n_{2}^{\prime
},\epsilon}^{(0)},\Phi_{n_{1},\,n_{2},\epsilon}^{(0)}\right)  _{\bot
}\right\vert _{m=0}\ . \label{ab22}%
\end{align}

As follows from (\ref{abe8}), for massless $\zeta=-1$ particles the
corresponding eigenvectors can be represented as
\begin{align}
&  \Psi_{\pm,\,m,\,\tilde{l}}^{(j,-1)}(t,\mathbf{r})=\exp\left[  \mp\frac
{i}{\hbar}\mathcal{E}t\right]  \sigma^{2}u_{\mp,n_{1},\,n_{2}}^{(j,+1)}%
(\varphi,\rho),\nonumber\\
&  u_{0,n_{1},\,n_{2}}^{(0,-1)}(\varphi,\rho)=\sigma^{2}u_{0,n_{1},\,n_{2}%
}^{(0,+1)}(\varphi,\rho). \label{ab24}%
\end{align}
The inner products of these solutions\textrm{\ }coincide with ones of
solutions $\Psi_{\pm,\,m,\,\tilde{l}}^{(j,+1)}$\ \ given by eq. (\ref{ab23}):
\[
\left(  \Psi_{\pm,\,m^{\prime},\,\tilde{l}^{\prime}}^{(j^{\prime},-1)}%
,\,\Psi_{\pm,\,m,\,\tilde{l}}^{(j,-1)}\right)  _{D}=\left(  \Psi
_{\pm,\,m^{\prime},\,\tilde{l}^{\prime}}^{(j^{\prime},+1)},\,\Psi
_{\pm,\,m,\,\tilde{l}}^{(j,+1)}\right)  _{D},\;\left(  u_{0,n_{1}^{\prime
},\,n_{2}^{\prime}}^{(0,-1)},u_{0,n_{1},\,n_{2}}^{(0,-1)}\right)  _{D}=\left(
u_{0,n_{1}^{\prime},\,n_{2}^{\prime}}^{(0,+1)},u_{0,n_{1},\,n_{2}}%
^{(0,+1)}\right)  _{D}\ .
\]

\subsubsection{$\left(  3+1\right)  $-dimensions}

Here we consider Dirac equation (\ref{abe1}) in $\left(  3+1\right)  $-dim.
Let us introduce projection operators $\hat{P}_{(\pm)}$ and two kinds of Dirac
bispinor $\Psi_{(\pm)},$
\begin{equation}
\mathcal{\hat{P}}_{(\pm)}=\left(  1\pm\alpha^{3}\right)  /2,\ (\mathcal{\hat
{P}}_{(\pm)})^{\dag}=\mathcal{\hat{P}}_{(\pm)},\ (\mathcal{\hat{P}}_{(\pm
)})^{2}=\mathcal{\hat{P}}_{(\pm)},\ \mathcal{\hat{P}}_{(+)}\mathcal{\hat{P}%
}_{(-)}=0,\ \mathcal{\hat{P}}_{(+)}+\mathcal{\hat{P}}_{(-)}=\mathbb{I}%
,\nonumber
\end{equation}
where $\mathbb{I}$ is unit $4\times4$ matrix,\ such that any $\Psi$ can be
represented as $\Psi=\Psi_{(+)}+\Psi_{(-)},\ \Psi_{(\pm)}=\mathcal{\hat{P}%
}_{(\pm)}\Psi.$ Then Dirac equation (\ref{abe1}) is reduced to the following
set of equations%
\begin{align}
&  (\hat{P}_{0}+\hat{P}_{3})\Psi_{(+)}=\mathcal{\hat{Q}}\Psi_{(-)},\ (\hat
{P}_{0}-\hat{P}_{3})\Psi_{(-)}=\mathcal{\hat{Q}}\Psi_{(+)};\nonumber\\
&  \mathcal{\hat{Q}}=({\mbox{\boldmath$\alpha$}}_{\perp}\mathbf{\hat{P}%
}_{\perp})+Mc\gamma^{0},\;\hat{P}_{0}=i\hbar\partial_{0}, \label{k3}%
\end{align}
where $\alpha^{i}=\gamma^{0}\gamma^{i}$. Due to the axial symmetry of the
problem, it is convenient to use the following representation for $\gamma
$-matrices (see \cite{AMW89}),
\begin{equation}
\gamma^{0}=\mathrm{diag}\left(  \sigma^{3},-\sigma^{3}\right)  ,\;\gamma
^{1}=\mathrm{diag}\left(  i\sigma^{2},-i\sigma^{2}\right)  ,\;\gamma
^{2}=\mathrm{diag}\left(  -i\sigma^{1},i\sigma^{1}\right)  ,\;\gamma
^{3}=\mathrm{antidiag}\left(  -I,I\right)  \,, \label{rep}%
\end{equation}
where $I$ is unit $2\times2$ matrix. Nevertheless, expressions for $\alpha
^{3}$ ,$\Sigma_{z}$, and $\gamma^{5}$ are the same in the representation
(\ref{rep}) and in the standard representation.

One can see that
\begin{equation}
\mathcal{\hat{Q}}^{2}=M^{2}c^{2}+\mathcal{\hat{Q}}_{\perp}^{2},\;\mathcal{\hat
{Q}}_{\perp}^{2}=({\mbox{\boldmath$\alpha$}}_{\perp}\mathbf{\hat{P}}_{\perp
})^{2}=\mathbf{\hat{P}}_{\perp}^{2}-\epsilon\hbar c^{-1}\left\vert
qB\right\vert \Sigma_{z}\ , \label{k5}%
\end{equation}
where $\Sigma_{z}=\mathrm{diag}\left(  \sigma^{3},\sigma^{3}\right)  $. In the
MSF, the operators $\hat{P}_{0}+\hat{P}_{3}$, $\hat{P}_{0}-\hat{P}_{3}$, and
$\hat{Q}$ mutually commute, such that (as it follows from (\ref{k3}))
bispinors $\Psi_{(\pm)}$\ obey the same equations:%
\begin{equation}
(\hat{P}_{0}^{2}-\hat{P}_{3}^{2}-\mathcal{\hat{Q}}^{2})\Psi_{(\pm)}(x)=0.
\label{k4}%
\end{equation}
Representing $\Psi_{(\pm)}$ via spinors $\psi$ and $\chi,$%
\begin{equation}
\Psi_{(-)}=\frac{1}{2}\left(
\begin{array}
[c]{c}%
\psi\\
-\sigma^{3}\psi
\end{array}
\right)  ,\;\Psi_{(+)}=\frac{1}{2}\left(
\begin{array}
[c]{c}%
\chi\\
\sigma^{3}\chi
\end{array}
\right)  , \label{k8}%
\end{equation}
we find the following equations for the spinors:
\begin{align}
&  \left[  \hat{P}_{0}^{2}-\hat{P}_{3}^{2}-\left(  \mathbf{\hat{P}}_{\perp
}^{2}-\epsilon\left\vert qB\right\vert \frac{\hbar}{c}\sigma^{3}\right)
-M^{2}c^{2}\right]  \psi(x)=0,\label{k9}\\
&  (\hat{P}_{0}+\hat{P}_{3})\chi(x)=\left(  \sigma^{3}%
\mbox{\boldmath$\sigma$\unboldmath}\mathbf{\hat{P}}_{\perp}+Mc\right)
\psi(x). \label{k12}%
\end{align}
We note that both solutions $\Psi_{(-)}$ and $\Psi_{(+)}$ enter into a
complete set of functions on the hypersurface $t=\mathrm{const}$.

The inner product of Dirac bispinors on the light-cone hypersurface
$x_{-}=\mathrm{const}$ has the form:%
\begin{equation}
\left(  \Psi,\Psi^{\prime}\right)  _{x_{-}}=\int\Psi^{\dag}\mathcal{\hat{P}%
}_{(-)}\Psi^{\prime}dx_{+}dx^{1}dx^{2}=\frac{2\pi}{\gamma}\int\left(
\Psi_{(-)},\,\Psi_{(-)}^{\prime}\right)  _{D}^{\perp}dx_{+}, \label{ip}%
\end{equation}
see \cite{31}, where the inner product of four-component spinors $\Psi$
and$\,\Psi^{\prime}$ on $xy$-plane is defined as%
\begin{equation}
\left(  \Psi,\,\Psi^{\prime}\right)  _{D}^{\perp}=\frac{1}{2\pi}\int%
_{0}^{\infty}d\rho\int_{0}^{2\pi}d\varphi\,\Psi^{\dag}(\varphi,\rho
)\,\Psi^{\prime}(\varphi,\rho)\ . \label{ipxy}%
\end{equation}
It is expressed only in terms of the components $\Psi_{(-)}$.\textrm{\ }At the
same time, a complete set of functions on the hypersurface{\Large \ }%
$x_{-}=\mathrm{const}$ consists only of $\Psi_{(-)}$.

In the case under consideration, the operators $\hat{P}_{0}$, $\,\hat{P}_{3}$,
$\hat{J}_{z}=\hat{L}_{z}+\Sigma_{z}/2$, and a spin operator $\hat{S}_{z}$
($z$-component of a polarization pseudovector)
\begin{equation}
\hat{S}_{z}=\frac{1}{2}\left(  \hat{H}^{\vartheta}\Sigma_{z}+\Sigma_{z}\hat
{H}^{\vartheta}\right)  =\gamma^{0}\Sigma_{z}Mc^{2}-\gamma^{5}c\hat{P}^{3},
\label{k13a}%
\end{equation}
are mutually commuting integrals of motion (all these operators commute with
the Hamiltonian $\hat{H}^{\vartheta}$) \cite{205,212}. In addition, the set
$\hat{P}_{0}$, $\,\hat{P}_{3}$, $\hat{J}_{z}$, $\Sigma_{z}$, and
$\mathcal{\hat{Q}}_{\perp}^{2}$ represents mutually commuting operators,
which, at the same time, commute with $\alpha^{3}$. This fact allows one to
find solutions $\Psi_{(-)}$ that are eigenvectors for the latter set. To this
end one has to subject spinors $\psi$ to the following equations:%
\begin{align}
&  \left(  \hat{P}_{0}+\hat{P}_{3}\right)  \psi(x)=\lambda Mc\psi
(x),\ \ \hat{J}_{z}\psi(x)=J_{z}\psi(x),\nonumber\\
&  c^{2}\left(  \mbox{\boldmath$\sigma$\unboldmath}\mathbf{\hat{P}}_{\perp
}\right)  ^{2}\psi(x)=\mathcal{E}_{\perp\left(  \sigma\right)  }^{2}%
\psi(x),\ \ \sigma^{3}\psi(x)=\sigma\psi(x), \label{k14}%
\end{align}
where $\mathcal{E}_{\perp\left(  \sigma\right)  }^{2}$ is given by
(\ref{ab14}). Thus, we obtain for $\Psi_{(-)}:$%
\begin{align}
&  \left(  \hat{P}_{0}+\hat{P}_{3}\right)  \Psi_{(-)}=\lambda Mc\Psi
_{(-)},\ \ \hat{J}_{z}\Psi_{(-)}=J_{z}\Psi_{(-)},\ \ J_{z}=\epsilon\hbar
(l_{0}-l+1/2),\nonumber\\
&  c^{2}\mathcal{\hat{Q}}_{\perp}^{2}\Psi_{(-)}=\mathcal{E}_{\perp\left(
\sigma\right)  }^{2}\Psi_{(-)},\ \ \Sigma_{z}\Psi_{(-)}=\sigma\Psi
_{(-)},\ \ \sigma=\pm1. \label{k15}%
\end{align}
In the light cone variables (\ref{b17}), we have the following representation
\begin{equation}
\hat{P}_{0}+\hat{P}_{3}=2i\hbar\frac{\partial}{\partial x_{+}},\ \ \hat{P}%
_{0}-\hat{P}_{3}=2i\hbar\frac{\partial}{\partial x_{-}}\ . \label{d11}%
\end{equation}
Then, we can represent eq. (\ref{k4}) for eigenfunctions of $\hat{P}_{0}%
+\hat{P}_{3}$ with the eigenvalues $\lambda$ in the form of the first order
Schrödinger-like equation
\begin{equation}
\left[  2i\hbar\lambda Mc\frac{\partial}{\partial x_{-}}-\mathcal{\hat{Q}}%
^{2}\right]  \Psi_{(-)\lambda}(x)=0, \label{Sch-like}%
\end{equation}
and we find a complete set of solutions $\Psi_{(-)}$ in the following form:
\begin{align}
&  \Psi_{(-)\lambda,\,m,\,\tilde{l},\sigma}^{(j)}=\exp\left\{  -\frac
{i}{2\hbar}\left[  \lambda Mcx_{+}+\left(  \frac{Mc}{\lambda}+\hbar
\widetilde{\omega}\left(  1-\sigma\epsilon\right)  \right)  x_{-}\right]
\right. \nonumber\\
&  \left.  -i\widetilde{\omega}n_{1}x_{-}\right\}  \Phi_{n_{1},\,n_{2},\sigma
}^{(j)}(\varphi,\,\rho)\left(
\begin{array}
[c]{c}%
\upsilon_{\sigma}\\
-\sigma\upsilon_{\sigma}%
\end{array}
\right)  , \label{k18}%
\end{align}
where $\widetilde{\omega}$ is given by (\ref{b20}), $\upsilon_{\sigma}$ by
eqs. (\ref{t-s}), $\Phi_{n_{1},\,n_{2},\sigma}^{(j)}$ by \ (\ref{ab12}), and
$\lambda>0$ for particles and $\lambda<0$ for antiparticles. We note that the
quantum number $\lambda$ is associated with the corresponding classical
quantity $\lambda$ from (\ref{b18}).

We note that the spin integral of motion $\hat{S}_{z}$ does not commute with
$\alpha^{3}$ such that solutions (\ref{k18}) are not eigenvectors of $\hat
{S}_{z}$. One can use the operator $\Sigma_{z}$ instead of $\hat{S}_{z}$ to
characterize the spin polarization. In spite of the fact that $\left[
\Sigma_{z},\hat{H}\right]  \neq0$ and, therefore, $\Sigma_{z}$ is not an
integral of motion with respect to the $t$-evolution, $\Sigma_{z}$ is an
integral of motion with respect of the evolution in the light-cone "time"
$x_{-}$. That is why the "spin polarization" $\sigma$ of solutions (\ref{k18})
is conserved with the time $x_{-}$. Taking all this into account, one can
calculate the light-cone inner product (\ref{ip}) of solutions (\ref{k18}):%
\begin{equation}
\left(  \Psi_{\left(  -\right)  \lambda^{\prime},\,m^{\prime},\,\tilde
{l}^{\prime},\sigma^{\prime}}^{(j^{\prime})},\Psi_{\left(  -\right)
\lambda,\,m,\,\tilde{l},\sigma}^{(j)}\right)  _{x_{-}}=\frac{\left(
4\pi\right)  ^{2}\hbar}{\gamma Mc}\delta_{\sigma^{\prime},\sigma}\delta\left(
\lambda^{\prime}-\lambda\right)  \left(  \Phi_{n_{1}^{\prime},\,n_{2}^{\prime
},\sigma}^{(j\,^{\prime})},\Phi_{n_{1},\,n_{2},\sigma}^{(j)}\right)  _{\bot},
\label{k19}%
\end{equation}
where the inner product $\left(  ,\right)  _{\bot}$ is given by eq. (\ref{f23}).

Let us consider the quantum motion of spinning particles in the
nonrelativistic limit. To this end it is more convenient, instead of solutions
(\ref{k18}), to use another set of solution $\Psi_{s}(x),$%
\begin{align}
&  \Psi_{s}(x)=\exp\left[  -\frac{i}{\hbar}(cp_{0}t+p_{3}z)\right]  \Psi
_{s}(x_{\perp})\,,\ s=\pm1,\nonumber\\
&  \Psi_{s}(x_{\perp})=N\left(
\begin{array}
[c]{c}%
\left[  1+\left(  p^{3}/c+s\widetilde{M}\right)  /M\right]  \psi_{p_{0}%
,s}(x_{\perp})\\
\left[  -1+\left(  p^{3}/c+s\widetilde{M}\right)  /M\right]  \psi_{p_{0}%
,s}(x_{\perp})
\end{array}
\right)  \,.\label{k20}%
\end{align}
where $\widetilde{M}=\sqrt{M^{2}+(p_{3})^{2}}$. These solutions are
eigenvectors of mutually commuting integrals of motion $\hat{P}_{0}$,
$\,\hat{P}_{3}$, $\hat{J}_{z}$, and $\hat{S}_{z}$,%
\begin{align}
&  \hat{P}_{0}\Psi_{s}(x)=p_{0}\Psi_{s}(x),\ \ \hat{P}_{3}\Psi_{s}%
(x)=p_{3}\Psi_{s}(x),\nonumber\\
&  \hat{J}_{z}\Psi_{s}(x)=J_{z}\Psi_{s}(x),\ \ J_{z}=\epsilon\hbar
(l_{0}-l+1/2),\ \ \hat{S}_{z}\Psi_{s}=s\widetilde{M}c^{2}\Psi_{s}%
,\;\label{k21}%
\end{align}
The spinors $\psi_{p_{0},s}(x_{\perp})$\ obey the equation
\begin{equation}
\left(  \mathbf{\mbox{\boldmath$\sigma$\unboldmath}P}_{\perp}+s\widetilde{M}%
c\sigma^{3}\right)  \psi_{p_{0},s}(x_{\perp})=p_{0}\psi_{p_{0},s}(x_{\perp
}).\label{k22}%
\end{equation}
One can see that at fixed $s$ and $p^{3}$, eq. (\ref{k22}) is similar to eq.
(\ref{abe7a}) in ($2+1)$-dim. such that its solutions will be used in what follows.

In the nonrelativistic limit, the spin operator $\hat{S}_{z}$ is reduced to
$\hat{S}_{z}^{\mathrm{NR}}=\gamma^{0}\Sigma_{z}Mc^{2}$ and $\widetilde{M}=M.$
Then for $s=+1$ eq. (\ref{k22}) coincides with eq. (\ref{abe7a}). We remark
that $\psi_{p_{0},-1}(x_{\perp})=\sigma^{3}\psi_{-p_{0},1}(x_{\perp})\,$. As a
result, we obtain wave functions of nonrelativistic spinning particles from
eq. (\ref{k20}):%
\begin{align}
\Psi_{\pm,p_{3},m,\,\tilde{l},s}^{(j)\mathrm{NR}}(x)  &  =\exp\left\{
-\frac{i}{\hbar}\left[  \pm\frac{\left(  p_{3}\right)  ^{2}t}{2M}%
+p_{3}z\right]  \right\}  \Psi_{\pm,m,\,\tilde{l},s}^{(j)\mathrm{NR}%
}(t,\mathbf{r}),\nonumber\\
\Psi_{\pm,m,\,\tilde{l},+1}^{(j)\mathrm{NR}}(t,\mathbf{r})  &  =\binom
{\Psi_{\pm,\,m,\,\tilde{l}}^{(j)\mathrm{up}}(t,\mathbf{r})}{0},\;\Psi
_{\pm,m,\,\tilde{l},-1}^{(j)\mathrm{NR}}(t,\mathbf{r})=\binom{0}{\Psi
_{\pm,\,m,\,\tilde{l}}^{(j)\mathrm{down}}(t,\mathbf{r})}, \label{k23}%
\end{align}
where spinors $\Psi_{\pm,\,m,\,\tilde{l}}^{(j)\mathrm{up}}$ and $\Psi
_{\pm,\,m,\,\tilde{l}}^{(j)\mathrm{down}}$ are respectively solution
(\ref{ab17}) and (\ref{ab19}) of the Pauli equation in $(2+1)$-dimensions with
MSF. Thus, wave functions of nonrelativistic spinning particles
(antiparticles) in $(3+1)$-dimensions obey the nonrelativistic Dirac equation
with the Hamiltonian $\hat{H}_{\pm}^{\mathrm{NR}}=\left(  \mathcal{\hat{Q}%
}_{\perp}^{2}+\hat{P}_{3}^{2}\right)  /2M$.

\section*{Appendix B. Mean ${\overline{(a_{1}\left(  t\right)  )}}_{(j,\pm
)}^{\ast}$}

To study the mean ${\overline{(a_{1}\left(  t\right)  )}}_{(j,\pm)}^{\ast},$
one has to calculate the matrix element $\left(  \Psi_{\pm,\,z_{1},z_{2}%
}^{(j,+1)}\left(  t\right)  ,\hat{a}_{1}^{\dag}\Psi_{\pm,\,z_{1},z_{2}%
}^{(j,+1)}\left(  t\right)  \right)  _{D}$. The latter can be reduced to a
matrix element with respect to the initial ICS $\mathbf{\Psi}_{\pm
,\,z_{1},z_{2}}^{(j,+1)}(\varphi,\,\rho)$ as follows:%
\begin{equation}
\left(  \Psi_{\pm,\,z_{1},z_{2}}^{(j,+1)}\left(  t\right)  ,\hat{a}_{1}^{\dag
}\Psi_{\pm,\,z_{1},z_{2}}^{(j,+1)}\left(  t\right)  \right)  _{D}=\left(
\mathbf{\Psi}_{\pm,\,z_{1},z_{2}}^{(j,+1)},\hat{a}_{1}^{\dag}\left(  \pm
t\right)  \mathbf{\Psi}_{\pm,\,z_{1},z_{2}}^{(j,+1)}\right)  _{D}, \label{rp5}%
\end{equation}
where%
\begin{equation}
\hat{a}_{1}^{\dag}\left(  \pm t\right)  =\exp\left[  \pm\frac{ic}{\hbar}%
\hat{\Pi}_{0}t\right]  \hat{a}_{1}^{\dag}\exp\left[  \mp\frac{ic}{\hbar}%
\hat{\Pi}_{0}t\right]  . \label{rp6}%
\end{equation}
The operator $\hat{a}_{1}^{\dag}\left(  t\right)  $ obeys the equation%
\begin{equation}
\frac{d\hat{a}_{1}^{\dag}\left(  t\right)  }{dt}=-\frac{ic}{\hbar}\left[
\hat{a}_{1}^{\dag}\left(  t\right)  ,\hat{\Pi}_{0}\right]  . \label{rp7}%
\end{equation}
Let us consider the commutator $\left[  \hat{a}_{1}^{\dag}\left(  t\right)
,\hat{\Pi}_{0}\right]  .$ First, we write%
\begin{align}
&  \left[  \hat{a}_{1}^{\dag}\left(  t\right)  ,\hat{\Pi}_{0}\right]  =\left[
\hat{a}_{1}^{\dag}\left(  t\right)  ,\hat{\Pi}_{0}^{2}\right]  \hat{\Pi}%
_{0}^{-1}+\hat{\Pi}_{0}^{2}\left[  \hat{a}_{1}^{\dag}\left(  t\right)
,\hat{\Pi}_{0}^{-1}\right]  ,\nonumber\\
&  \hat{\Pi}_{0}^{-1}=\frac{2}{\sqrt{\pi}}\int_{0}^{\infty}e^{-\hat{\Pi}%
_{0}^{2}\tau^{2}}d\tau, \label{rp7a}%
\end{align}
where the identity $\left[  \hat{a}_{1}^{\dag}\left(  t\right)  ,\hat{\Pi}%
_{0}\right]  =\left[  \hat{a}_{1}^{\dag}\left(  t\right)  ,\hat{\Pi}_{0}%
^{2}\hat{\Pi}_{0}^{-1}\right]  $ is used. Then, we represent the commutator
$\left[  \hat{a}_{1}^{\dag}\left(  t\right)  ,\hat{\Pi}_{0}^{-1}\right]  $ as
follows:%
\begin{align}
&  \left[  \hat{a}_{1}^{\dag}\left(  t\right)  ,\hat{\Pi}_{0}^{-1}\right]
=\frac{2}{\sqrt{\pi}}\int_{0}^{\infty}\left(  \hat{a}_{1}^{\dag}\left(
t\right)  -\hat{b}_{1}^{\dag}\left(  \tau^{2}\right)  \right)  e^{-\hat{\Pi
}_{0}^{2}\tau^{2}}d\tau,\nonumber\\
&  \hat{b}_{1}^{\dag}\left(  \tau^{2}\right)  =e^{-\hat{\Pi}_{0}^{2}\tau^{2}%
}\hat{a}_{1}^{\dag}\left(  t\right)  e^{+\hat{\Pi}_{0}^{2}\tau^{2}}.
\label{rp9}%
\end{align}
The operator $\hat{b}_{1}^{\dag}\left(  \tau^{2}\right)  $ obeys the equation%
\begin{equation}
\frac{d\hat{b}_{1}^{\dag}\left(  \tau^{2}\right)  }{d\tau^{2}}=\left[  \hat
{b}_{1}^{\dag}\left(  \tau^{2}\right)  ,\hat{\Pi}_{0}^{2}\right]
=-\frac{2\hbar\left\vert qB\right\vert }{c}\hat{b}_{1}^{\dag}\left(  \tau
^{2}\right)  , \label{sol2}%
\end{equation}
and coincides with $\hat{a}_{1}^{\dag}\left(  t\right)  $ at $\tau^{2}=0$.
Such a solution of eq. (\ref{sol2}) reads:%
\begin{equation}
\hat{b}_{1}^{\dag}\left(  \tau^{2}\right)  =\hat{a}_{1}^{\dag}\left(
t\right)  \exp\left(  -2\hbar\left\vert qB\right\vert \tau^{2}/c\right)  .
\label{sol1}%
\end{equation}
Substituting (\ref{sol1}) into (\ref{rp9}) and calculating the integral, we
obtain:%
\[
\left[  \hat{a}_{1}^{\dag}\left(  t\right)  ,\hat{\Pi}_{0}^{-1}\right]
=\hat{a}_{1}^{\dag}\left(  t\right)  \left[  \hat{\Pi}_{0}^{-1}-\left(
\hat{\Pi}_{0}^{2}+2\hbar\left\vert qB\right\vert /c\right)  ^{-1/2}\right]  .
\]
Using this result in (\ref{rp7a}), we find%
\[
\left[  \hat{a}_{1}^{\dag}\left(  t\right)  ,\hat{\Pi}_{0}\right]  =\hat
{a}_{1}^{\dag}\left(  t\right)  \left(  \hat{\Pi}_{0}-\sqrt{\hat{\Pi}_{0}%
^{2}+2\hbar\left\vert qB\right\vert /c}\right)  ,
\]
such that eq. (\ref{rp7}) has the following solution:%
\begin{equation}
\hat{a}_{1}^{\dag}\left(  t\right)  =\hat{a}_{1}^{\dag}e^{i\Omega\left(
\hat{N}_{1}\right)  t},\;\Omega\left(  \hat{N}_{1}\right)  =2\left\vert
qB\right\vert \left(  \sqrt{\hat{\Pi}_{0}^{2}+2\hbar\left\vert qB\right\vert
/c}+\hat{\Pi}_{0}\right)  ^{-1}, \label{rp10}%
\end{equation}
where $\Omega\left(  \hat{N}_{1}\right)  $ can be interpreted as the frequency
operator. With account taken of (\ref{rp10}) in eqs. (\ref{rp5}), and using
relations (\ref{s16}) and (\ref{s20}), we finally obtain expression
(\ref{rp11}):

\end{document}